\documentstyle[psfig]{aa}

\def\ltsima{$\; \buildrel < \over \sim \;$}
\def\simlt{\lower.5ex\hbox{\ltsima}}
\def\gtsima{$\; \buildrel > \over \sim \;$}
\def\simgt{\lower.5ex\hbox{\gtsima}}
\def\gsimeq
{\hbox{\raise0.5ex\hbox{$>\lower1.06ex\hbox{$\kern-1.07em{\sim}$}$}}}
\def\lsimeq
{\hbox{\raise0.5ex\hbox{$<\lower1.06ex\hbox{$\kern-1.07em{\sim}$}$}}}
\def\pn{\par\noindent}

\def\cha{{\it Chandra}~}
\def\xmm{{\it XMM--Newton}~}
\def\aap{{A\&A}}

\def\aj{{AJ}}

\def\apj{{ApJ}}
\def\apjl{{ApJ}}
\def\apjs{{ApJS}}
\def\pasj{{PASJ}}

\def\mnras{{MNRAS}}

\begin{document}

%\thesaurus{03(13.25.2; 11.19.1; 11.09.1: Mkn3)}
%(03) 13.25.2 -- 11.09.1 -- 11.19.1}

%  \title{X-ray Spectral Properties of a Complete and Distance-Limited \\
%Sample of Seyfert Galaxies observed with XMM--Newton}

\title{X-ray Spectral Survey with XMM--Newton of a Complete Sample of Nearby 
Seyfert Galaxies\footnote{}}

\author{M. Cappi\inst{1}\and
F. Panessa\inst{2}\and
L. Bassani\inst{1}\and
M. Dadina\inst{1}\and
G. DiCocco\inst{1}\and
A. Comastri\inst{3}\and
R. Della Ceca\inst{4}\and
A. V. Filippenko\inst{5}\and
F. Gianotti\inst{1}\and
L. C. Ho\inst{6}\and
G. Malaguti\inst{1}\and
J. S. Mulchaey\inst{6}\and
G. G. C. Palumbo\inst{7}\and
E. Piconcelli\inst{8}\and
W. L. W. Sargent\inst{9}\and
J. Stephen\inst{1}\and 
M. Trifoglio\inst{1}\and
K. A. Weaver\inst{10}}

\institute{
INAF-IASF Sezione di Bologna, Via Gobetti 101, I-40129 Bologna, Italy
\and
Instituto de Fisica de Cantabria (CSIC-UC), Avda de los Castros, 39005, 
Santander, Spain
\and
INAF- Osservatorio Astronomico di Bologna, Via Ranzani 1, I-40127 Bologna, Italy
\and
INAF- Osservatorio Astronomico di Brera, Via Brera 28, I-20121 Milano, Italy
\and
Department of Astronomy, University of California, Berkeley, CA 94720-3411, USA
\and
Carnegie Observatories, 813 Santa Barbara Street, Pasadena, CA 91101, USA
\and
Dipartimento di Astronomia, Universita' degli Studi di Bologna, Via Ranzani 1, I40127
Bologna, Italy
\and
XMM--Newton Science Operation Center/RSSD-ESA, Apartado 50727, E-28080 Madrid, Spain
\and
Department of Astronomy, California Institute of Technology, Pasadena, CA 91125, USA
\and
Laboratory for High Energy Astrophysics, NASA's Goddard Space Flight Center, 
Greenbelt, MD 20771, USA
}
\offprints{M. Cappi; e-mail:{\tt cappi@bo.iasf.cnr.it} 
\pn \Large $^{\star}$ Appendix A and B are only available in electronic 
form at http://www.edpsciences.org}

\date{Received / Accepted}

\markboth{Cappi et al.: X-Ray survey of nearby Seyfert galaxies}{}
\titlerunning{X-ray survey of nearby Seyfert galaxies with XMM--Newton}
%\authorrunning{M. Cappi et al.}

\abstract{
Results obtained from an X-ray spectral survey of nearby Seyfert galaxies 
using {\it XMM--Newton} are reported. 
The sample was optically selected, well defined, complete in $B$ magnitude, and 
distance limited: it consists of the nearest ($D \lsimeq 22$ Mpc) 27 Seyfert 
galaxies (9 of type 1, 18 of type 2) taken from the Ho et al. (1997) sample. 
This is one of the largest atlases of hard X-ray spectra of low-luminosity 
active galaxies ever assembled. All nuclear sources except two Seyfert
2s are detected between 2 and 10 keV, half for the first time ever, and average spectra 
are obtained for all of them. Nuclear luminosities reach values down to 10$^{38}$ 
erg~s$^{-1}$. The shape of the distribution of X-ray parameters is affected by the 
presence of Compton-thick objects ($\gsimeq$ 30\% among type 2s). 
The latter have been identified either directly from their intense FeK line and 
flat X-ray spectra, or indirectly with flux diagnostic diagrams which use isotropic 
indicators. After taking into account these highly absorbed sources, we find that 
(i) the intrinsic X-ray spectral properties (i.e., spectral shapes and 
luminosities above 2~keV) are consistent between type 1 and type 2 Seyferts, 
as expected from ``unified models'', (ii) Seyfert galaxies as a whole 
are distributed fairly continuously over the entire range of $N_{\rm H}$, 
between 10$^{20}$ and 10$^{25}$ cm$^{-2}$, and (iii) while Seyfert 1s 
tend to have lower $N_{\rm H}$ and Seyfert 2s tend to have the highest, 
we find 30\% and 10\% exceptions, respectively. 
Overall the sample is of sufficient quality to well represent the average intrinsic 
X-ray spectral properties of nearby active galactic nuclei, including a proper 
estimate of the distribution of their absorbing columns.
Finally, we conclude that, with the exception of a few cases, the present study agrees 
with predictions of unified models of Seyfert galaxies, and extends 
their validity down to very low luminosities. 
} 

\maketitle

\keywords{X-rays: galaxies -- Galaxies: Seyfert -- Galaxies: active}

\section{Introduction}

X-ray studies are crucial in understanding active galactic nuclei (AGNs) 
because of the unambiguous association of high-energy 
emission with genuine nuclear activity and the important diagnostics 
provided in this band for studying accretion mechanisms.

%After $ASCA$ and $BeppoSAX$ results, it has become clear that X-ray observations 
%of Seyfert galaxies 
%are key to verifying the predictions and, thus, the validity of unified models 
%of active galactic nuclei (AGNs). 

Hard X-ray selected samples of nearby Seyfert galaxies available from studies with
$GINGA$, $ASCA$, and $BeppoSAX$ (Awaki et al. 1991, Smith \& Done 1996, Turner
et al. 1997a, 1997b, 1998, Bassani et al. 1999) have been used to
(successfully) verify the validity of unified models of AGNs. These models try
to explain the observed differences between broad (Seyfert 1-like) and narrow
(Seyfert 2-like) emission-line active galaxies by invoking obscuration (from an
optically and geometrically thick torus) and viewing-angle effects rather than
intrinsic physical differences (Antonucci 1993). The largest compilations of
hard X-ray spectra available to date have, however, been severely biased toward
the most X-ray luminous, and less absorbed, AGNs.  Studies by Maiolino et
al. (1998) and Risaliti, Maiolino, \& Salvati (1999) reduced these selection
effects by applying a careful analysis for a sample of type 2 Seyferts limited
in [O~III] flux. They discovered the existence of a large fraction
($\gsimeq$50--60\%) of highly obscured AGNs at low redshifts, a result which
confirms the bias against heavily obscured systems affecting previous surveys.

%Some secondary selection effects favored also type 2 objects with either broad 
%lines visible in 
%spectropolarimetry or conical morphologies in emission-line gas to be observed 
%in X-rays, and being 
%confirmed absorbed type 1s.

With the advent of new sensitive X-ray telescopes, there is now hope of probing 
the applicability of standard accretion-disk theories down to very low
nuclear luminosities and, possibly, for AGNs with smaller black hole masses.
Recent studies performed with the last generation X-ray observatories 
have extended the X-ray spectral analysis down to 
lower luminosities, sometimes also comparing Seyferts with LINERs (Heckman
1980) and/or H~II-starburst galaxies (e.g., Terashima et al. 2002, 
Georgantopoulos et al. 2002). Snapshot surveys with 
$Chandra$ have been able to detect for the first time point-like nuclear 
sources in increasingly larger samples of nearby galaxies at very low 
luminosities (down to less than 10$^{38}$ erg~s$^{-1}$; Ho et al. 2001, 
Terashima \& Wilson 2003). Such studies seem to suggest that the standard 
unified model may not hold down to such low luminosities because 
low luminosity sources have X-ray luminosities a factor of 10 below 
the $L_{\rm X}-L_{\rm H_\alpha}$ relation for more luminous AGNs.

Moreover, deep $Chandra$ and {\it XMM-Newton} surveys indicate that the bulk 
of the X-ray background (XRB) originates at relatively low redshift ($z \approx 1$)
and is due to a combination of unobscured (type 1) and obscured (type 2) 
Seyfert galaxies (Hasinger et al. 2001, Mateos et al. 2005) as expected 
by synthesis models based on AGN unified schemes (Setti \& Woltjer 1989, 
Comastri et al. 1995, Gilli, Salvati \& Hasinger 2001). Given that a large fraction of 
the XRB is due to relativly low luminosity sources (lower than 
10$^{44}$ erg~s$^{-1}$) and that the $N_{\rm H}$ distribution is 
essentially a free parameter in AGN models, a precise knowledge of the true 
column density distribution of nearby Seyfert galaxies, especially 
in the low luminosity regime, is essential.

%Interest on understanding Seyfert galaxies locally, at both high and 
%low-luminosities, also rests 
%in the fact that Seyfert-like galaxies are probably 

With the aim of sorting out some of the questions raised by the above
arguments, we decided to perform an X-ray survey on a well defined, 
bias-free sample of Seyfert galaxies. The present sample contains 
the 27 nearest Seyfert galaxies (with $D < 22$ Mpc) of the sample 
presented by Ho, Filippenko \& Sargent (1997, HFS97a).
The X-ray survey reported here are the results obtained using the EPIC CCDs 
on-board {\it XMM-Newton}. This survey is part of a
larger program (Panessa 2004) aimed at characterizing and understanding the
multi-wavelength properties of all 60 sources with a Seyfert classification
taken from HFS97a.
 
Compared to previous studies, the strength of using {\it XMM--Newton} rests 
on two main facts: its high throughput (especially at energy $E > 2$~keV) allows 
a search for spectral components with absorption columns up to $N_{\rm H} \approx 
10^{22}$--$10^{24}$ cm$^{-2}$, and its spatial resolution (half-power radius 
$\sim7''$) minimizes any strong contamination from off-nuclear sources to the 
soft ($E < 2$~keV) and/or hard ($E \gsimeq 2$~keV) energy band.

A description of the sample, observations, and data reduction can be found
in \S~2. Section 3 summarizes the spatial and timing analysis. Spectra 
are shown in \S~4 together with a summary of the source spectral parameters 
in tabular form. Spectral properties are detailed in \S~5, and the results 
and conclusions are summarized in \S~6. A discussion of the spectral 
properties of each individual object is deferred to the Appendix. 
%We adopt a Hubble constant of $H_0 = 75$ km s$^{-1}$ Mpc$^{-1}$.

\section{The Sample, Observations, and Data Reduction}

The sample studied in this paper is drawn from 52 nearby Seyfert 
galaxies given in HFS97a (see also Ho \& Ulvestad 2001, HU01 hereinafter). Originally derived 
from the Palomar optical spectroscopic survey of nearby galaxies\footnote{The 
Palomar survey presented in Ho, Filippenko, \& Sargent (1995) 
includes high-quality spectra of 486 bright ($B_T \lsimeq 12.5$ mag), 
northern ($\delta >$0$^{\circ}$) galaxies for which a comprehensive, 
homogeneous catalog of spectral classifications have been obtained (HFS97a). 
The Palomar survey is complete to  $B_T = 12.0$ mag and 80\% complete to 
$B_T = 12.5$ mag (Sandage, Tammann, \& Yahil 1979).}
(Filippenko \& Sargent 1985, Ho, Filippenko, \& Sargent 1995), this sample 
has the advantage of having uniform and high-quality data that allowed the 
Seyfert classifications to be determined with well-defined and objective 
criteria (HFS97a). It is among the most complete and least biased samples 
of Seyfert galaxies available to date (see Appendix in HU01).

For the purpose of this study, we choose the nearest AGNs in the 
HFS97a sample --- the 30 Seyfert galaxies which are within 
a distance of 22 Mpc.
Three of these have been excluded from the present analysis: the Seyfert 2 
galaxy NGC 185 for having line intensity ratios probably produced by stellar 
processes rather than an AGN (HU01), the Seyfert 1.9 
galaxy NGC 3982 for lack of available {\it XMM--Newton} data at the time of writing, 
and the Seyfert 1.9 galaxy NGC 4168 which no longer meets the distance 
criterion\footnote{Its original distance measurement 
($D = 16.8$~Mpc; Tully 1988) has been updated recently to $D = 31.7$~Mpc 
(Caldwell, Rose, \& Concannon 2003).}.

%NGC 4579 and NGC 4395 are both reclassified here type 1 objects 
%(instead of type 1.9 and 1.8 respectively, as given in HFS97a) after 
%the detections of broad 
%permitted lines by Barth et al. (1996, 2001) and Filippenko, 
%Ho \& Sargent (1993).

Two objects, NGC 4395 and NGC 4579, have been classified by 
Ho et al. (1997b) as S1.8 and S1.9/L1.9, respectively.
However, here we reclassify these sources as type 1.5 objects.
A broad component is indeed clearly present in a number of optical 
(Filippenko \& Sargent  1989) and ultraviolet 
(Filippenko, Ho, \& Sargent 1993) emission lines of NGC 4395.
In addition, the UV spectrum of NGC 4395 resembles those of 
Seyfert 1s (Schmitt \& Kinney 1996).
Extremely broad permitted lines have been detected in NGC 4579.
The FWHM of C~IV is over 6000 km s$^{-1}$ in this object, meeting 
the criteria of a standard broad-line region (BLR), although it is 
fainter than the BLR in bright Seyfert nuclei (Barth et al. 1996, 2001).
In addition, two objects (NGC 4639, NGC 4051) have been reclassified
as Seyfert 1.5 (from 1.0 and 1.2, respectively); they have prominent
narrow-line regions (NLRs) along with their BLRs. 

The final sample consists of 27 Seyfert galaxies which include 9 type 1s
(specifically type 1.5) and 18 type 2s. (Here, the ``type 2'' category is 
defined to include type 1.9 as well, because the broad H$\alpha$ line
is very hard to detect in spectra having low signal-to-noise ratios; 
moreover, such nuclei may be largely obscured, as are many of the ``pure''
Seyfert 2s.) The sample properties are listed in Table 1. 

%The fraction of type 2 versus type 1 is 2:1 in this sample, while it is 
%3:1 in the parent sample of 52 galaxies, and is 4:1 in the subsample of 
%the nearest 10 Seyferts.

A log of all {\it XMM--Newton} observations is shown in Table 2. 
Seventeen objects were observed as part of the EPIC Guaranteed Time
observation program with exposure time ranging between 5 and 50~ks, 
with typical values around 15~ks. Remaining objects were taken from the 
{\it XMM--Newton} Science Archive. All sources were observed with 
the EPIC CCDs (MOS and pn) as the prime instrument. 

Observation dates, exposure times, and filters used during the observations 
are listed in Table 2. 
The raw observation data files (ODFs) were reduced and analyzed using 
the standard Science Analysis System (SAS) software package (version 5.3, 
released in June 2002; Saxton 2002) with associated latest calibration files. 
We used $epproc$ and $emproc$ tasks for the pipeline processing of the ODFs to 
generate the corresponding event 
files and remove dead and hot pixels. Time intervals with very high background 
rates were identified in light curves at energy $>$10~keV and removed.
Only patterns $\leq$12 for ``MOS'' and $\leq$4 for ``pn'' were considered 
in the analysis and a standard selection filter of FLAG=0 was applied.

\section{Spatial and Timing Analysis}

Images and light curves were analyzed in the 0.5--2 keV (soft) and 
2--10 keV (hard) energy bands, for MOS and pn separately. 
Despite the short exposures (as low as $\sim$5~ks; see Table 2), 
all targets are detected with a minimum of 20 counts per 
detector in either the soft or the hard energy band.
Flux limits reached are on the order of $F_{(0.5-2 ~\rm keV)}$ $\approx$ 
10$^{-14}$ erg~cm$^{-2}$ s$^{-1}$ and $F_{(2-10 ~\rm keV)}$ $\approx$ 
10$^{-13}$ erg~cm$^{-2}$ s$^{-1}$. These flux limits translate into 
minimum luminosities detectable of $\sim$10$^{38}$ erg~s$^{-1}$ and 
$\sim$10$^{39}$ erg~s$^{-1}$ for the nearest and farthest Seyferts, 
respectively. 

We find that 25 out of 27 sources, and in particular all type 1 Seyferts, 
have a compact dominant nucleus coincident with the optical nuclear position 
reported in Table 1. The only sources that did not display a dominant 
nucleus were NGC 1058 and NGC 4472, for which upper limits are calculated. 

In agreement with recent $Chandra$ studies (e.g., Ho et al. 2001, Terashima 
\& Wilson 2003), a wide variety of morphologies is seen. In many cases a 
bright point-like nucleus is present; in other cases, the nuclear (arcmin 
size) regions are characterized by the presence of structures, off-nuclear 
sources, and/or diffuse emission. An atlas of {\it XMM--Newton} 
and $Chandra$ images of the all sample is given by Panessa (2004).
Adopting the classification scheme proposed by Ho et al. (2001), which separates 
the morphologies in four classes (according to the predominance of the nuclear 
emission with respect to the 
surrounding structures), we find that the most common morphology is that 
of a single compact nucleus centered on the position of the optical nucleus ($\sim$60\% 
of the sources), followed by those having a nucleus comparable in brightness to off-nuclear 
sources in the galaxy ($\sim$25\%), and a few percent 
have their nuclei embedded in diffuse soft 
emission or no core emission. There is also good agreement between 
the {\it XMM--Newton} and $Chandra$ 
classifications, which guarantees that the {\it XMM--Newton} 
point-spread function (PSF) is effective 
enough to exclude contamination from off-nuclear sources.

Analysis of the soft and hard X-ray light curves indicates that most sources 
do not exhibit significant flux variations, except for the few brightest Seyfert 
1 galaxies (i.e., NGC 3227, NGC 4051, NGC 4151, NGC 4395) for which the detailed 
timing analysis is deferred to specific papers presented in the litterature and cited 
in the Appendix. However, this is not inconsistent with the expectation 
that low luminosity sources should exhibit higher variability amplitude 
(Nandra et al. 1997) because, given the low statistics, flux 
variations up to a factor of a few cannot be excluded for most sources.
We thus ignored here any potential spectral variations and considered only the 
source {\it average} X-ray spectra which are the subject of our study. 
This assumption should, however, be kept in mind.

\section{Spectral Analysis}

Source spectra were extracted from circular regions with radii 
of $50''$ and $25''$ for sources brighter and fainter than  $F_{(0.5-10 ~\rm keV)}$ 
$\approx$ 10$^{-13}$ erg~cm$^{-2}$ s$^{-1}$, 
respectively, except where noted in Appendix B.
These extraction regions correspond to energy encircled fractions of
$\sim$90\% and $\sim$80\%, respectively. When available, we have looked at 
$Chandra$ images in order to check for possible contamination due to 
off-nuclear sources or diffuse emission unresolved by {\it XMM--Newton}. 
Sources which may have been affected by this type of contamination are 
marked with ``${\dagger}$'' (only soft) or ``${\ddagger}$'' (both soft plus hard) 
in Table 3, which lists the best-fit spectral parameters. 
The background was estimated using standard 
blank-sky files or, when unusually high (as in the case of NGC 5033), 
locally from source-free circular regions placed in offset positions close 
to the source. 

Spectral channels were binned to a minimum of 20 counts per bin 
and spectra were fitted using data from the three CCD detectors 
(MOS1/2 and pn) simultaneously.
The pn normalization is fixed to 1, while the MOS1/2 normalizations are 
left free to vary in the range 0.95--1.05 to account for the
remaining calibration uncertainties in their 
cross-normalizations\footnote{See 
http://xmm.vilspa.esa.es/external/xmm\_sw\_cal/calib/index.shtml}. 
Statistical errors are in any case typically much larger 
than current calibration uncertainties.
Data were fitted in the range 0.3--10 keV for MOS1/2 and 
0.5--10 keV for the pn.

Spectral analysis was performed to first identify the underlying 
continuum when possible, and then additional components and features 
were included to best reproduce the data. 
Hence, each spectrum was initially fitted with a single
model consisting of a 
power law plus absorption fixed at the Galactic value as 
quoted in column (2) of Table 3, 
plus intrinsic absorption as quoted in column (3).

In many cases this simple parametrization is not sufficient to model 
the whole 0.3--10~keV spectrum. Residuals often show, for example, 
a soft excess on top of the (absorbed or non-absorbed) power law. 
The soft-excess component is clearly more complex than a single 
power law, often exhibiting emission or absorption structures, or both. 
The soft excess is fitted here using a simple, and 
approximate, description/parameterization in terms of a scattered 
power-law component (with index given in column (5)) plus a thermal 
plasma model (with temperature $kT$ given in column (6), and metallicity 
fixed to the solar value). The possible presence of a narrow emission 
line centered at 6.4~keV originating from neutral iron has also been 
checked, and modeled with a single Gaussian line (with equivalent width, EW, 
given in column (7)). Best-fit spectral parameters are reported in Table 3. 
Errors given in the table are calculated at 90\% confidence for two 
interesting parameters ($\Delta \chi ^2 = 4.61$), and we applied a 
single-digit approximation. 

It is stressed that some source spectra, in particular those with 
high-quality statistics, do clearly require more complex modeling 
of the continuum and additional narrow absorption 
and/or emission features than used above in our simplified procedure.
For example, NGC 3227, NGC 4051, NGC 4151, NGC 4395,
and NGC 5273 show additional continuum curvatures in the residuals which indicate 
either multiple or ionized absorption. The same is true for NGC 1068 and 
NGC 4051 for which a reflection continuum in the data is likely to be significant. 
Other sources such as NGC 1068, NGC 3031, NGC 3079, NGC 4051, NGC 4151,
and NGC 5273 also show evidence for additional 
absorption and/or emission structures at soft and/or hard energies.
Given the purpose of this work (to obtain a proper, uniform, average 
description of the spectra in terms of absorption, photon-index 
continuum, flux, and Fe~K line intensity), 
we do not attempt to fit all these extra components in a systematic way.
Rather, we address these issues case-by-case in Appendix B, where 
references from the literature are also quoted, and we take these 
caveats into account when interpreting the overall spectral results.

%where we list the galaxy name (col. 1), 
%the Galactic absorption along the line of sight (col. 2) in units of 
%10$^{20}$ cm$^{-2}$, the measured 
%absorption column density (col. 3) in units of 10$^{22}$ cm$^{-2}$, the 
%power-law photon index (col. 4), 
%and, for the soft component, either the temperature of the thermal component (kT in keV), or 
%the ionization parameter of the warm absorber ($\xi$), or the photon index of the soft 
%power-law component ($\Gamma_{SX}$), or a combination of these (col. 5).
%Also reported are the equivalent width of the FeK line, in keV (col. 6), the 
%reduced chi-square and degrees of freedom (col. 7), the observed soft 
%(col. 8) and hard (col. 9) 
%fluxes in units of 10$^{-13}$erg~cm$^{-2}$s$^{-1}$, and finally, the soft
% (col. 10) and hard (col. 11)
%luminosities corrected for the absorption given in col. 3.    
%Values marked with ``$^{\dagger}$'' indicate that the current data are contaminated 
%(by more than 10\%) by off-nuclear point-sources 
%and/or diffuse emission. 

\section{X-ray Properties}

\subsection{Observed values of $\Gamma$, $N_{\rm H}$, and $L_{2-10~\rm keV}$}

%Photon indices, absorption column densities and hard X-ray luminosities}

Figure 1 (upper panel) reports the distribution of best-fit photon indices.
These vary from object to object over the range $0.5 < \Gamma < 2.3$ (the value 
$\Gamma = 2.7$ for NGC 4472 is excluded here because it is due to 
diffuse emission from the galaxy; see Appendix B).
The weighted mean for the total sample is $\Gamma$ $\approx$ 1.60 
$\pm$ 0.04, with a dispersion $\sigma$ $\approx$ 0.44.  
The distribution for type 1 objects (middle panel of Fig. 1) has a mean 
value of 1.56 $\pm$ 0.04 and a dispersion $\sigma$ $\approx$ 0.24. The 
rather flat ($\Gamma \lsimeq 1.5$) spectrum of 4 (out of 9) type 1 
objects can be ascribed to either the presence of a warm absorber, a 
complex absorber, and/or a reflection 
component, all of which produce a flattening of the continuum.
The distribution of $\Gamma$ for type 2 objects (lower panel of 
Fig. 1) is somewhat broader. The weighted mean for this class 
is 1.61 $\pm$ 0.06 with a dispersion of $\sim$ 0.5. A Kolmogorov-Smirnov
(KS) test gives a probability of $\sim$0.43, consistent 
with the same parent population.

Figure 2 (left panel) shows the observed column density distribution 
obtained for the total sample (top panel), type 1 
(middle panel), and type 2 (bottom panel) Seyfert galaxies. Arrows indicate upper limits. 
The observed total distribution varies over the range of Galactic 
absorptions ($N_{\rm H}$ $\approx$ 10$^{20}$ cm$^{-2}$) to high 
column densities ($N_{\rm H}$ $\lsimeq$ 10$^{24}$ cm$^{-2}$), 
with most of the measurements being upper-limits of $N_{\rm H}$ $\approx$ 
10$^{21-22}$ cm$^{-2}$.
Type 1 objects are known to be less absorbed than type 2s, but three 
of them show a column density higher than 10$^{22}$
cm$^{-2}$. The nature of the absorbing material in these objects is 
likely associated with highly ionized gas (NGC 3516, Netzer et al. 2002) and/or 
to dense and variable absorbing columns (NGC 3227, Lamer, Uttley, \& McHardy 
2003; NGC 4151, Puccetti et al. 2003). 

Past hard X-ray surveys of Seyfert 2 galaxies have revealed 
that the column density distribution of this class is significantly 
shifted toward large columns; most ($\sim$ 75\%) of type 2 Seyferts 
are heavily obscured, with $N_{\rm H}$ $>$ 10$^{23}$ cm$^{-2}$ 
(Risaliti et al. 1999). The distribution observed here and shown 
in the lower (left side) panel of Figure 2 apparently deviates from past results, 
containing mostly mildly absorbed objects. However, this distribution does not 
take into account the possible presence of heavily absorbed sources, not recognised 
as such because of the absence of a low energy cut-off.

Histograms of the observed hard X-ray luminosities are shown in Figure 3 (left side).
A wide range of luminosities is covered, from objects with $L_{2-10 \rm keV}$ 
$\approx$ 10$^{38}$ erg~s$^{-1}$, 
comparable to those of bright binary systems or ultraluminous X-ray sources 
(ULXs), to those with luminosities $L_{2-10~\rm keV}$ $\approx$ 
10$^{43}$ erg~s$^{-1}$, typical of bright Seyferts. 
The upper limits on the hard luminosity for the two type 2 Seyferts 
(NGC 1058 and NGC 4472) are indicated with arrows. It has been shown 
in previous soft and hard X-ray surveys that Seyfert 2 galaxies are 
generally weaker than their type 1 counterparts. Our measurements 
confirm, at first glance, this evidence. A KS test to compare the 
two distributions yields a probability of 0.009 that they are drawn 
from the same parent population. The mean hard X-ray luminosity of type 1 
objects is $L_{2-10~\rm keV}$ $\approx$ $10^{41}$ erg~s$^{-1}$ 
($\sigma$ $\approx$ 0.8), while for type 2 objects it is $L_{2-10~\rm keV}$ 
$\approx$ $10^{39.8}$ erg~s$^{-1}$ ($\sigma$ $\approx$ 0.9).

A Gaussian iron K emission line has been detected in 8 out of 9 
type 1s and in 6 out of 18 type 2s. Their mean equivalent widths 
are $\approx$ 215$\pm$80 eV and $\approx$700 $\pm$ 220 eV, respectively, which 
are consistent with previous works (Nandra et al. 1997, Turner et al. 1997) 
when the large errors and dispersions are taken into account.
The larger rate of detected lines in type 1s with respect to type 2s is at first 
surprising but is consistent with the fact that type 1s are typically brigther in the sample.
As mentioned in \S 4, the present estimates are to be taken as only rough parameterizations 
because they do not take into account in detail of the multiple lines often present 
(i.e. in NGC1068, NGC3031, NGC4579), nor the possibility of line broadening (NGC3031, 
NGC4051, NGC4151, NGC4395) or variability. 

\subsection{Identification of heavily absorbed sources}

Several of the above differencies in terms of $\Gamma$, $N_{\rm H}$ and $L_{2-10~\rm keV}$, 
could well be due, in part or totally, to our inability to directly measure 
heavily absorbed sources. In the hard band, the effective area of the 
pn peaks at energies around 5--6~keV and has an exponential roll-over at higher energies.
This implies that we are only weakly sensitive to measurements of absorption columns 
of $N_{\rm H}$ $\gsimeq$ 10$^{23}$ cm$^{-2}$. In particular, in Compton-thick sources 
with $N_{\rm H}$ $>$ 10$^{24}$ cm$^{-2}$, the transmitted component 
is completely suppressed below 10~keV and the spectrum observed in the 
2--10~keV band is dominated by flattened reprocessed components 
from a cold and/or warm scatterer (Matt et al. 2000). The galaxy may, 
thus, be erroneously classified as a source with a flat spectral shape, 
low absorption and, thus, low luminosity while actually being 
intrinsically steep, heavily obscured, and luminous (see, for example, 
the prototypical case of the Seyfert 2 galaxy NGC 1068; Matt et al. 1997, 
Iwasawa, Fabian \& Matt 1997). 
To take these factors into account, we apply three independent 
tools to unveil the presence of 
heavy obscuration: (i) X-ray spectral diagnostics such as 
flat slope and large Fe~K$\alpha$ EW; (ii) flux diagnostic diagrams; 
and (iii) the $N_{\rm H}$ vs. $F_{2-10~\rm keV}$/$F_{\rm [O~III]}$ diagram.

%For example, in the case of the archetype Seyfert 2 galaxy NGC1068, the slope, absorption 
%column and luminosity measured using our data below 10 keV are $\Gamma$ $\sim$ 
%1.0, $N_{\rm H}$ $<$ 10$^{21}$ 
%cm$^{-2}$, and L$_{2-10 keV}$$\sim$1.2$\times$10$^{41}$ erg/s while the 20-100 keV data from 
%$BeppoSAX$ require that the primary emission is 
%steeper ($\Gamma$ $\sim$1.8), the absorption column is greater than 10$^{25}$ cm$^{-2}$ and 
%the implied nuclear luminosity about two order of magnitude greater than observed 
%(Matt et al. 1997, Iwasawa et al. 1997).
%It is therefore possible that other sources in our sample, in particular either 
%the one with a flat slope and low absorption or the weakest ones for which few counts are 
%detected above a few keV, could actually be heavily absorbed objects of the kind of NGC 1068. 

In three type 2 objects (NGC 1068, NGC 3079, and NGC 5194), the EW of the 
detected iron K line is higher than $\sim$1~keV. Such a high value 
of the EW is expected in highly obscured objects since it is measured 
against a much-depressed continuum  ($N_{\rm H}$ $\approx$ 10$^{23}$--10$^{24}$ 
cm$^{-2}$; Leahy \& Creighton 1993) or against a pure reflection component 
($N_{\rm H}$ $>$ 10$^{24-25}$ cm$^{-2}$; Makishima 1986, Bassani et al. 1999). 

In 2 Seyfert 2 galaxies (NGC 2685 and NGC 3486), the photon index is rather 
flat ($\lsimeq$1) and may also be indicative of Compton-thick sources. 
However, the lack of any strong line makes this criterion alone 
not sufficient to classify the sources as Compton-thick candidates. 
In total, the spectral analysis is able to directly assess three candidate 
Compton-thick sources, namely NGC 1068, NGC 3079, and NGC 5194. This is 
consistent with studies which have been able to obtain hard ($E > 10$~keV) 
X-ray spectra of these sources with $BeppoSAX$ (Matt et al. 1997, Iyomoto 
et al. 2001, Fukazawa et al. 2001).

Another way of evaluating the true amount of absorption is through 
flux diagnostic diagrams 
(e.g., Bassani et al. 1999, Panessa \& Bassani 2002, Panessa 2004, 
Guainazzi et al. 2005). 
These make use of independent indicators of the intrinsic brightness of the 
source such as the [O~III] $\lambda$5007 flux and the infrared emission, 
to be compared with the hard X-ray flux. 

By studying a large sample of Seyfert 2 galaxies, Bassani et al. (1999) have 
found that the ratio $F_{\rm 2-10~keV}$/$F_{\rm [O~III]}$ is effective in 
separating Compton-thin from Compton-thick sources, the latter having ratios 
lower than $\sim$1. This is because the [O~III] flux is considered 
to be a good isotropic indicator; it is mostly produced far from the nucleus,  
in the NLR, by photoionizing photons from the AGN.
Applying this criterion to our sample, and using the [O~III] 
measurements reported in HFS97a, we identify 5 Compton-thick sources, i.e. with 
$F_{\rm 2-10~keV}$/$F_{\rm [O~III]} <$ 1: NGC 676, NGC 1068, NGC 3079, NGC 
3185, and NGC 5194. Of note is the fact that the three candidates 
Compton-thick with a strong line are confirmed.

The effectiveness of identifying Compton-thick vs. Compton-thin sources 
through the ratio $F_{\rm 2-10~keV}$/$F_{\rm [O~III]}$
is also exemplified by the third of our diagnostics: the $N_{\rm H}$ vs. 
$F_{2-10~\rm keV}$/$F_{\rm [O~III]}$ 
diagram shown in Fig. 4. Assuming, as mentioned above, that the 
[O~III] luminosity is an isotropic indicator of the intrinsic 
luminosity, one expects that the ratio of $F_{2-10~\rm keV}$/$F_{\rm [O~III]}$ 
decreases as $N_{\rm H}$ increases, following a path as indicated by the 
dashed region in Fig. 4. The relation was obtained assuming that the 
observed $F_{2-10~\rm keV}$ changes according to the $N_{\rm H}$ value 
given on the ordinate, and starting from the average 
$F_{2-10~\rm keV}$/$F_{\rm [O~III]}$ ratio observed in type 1 
Seyferts and assuming a 1\% scattered component. The width of the 
shaded area (from lower left to upper right) was drawn considering 
the lower and higher $F_{2-10~\rm keV}$/$F_{\rm [O~III]}$ 
ratios obtained for the type 1 objects of the present sample. 
The shaded region (from upper left to lower right) obtained by 
Maiolino et al. (1998) is also shown for comparison and is 
consistent with the present results, though slightly shifted to 
lower flux ratios.

Compton-thick AGNs should occupy the {\it observed} low $N_{\rm H}$ and 
low $F_{2-10~\rm keV}$/$F_{\rm [O~III]}$ 
part of the diagram, but after correction for their 
{\it intrinsic} high $N_{\rm H}$, they should occupy 
the high $N_{\rm H}$ and low $F_{2-10~\rm keV}$/$F_{\rm [O~III]}$ 
region of the predicted distribution (as indicated 
by the arrows in Fig. 4). 
We find that this is indeed the case for the previously identified 
Compton-thick objects of our sample (NGC 676, NGC 1068, NGC 3079, NGC 3185, 
and NGC 5194). Other sources (NGC 3941, NGC 4698, and NGC 4501) are 
located in the same area of the plot, but we found no independent 
confirmation to classify them as secure Compton-thick candidates.
Moreover, they have $F_{2-10~\rm keV}$/$F_{\rm [O~III]}$ ratios
exceeding 1, in the Compton-thin regime. Interestingly, two of 
these sources (NGC 4698 and NGC 4501) have been identified in 
the literature as ``anomalous'' cases of Seyfert 2 galaxies with no 
intrinsic absorption (Georgantopoulos \& Zezas 2003, Terashima et al. 2002).

By combining the X-ray spectral properties and the diagnostic 
diagrams using isotropic indicators, we identify Compton-thin and
Compton-thick objects as indicated in Table 4. 
In total, a subsample of 5 Compton-thick candidate objects have 
been confidently recognized (5 from diagnostic diagrams, 3 of which
have direct spectral information as well). 

\subsection{$\Gamma$, $N_{\rm H}$ and L$_{2-10~\rm keV}$: after correction for 
heavily absorbed sources}

As a consequence of the above considerations, we correct the above 
distributions by adopting, for all Compton-thick candidates, a value 
of $\Gamma$=1.62, $N_{\rm H}$ = 2 $\times 10^{24}$ cm$^{-2}$ 
and increasing $L_{2-10~\rm keV}$ by a factor of 100. 
The first value is taken equal to the average value of $\Gamma$ in type 2s 
after removal of the 5 Compton-thick candidates. The second is taken 
as a lower-limit characteristic value of optically thick matter.
The latter value/shift in luminosity is drawn from Fig. 4. This value 
corresponds, as a rough estimate, to the factor that is necessary to bring the 
average $F_{2-10~\rm keV}$/$F_{\rm [O~III]}$ ratio measured for the 
5 Compton~thick sources ($\simeq 10^{-0.44}$) equal to that of 
type 1 sources ($\simeq 10^{1.54}$), as indicated from the arrows in Fig. 4. 
New histograms are shown in the 
right-hand panels of Fig. 1, 2 and 3, where the Compton-thick candidates 
are marked with right-sided arrows.
Although average values do not change significantly before 
and after this correction, the shapes of 
the distributions are quite different. 
The fraction of Compton-thick sources is at least a third 
of all type 2 Seyferts, and a  
fifth of all Seyferts. This is similar to what was 
found by Bassani et al. (1999) using data from 
the literature and is slightly lower than, but in substantial 
agreement with the survey by Risaliti et al. (1999) when errors on the 
percentages are considered. However, the present study adds to the 
significance of this result because for the 
first time it is derived using an unbiased 
(distance-limited) optical sample, applying a uniform 
optical and X-ray analysis to the data (with data from a single 
satellite), and excluding any 
severe contamination from nearby off-nuclear sources. 

%We do not find anymore any significant tail versus low 
%values of $\Gamma$ increasing the similitude between type1 and type 2 Seyferts. 

These findings are, also for the first time, extended down to 
very low luminosities. Finally, unlike previous studies, no more 
significant difference between the luminosity distributions 
of type 1 and type 2 Seyferts are found (\S~5.1). The probability of the two 
classes being drawn from the same parent population is now 0.16. 
The previous differences in luminosities (\S~5.1) are therefore consistent 
with being ascribed almost entirely to absorption effects.

We have checked that the above results are robust even with slightly different 
assumptions in correcting for Compton-thick candidates.
Assuming a steeper value of $\Gamma$=1.8 or 1.9, more typical of the {\it intrinsic}
(reflection-corrected) spectrum of unobscured AGNs (Nandra \& Pounds 1994), or 
cancelling out all 5 Compton-thick candidates, similarity between the index distributions 
of type 1 and type 2s was always recovered.
For the luminosity distributions, if we assumed conservatively a correction factor 
of about 30, i.e. corresponding to the lowest observed value of 
$F_{2-10~\rm keV}$/$F_{\rm [O~III]}$ ratio in 
type 1s, the KS probability that type 1 and type 2 luminosity distributions 
are drawn from the same parent population becomes 3$\times$10$^{-2}$ (from a 
probability of about 10$^{-4}$ when luminosities were not corrected, see \S5.1).
This indicates that the above results are, of course, sensitive to the assumptions 
made to correct for the Compton-thick sources but that even our 
most conservative choice of correction brings the luminosities of type 2s consistent with 
that of type 1s.

%\section{Multiwavelength properties}

%\subsection{X-ray vs. H$_{\alpha}$ luminosity}

%\subsection{X-ray vs. radio luminosity}

%\subsection{SEDs}

\section{General results and summary}

The optical spectroscopic survey of Ho, Filippenko \& Sargent (1995, 1997) 
has provided a new, comprehensive catalog of 52 Seyfert galaxies, the
most complete and least biased available to date. 
We have performed an X-ray spectral survey of the 27 nearest 
($D \lsimeq 22$ Mpc) Seyfert galaxies in 
that survey using the EPIC CCDs on-board {\it XMM--Newton}. 
This paper presents the observational material, along with a 
compilation of X-ray spectral parameters to 
be used in subsequent analysis.

We have detected in the hard X-ray band all but two of the observed 
Seyfert nuclei. The sample extends to significantly lower X-ray 
luminosities than many previous surveys. 

Nuclear X-ray spectra have been obtained forming one of the 
largest atlases of low-luminosity Seyfert galaxies ever assembled.
Simple models have been applied in order to characterize
the spectral shape, the presence of absorption in excess of the Galactic
value and the presence of an Fe emission line.
The distribution of spectral parameters, in particular for type 1 Seyferts,
is found to be within the range of values observed in luminous AGNs.
At a first glance, the observed column density distribution for type 2
Seyferts appears to be shifted toward low absorbing column densities 
($N_{\rm H}$ $<$ 10$^{22}$ cm$^{-2}$).

Moreover, the observed 2--10~keV luminosity distribution 
of type 2 Seyfert galaxies appears to be significantly shifted
toward low luminosities with respect to type 1 objects.
However, the presence of Compton-thick sources in our sample
may affect the estimate of these parameters.
Therefore, indirect arguments have been used to infer their presence, such as 
evidence of a flat power-law spectrum in the X-ray band, the presence
of a strong Fe K$\alpha$ line at 6.4 keV, and flux diagnostic diagrams
which employ isotropic indicators of the nuclear unobscured emission.   
Results obtained by combining spectral and flux diagnostic tools 
indicate that the fraction of heavily obscured objects is large, 
at least one third of all objects in the sample, in good agreement 
with previous estimates performed on a flux-limited sample (Risaliti et al. 1999).
Interestingly, Seyfert galaxies as a whole possess the entire range of 
$N_{\rm H}$, from 10$^{20}$ cm$^{-2}$ to 10$^{24}$ cm$^{-2}$, fairly continuously.
This is similar to what was found in the deepest X-ray surveys available 
to date (Mateos et al. 2005), and it is 
consistent with local absorption distributions adopted by, e.g.,  
La Franca et al. (2005) to fit the hard X-ray luminosity functions of 
AGNs.

With the present work we are able to probe much lower luminosities
and still find that the fraction of absorbed objects remains significantly
high. In light of these findings, the column density
and luminosity distributions have been revisited.
In particular, it has been shown that the dichotomy
observed in the luminosity of type 1 and type 2 Seyferts is mainly due
to absorption effects. 

We point out a note of caution, however, regarding our ability to identify 
the Compton-thick cases for the lowest-luminosity sources of the sample. 
It may be that the statistics are so low that one cannot firmly exclude the
possibility that some of them are indeed Compton-thick objects. 
Correcting for this would of course increase the fraction of 
Compton-thick sources, and bring the present results to meet 
with Risaliti et al. (1999) estimates. On the other 
hand, observational criteria such as those above (flat spectra, 
2--10 keV/[O~III] ratios, etc.) that are ``calibrated'' for higher-luminosity 
objects may not hold for the very low-luminosity AGNs. These may 
indeed have different spectral energy distributions (e.g., 
Ho 1999), and hence their X-ray/optical 
emission-line ratios may not be the same as those of traditional AGNs.   
Because of these effects, the fraction of Compton-thick sources may vary 
accordingly, i.e. it could be different in this sample with respect to 
other previous works because lower-luminosities are probed here.
We note, nevertheless, that in these data, we find neither a 
significant relation between $N_{\rm H}$ and $L_{\rm 2-10~ keV}$, nor between 
$N_{\rm H}$ (or $L_{\rm 2-10~ keV}$) and the source sub-classification.
 
Another result of this survey is the realization (and 
statistical quantification) of a number of exceptions to the baseline 
standard model of Seyfert galaxies. First, at least two objects (M81 
and NGC 4579) show a complex of three distinct emission lines at 
$E \approx 6.4$, 6.7, and 6.9 keV.
Detailed modeling is presented by Dewangan et al. (2004) 
and Page et al. (2004), but the origin of these lines is not clear. The
narrow lines in both objects are like those of LINERs, suggesting possibly 
different, perhaps transient, emission processes in their nuclear 
regions (e.g., an advection-dominated accretion flow instead of a thick 
disk, with or without an outflow or jet; Pellegrini et al. 2000, Blandford 
\& Begelman 1999).

The second exception rests on the fact that at least two, perhaps
three, of the Seyfert 2 galaxies show no absorption at all, with 
stringent upper limits. This is clearly in contrast with a standard 
unified model, but agrees well with previous findings by 
Pappa et al. (2001), Panessa \& Bassani (2002), Barcons et al. (2003), 
and Mateos et al. (2005).

The third exception is that our analysis has confirmed that three (out of 9) 
type 1 galaxies (NGC 4151, NGC 3227, and NGC 4395) show evidence for 
X-ray absorption, an apparent inconsistency between their optical classification 
and the X-ray one which has been extensively debated in the literature (Mateos et al. 2005,
and refences therein). This is a (30$\pm$17)\% fraction of absorbed type 1 Seyferts that 
is consistent with latest estimates based on wide field X-ray surveys 
(Piconcelli et al. 2003, Perola et al. 2004). 
All three sources are type 1.5 sources, so there might 
be an orientation effect where these objects are being viewed from just 
outside the ionization cone. Other plausible explanations in terms of 
either effects due to variability 
in the absorption column density, and/or geometry, and/or ionization 
state (Malizia et al. 1997, Piro 1999, Risaliti, Elvis \& Nicastro 2002, 
Matt, Guainazzi \& Maiolino 2003), or unusual dust-to-gas ratios have been proposed 
to explain such differences (Maiolino et al. 2001a, 2001b). 

In conclusion, we have given an unbiased estimate of the 
average intrinsic X-ray properties and column density 
distribution of Seyfert galaxies at low redshifts.
This is crucial to validate unified models of AGNs 
and for synthesis models of the X-ray background. 
The results obtained here are in agreement with the 
predictions of unified models, except for a few particular cases 
(10\% of unabsorbed Seyfert 2s and 30\% of absorbed Seyfert 1s, in agreement 
with previous works by Panessa \& Bassani (2002) and Perola et al. (2004))
which do not fit easily into the standard picture, but which we are able 
to quantify. Most significantly, these predictions are positively 
tested using a complete sample with datasets of unprecedented quality.
The first-ever extension to low luminosities also suggests that the same 
physical processes are governing emission in low-luminosity AGNs 
as in more luminous sources, although a larger sample is required 
to verify this conclusion.

A description of the multi-waveband correlations (e.g., X-ray 
luminosities vs. H$_{\alpha}$ and radio luminosities, spectral
energy distributions, etc.) and their astrophysical implications 
is deferred to forthcoming papers.

\acknowledgements

This paper is based on observations obtained with {\it XMM--Newton}, 
an ESA science mission with instruments and contributions directly 
funded by ESA Member States and the USA (NASA). The research has made 
use of data obtained through the High Energy Astrophysics Science 
Archive Research Center Online Service, provided by the NASA/Goddard 
Space Flight Center, and of the NASA/IPAC Extragalactic Database (NED), 
which is operated by the Jet Propulsion Laboratory, California Institute 
of Technology, under contract with the National Aeronautics and Space 
Administration. M.C. acknowledges financial support from the contract 
ASI-CNR/IASF:I/R/042/02. A.V.F. is supported by NSF grant AST--0307894; 
he is also grateful for a Miller Research Professorship at U.C. Berkeley, 
during which part of this work was completed.

{}

\onecolumn

\vfill\eject

\centerline{\bf Table 1: The complete, distance-limited sample of Seyfert galaxies}
\begin{flushleft}
\begin{tabular}{llllrrcc}
\hline
\hline
\multicolumn{1}{c}{Galaxy} &
\multicolumn{1}{c}{Other} &
\multicolumn{1}{c}{Seyfert} &
\multicolumn{1}{c}{Hubble} &
\multicolumn{1}{c}{Distance} &
\multicolumn{1}{c}{$B_{T}$} &
\multicolumn{1}{c}{R.A.} &
\multicolumn{1}{c}{Decl.} \\
\multicolumn{1}{c}{Name} &
\multicolumn{1}{c}{Name} &
\multicolumn{1}{c}{Type} &
\multicolumn{1}{c}{Type} &
\multicolumn{1}{c}{(Mpc)} &
\multicolumn{1}{c}{(mag)} &
\multicolumn{1}{c}{(J2000)} &
\multicolumn{1}{c}{(J2000)} \\
\multicolumn{1}{c}{(1)} &
\multicolumn{1}{c}{(2)} &
\multicolumn{1}{c}{(3)} &
\multicolumn{1}{c}{(4)} &
\multicolumn{1}{c}{(5)} &
\multicolumn{1}{c}{(6)} &
\multicolumn{1}{c}{(7)} &
\multicolumn{1}{c}{(8)} \\
\hline
\hline
NGC 676  &        & S2:     & S0/a: spin & 19.5 & 10.50 &   01 48 57.4  &  +05 54 25.7 \\   
NGC 1058 &        & S2      & Sc         &  9.1 & 11.82 &   02 43 30.2  &  +37 20 27.2 \\    
NGC 1068 &        & S1.9    & Sb         & 14.4 &  9.61 &   02 42 40.7  &  -00 00 47.6 \\    
NGC 2685 & Arp336 & S2/T2:  & SB0+ pec   & 16.2 & 12.12 &   08 55 34.8  &  +58 44 01.6 \\   
NGC 3031 & M81    & S1.5/L1.5    & Sab   &3.5$^a$& 7.89 &   09 55 33.2  &  +69 03 55.0 \\
NGC 3079 &        & S2      & SBc spin   &17.3$^b$&11.54&   10 01 58.5  &  +55 40 50.1 \\   
NGC 3185 &        & S2:     & SB0/a      & 21.3 & 12.99 &   10 17 38.7  &  +21 41 17.2 \\  
NGC 3227 & Arp94  & S1.5    & SABa pec   & 20.6 & 11.10 &   10 23 30.6  &  +19 51 53.9 \\   
NGC 3486 &        & S2      & SABc       &  7.4 & 11.05 &   11 00 24.1  &  +28 58 31.6 \\    
NGC 3941 &        & S2:     & SB0        & 12.2$^c$ & 11.25&11 52 55.4  &  +36 59 10.5  \\   
%NGC 3982 &       & S1.9    & SABb:      & 20.5 & 11.78 &   11 56 28.1  &  +55 07 31  \\   
NGC 4051 &        & S1.5$^*$    & SABbc      & 17.0 & 10.83 &   12 03 09.6  &  +44 31 52.8  \\   
NGC 4138 &        & S1.9    & S0+        & 13.8$^c$ & 12.16&12 09 29.9  &  +43 41 06.0  \\   
NGC 4151 &        & S1.5    & SABab:     & 20.3 & 11.50 &   12 10 32.6  &  +39 24 20.6  \\   
%NGC 4168 &       & S1.9:   & E2         & 31.7$^*$ & 12.11 &   12 12 17.3  &  +13 12 18 \\  
NGC 4258 & M106   & S1.9    & SABbc      &  7.2$^c$ &  9.10&12 18 57.5  &  +47 18 14.3  \\    
NGC 4388 &        & S1.9    & Sb: spin   & 16.7 & 11.76 &   12 25 46.7  &  +12 39 40.9  \\   
NGC 4395 &        & S1.5$^*$ & Sm:    &  4.1$^d$ & 10.64 &   12 25 48.9  &  +33 32 47.8  \\    
NGC 4472 & M49    & S2::    & E2         & 16.7 & 9.37  &   12 29 46.8  &  +07 59 59.9  \\   
NGC 4477 &        & S2      & SB0?       & 16.8 & 11.38 &   12 30 02.2  &  +13 38 11.3  \\   
NGC 4501 & M88    & S2      & Sb         & 16.8 & 10.36 &   12 31 59.3  &  +14 25 13.4  \\  
NGC 4565 &        & S1.9    & Sb? spin   &  9.7$^c$ & 10.42&12 36 21.1  &  +25 59 13.5  \\    
NGC 4579 & M58    &S1.5/L1.5$^*$  & SABb     & 16.8 & 10.48 &   12 37 43.4  &  +11 49 04.9   \\  
NGC 4639 &        & S1.5$^*$ & SABbc      & 16.8 & 12.24 &   12 42 52.5  &  +13 15 24.1  \\   
NGC 4698 &        & S2      & Sab        & 16.8 & 11.46 &   12 48 22.9  &  +08 29 14.8  \\   
NGC 4725 &        & S2:     & SABab pec  & 13.0$^c$ & 10.11&12 50 26.7  &  +25 30 02.3  \\   
NGC 5033 &        & S1.5    & Sc         & 18.7 & 10.75 &   13 13 27.5  &  +36 35 37.8  \\   
NGC 5194 & M51    & S2      & Sbc pec    &  8.4 &  8.96 &   13 29 52.4  &  +47 11 40.8  \\    
NGC 5273 &        & S1.5    & S0         & 16.5$^c$ & 12.44&13 42 08.3  &  +35 39 15.2  \\   
\hline
\hline
\end{tabular}
\end{flushleft}
Notes: Col. (1): galaxy name. Col. (2): other name. Col. (3): optical classification as 
given by HFS97a: ``S'' represents Seyfert, ``L'' represents LINER, and ``T'' represents 
objects with LINER plus H~II-region spectra. ``2'' implies that no broad H$\alpha$ 
is detected; ``1.9'' implies that broad H$\alpha$ is present, but not broad H$\beta$; 
``1.5'' implies that both broad H$\alpha$ and broad H$\beta$ 
are detected, with substantial contributions from both the BLR and NLR 
(Osterbrock 1981). Objects with a changed classification with respect to the 
original given by HFS97a (see also Ho et al. 1997b and HU01) are 
denoted by ``*'' after their name. 
Col. (4): Hubble type as listed in HFS97a. Col. (5): distance from 
Tully (1988), except when marked with (a) Paturel et al. (2002), (b) Cecil et
al. (2002), (c) Tonry et al. (2001), and (d) Thim et al. (2004). Col. (6):
total apparent $B_T$ magnitude of the galaxy. Col. (7)--(8): nuclear optical 
position in epoch J2000 as given by HU01.
N.B: Two sources (NGC 3031 and NGC 4579) have classifications which 
fall near the somewhat arbitrary boundary between a Seyfert 
and a LINER, depending on the adopted criteria. Because they are on the 
Seyfert side in HFS97a, they are included here and classified as ``S/L''.  
In objects like NGC 676, NGC 2685, NGC 3185, NGC 4472, and NGC 4725, the starlight 
subtraction process has been particularly difficult, leading to uncertain classifications.
Following HFS97b, we marked these sources with a quality rating ``:'' 
(uncertain) or ``::'' (highly uncertain).

\small

\vfill\eject

\centerline{\bf Table 2: Observation Summary}
\begin{tabular}{lcccc}
\hline
\hline
\multicolumn{1}{c}{Galaxy} &
\multicolumn{1}{c}{Start} &
\multicolumn{1}{c}{Obs.} &
\multicolumn{1}{c}{Exposure (s)} &
\multicolumn{1}{c}{Filter} \\
\multicolumn{1}{c}{Name} &
\multicolumn{1}{c}{Date (UT)} &
\multicolumn{1}{c}{orbit} &
\multicolumn{1}{c}{M1/M2/pn} &
\multicolumn{1}{c}{M1/M2/pn} \\
\multicolumn{1}{c}{(1)} &
\multicolumn{1}{c}{(2)} &
\multicolumn{1}{c}{(3)} &
\multicolumn{1}{c}{(4)} &
\multicolumn{1}{c}{(5)} \\
\hline
NGC 676  & 2002-07-14 & 475 & 17296/17578/15757 & thin/thin/thick \\  
NGC 1058 & 2002-02-01 & 393 & 11749/11798/6574  & thin/thin/med \\    
NGC 1068 & 2000-07-29 & 117 & 38351/34735/33521 & med/med/med\\     
NGC 2685 & 2001-10-15 & 339 & 6397/6687/4841    & thin/thin/thin \\  
NGC 3031 & 2001-04-22 & 251 & -/-/84400$^*$         & -/-/med$^*$\\  
NGC 3079 & 2001-04-13 & 246 & 21434/21776/10922 & med/med/thin\\  
NGC 3185 & 2001-05-07 & 258 & 11791/12117/8041  & thin/thin/thick \\ 
NGC 3227 & 2000-11-28 & 178 & 36514/36517/32579 & med/med/med\\   
NGC 3486 & 2001-05-09 & 259 & 4600/4663/5061    & thin/thin/med \\   
NGC 3941 & 2001-05-09 & 259 & 6233/6286/5302    & thin/thin/med \\  
%NGC 3982 &           &     & XMM public  & in September 2005 \\  
NGC 4051 & 2002-11-22 & 541 & 47814/48311/41889 & med/med/med  \\  
NGC 4138 & 2001-11-26 & 360 & 13575/13567/8856  & thin/thin/med \\  
NGC 4151 & 2000-12-21 & 190 & 29233/29253/21241 & med/med/med \\  
NGC 4168 & 2001-12-04 & 364 & 21886/21916/16009 & thin/thin/med \\  
NGC 4258 & 2000-12-08 & 183 & 20224/20246/13535 & med/med/med \\   
NGC 4388 & 2002-07-07 & 472 & 8768/8798/3866    & med/med/thin \\  
NGC 4395 & 2002-05-31 & 562 & 37096/30453/10873 & thin/thin/thin \\   
NGC 4472 & 2002-06-05 & 456 & 15181/15300/11043 & thin/thin/thin \\  
NGC 4477 & 2002-06-08 & 457 & 13329/13393/8457  & thin/thin/med \\  
NGC 4501 & 2001-12-04 & 364 & 12645/12637/2565  & thin/thin/med \\ 
NGC 4565 & 2001-07-01 & 286 & 14100/14113/8912  & thin/thin/med \\   
NGC 4579 & 2003-06-12 & 642 & 19568/20015/15396 & thin/thin/thin \\  
NGC 4639 & 2001-12-16 & 370 & 13672/13683/8863  & thin/thin/med \\  
NGC 4698 & 2001-12-16 & 370 & 14321/14427/8961  & thin/thin/med \\  
NGC 4725 & 2002-06-14 & 460 & 17062/17076/11941 & thin/thin/med \\  
NGC 5033 & 2001-07-02 & 286 & 7372/7232/869     & thin/thin/med \\  
NGC 5194 & 2003-01-15 & 568 & 20429/20465/17218 & thin/thin/thin \\   
NGC 5273 & 2002-06-14 & 460 & 15414/15474/9263  & thin/thin/med \\  
\hline
%NGC 3982 &- C priority - & -              & - \\ 
\hline
\hline
\end{tabular}

Notes: Col. (1): galaxy name. Col. (2): observation start date. Col. (3): observation 
orbital period. Col. (4): 
cleaned exposure of MOS1/MOS2/pn. Col. (5): filters used for MOS1/MOS2/pn;
``*'' means that MOS1 was not considered because it was operated in 
fast uncompressed mode, and MOS2 was not considered because it was 
operated in full-frame mode, so the data could thus be affected by pile-up.

\vfill\eject

\centerline{\bf Table 3: Best-fit parameters for the X-ray spectral analysis}
\
\hspace{-2cm}{
\begin{tabular}{lrrrrrrlrrrr}
\hline
\hline
\multicolumn{1}{l}{Galaxy name} &
\multicolumn{1}{r}{$N_{\rm H_{Gal}}$} &
\multicolumn{1}{r}{$N_{\rm H}$} &
\multicolumn{1}{r}{$\Gamma_{HX}$} &
\multicolumn{1}{r}{$\Gamma_{SX}$} &
\multicolumn{1}{r}{$kT$} &
\multicolumn{1}{r}{EW(Fe~K)} &
\multicolumn{1}{l}{$\chi^{2}_{red}$/$\nu$} &
\multicolumn{1}{r}{$F_{SX}$} &
\multicolumn{1}{r}{$F_{HX}$} &
\multicolumn{1}{r}{log $L^{int.}_{SX}$} &
\multicolumn{1}{r}{log $L^{int.}_{HX}$} \\
%\multicolumn{1}{l}{} &
%\multicolumn{1}{r}{(10$^{20}$)} &
%\multicolumn{1}{r}{(10$^{22}$)} &
%\multicolumn{1}{r}{} &
%\multicolumn{1}{r}{(keV/cgs/-)} &
%\multicolumn{1}{r}{(eV)} &
%\multicolumn{1}{l}{} &
%\multicolumn{1}{r}{($\times$10$^{-13}$)} &
%\multicolumn{1}{r}{($\times$10$^{-13}$)} &
%\multicolumn{1}{r}{log} &
%\multicolumn{1}{r}{log} \\
\multicolumn{1}{l}{(1)} &
\multicolumn{1}{r}{(2)} &
\multicolumn{1}{r}{(3)} &
\multicolumn{1}{r}{(4)} &
\multicolumn{1}{r}{(5)} &
\multicolumn{1}{r}{(6)} &
\multicolumn{1}{r}{(7)} &
\multicolumn{1}{l}{(8)} &
\multicolumn{1}{r}{(9)} &
\multicolumn{1}{r}{(10)} &
\multicolumn{1}{r}{(11)} &
\multicolumn{1}{r}{(12)} \\
\hline
NGC 676  &4.4 &$\leq$ 0.1       &1.9$\pm$0.3    & - & -                                &-            &1.2/12   &     0.1 &     0.2 &   38.9 &   39.0 \\ 
NGC 1058$^{\ddagger}$&6.7&$\leq$ 0.6&1.3$\pm$0.9  & - & -                                &-            &0.2/2    &$<$  0.1 &$<$  0.4 &$<$38.1 &$<$38.6 \\ 
NGC 1068$^{\dagger*}$&3.5&$\leq$ 0.1&1.0$\pm$0.1&3.5$\pm$0.5& 0.7$\pm$0.2              &1200$\pm$500 &4.3/1399 &   111.3 &    46.2 &   41.4 &   41.1 \\ 
NGC 2685 &4.1 &$\leq$ 0.3       &0.5$\pm$0.2    & - &-                                 &-            &1.5/8    &     0.2 &     2.7 &   38.8 &   39.9 \\ 
NGC 3031$^{*}$ &4.1 &$\leq$ 0.1 &1.9$\pm$0.1    &-&0.6$\pm$0.1                         &40$\pm$20    &1.2/1651 &    87.0 &   120.0 &   40.2 &   40.3 \\ 
NGC 3079$^{\dagger}$&0.8&0.05$\pm$0.03&1.7$\pm$0.1&-&0.7$\pm$0.1                       &1480$\pm$500 &1.1/264  &$<$  2.5 &     3.3 &$<$40.0 &   40.1 \\ 
NGC 3185 &2.1 &$\leq$ 0.2       &2.1$\pm$0.1    &-&-                                   &-            &0.6/5    &     0.2 &     0.2 &   39.0 &   39.0 \\ 
NGC 3227 &2.2 & 6.8$\pm$0.3     &1.5$\pm$0.1    &$\equiv$$\Gamma_{HX}$&-               &190$\pm$40  &1.1/1863  &     4.0 &    81.4 &   41.2 &   41.7 \\ 
NGC 3486 &1.9 &$\leq$ 0.3       &0.9$\pm$0.2    &-& -                                  &-            &1.1/6    &     0.2 &     1.1 &   38.0 &   38.9 \\ 
NGC 3941 &1.9 &$\leq$ 0.1       &2.1$\pm$0.3    &-& -                                  &-            &1.8/8    &     0.4 &     0.4 &   38.8 &   38.9 \\ 
NGC 4051$^{*}$ &1.3 &$\leq$0. 3 &1.2$\pm$0.1    &3$\pm$0.5&0.2$\pm$0.1                 &240$\pm$40   &2.2/2006 &    47.3 &    62.7 &   41.2 &   41.3 \\ 
NGC 4138 &1.4 &8.0$\pm$1        &1.5$\pm$0.1    &$\equiv$$\Gamma_{HX}$&0.3$\pm$0.1     &83$\pm$30    &1.0/408  &     0.6 &    55.0 &   40.9 &   41.3 \\ 
NGC 4151$^{*}$ &2.0 &7.5$\pm$1  &1.6$\pm$0.2    &1.8$\pm$0.5&-                         &300$\pm$30   &1.9/2785 &    29.7 &   451.0 &   42.0 &   42.5 \\ 
NGC 4258$^{\dagger}$&1.2&8.7$\pm$0.3&1.7$\pm$0.1&1.9$\pm$0.1&0.5$\pm$0.1               &27$\pm$20    & 1.0/899   &$<$  3.4 &    83.7 &$<$40.7 &   40.9 \\ 
NGC 4388 &2.6 &27$\pm$2         &1.3$\pm$0.2    &$\equiv$$\Gamma_{HX}$&0.3$\pm$0.1     &450$\pm$70   & 1.0/283 &     2.2 &    76.2 &   41.2 &   41.8 \\ 
NGC 4395 &1.3 &5.3$\pm$0.3      &1.2$\pm$0.1    &$\equiv$$\Gamma_{HX}$&0.2$\pm$0.1     &100$\pm$25   & 0.97/535&     1.0 &    61.6 &   39.2 &   39.8 \\ 
NGC 4472$^{\ddagger}$&1.7&0.3$\pm$0.1&2.7$\pm$0.3&-&0.8$\pm$0.2                        &-            &1.8/483  &$<$ 19.8 &$<$  3.8 &$<$40.8 &$<$40.1 \\ 
NGC 4477 &2.6 &$\leq$ 2         &1.9$\pm$0.3    &-&0.4$\pm$0.1                         &-            &1.0/106  &$<$  2.0 &     1.2 &$<$40.0 &   39.6 \\ 
NGC 4501 &2.5 &$\leq$ 0.2       &1.5$\pm0.3$    &-&0.4$\pm$0.1                         &-            &1.1/28   &     0.9 &     1.1 &   39.5 &   39.6 \\ 
NGC 4565 &1.3 &0.12$\pm$0.04    &1.8$\pm$0.2    &-&-                                   &-            &1.2/70   &     1.1 &     2.4 &   39.2 &   39.4 \\ 
NGC 4579 &2.4 &$\leq$0.02       &1.7$\pm$0.1    &-&0.6$\pm$0.1                         &170$\pm$50   &1.1/1043 &    25.3 &    38.5 &   41.0 &   41.1 \\ 
NGC 4639 &2.4 &$\leq$ 0.01      &1.8$\pm$0.1    &-&-                                   &-            &1.0/176  &     3.0 &     4.9 &   40.0 &   40.2 \\ 
NGC 4698 &1.9 &$\leq$ 0.4       &2.0$\pm$0.2    &-&-                                   &-            &1.2/17   &     0.3 &     0.4 &   39.1 &   39.2 \\ 
NGC 4725 &0.1 &$\leq$ 5         &1.9$\pm$0.5    &-&0.3$\pm$0.1                         &-            &0.84/44  &     0.7 &     0.4 &   39.2 &   38.9 \\ 
NGC 5033 &0.1 &$\leq$ 0.03      &1.7$\pm$0.1    &-&-                                   &466$\pm$215  &0.97/335 &    15.0 &    28.7 &   40.8 &   41.1 \\ 
NGC 5194 &1.6 &$\leq$ 0.03      &0.6$\pm$0.1    &-&0.5$\pm$0.1                         &986$\pm$210  &1.65/361 &     7.3 &     4.8 &   39.8 &   39.6 \\	
NGC 5273 &0.1 &0.9$\pm$0.1     &1.4$\pm$0.1    &$\equiv$$\Gamma_{HX}$&0.2$\pm$0.1     &226$\pm$75   &1.1/1009 &    13.2 &    67.1 &   41.0 &   41.4 \\ 
% to be inserted: NGC 3982 &-   &-                &-              &-         &-     &-   &   	      & 	  	 &	       &             \\ 
%% excluded for D>30 Mpc:   NGC 4168 &2.6 &$\leq$ 0.03    &2.0$\pm$0.2 &-   &- &1.1/51 & 4.3E-14 &   5.1E-14 &    4.6E-14 &   5.1E-14 & 1.25 & 39.69 & 39.77  \\ 
\hline
\hline
\hline
\end{tabular}
\par\noindent
Notes: SX = 0.5--2 keV, HX = 2--10 keV.
Galaxies marked with $^{\dagger}$ and $^{\ddagger}$ indicate that the 
nuclear regions are contaminated (by more than 10\%) by off-nuclear 
point-sources and/or diffuse emission in only the soft band (${\dagger}$) or 
in both the soft and hard bands (${\ddagger}$). 
Galaxies marked with ``*'' indicate sources with very complex spectra 
for which only a rough parametrization is given here.
See Appendix B for more details on individual sources. 
Col. (1): galaxy name. 
Col (2): Galactic absorption along the line of sight, in units of 10$^{20}$ cm$^{-2}$. 
Col. (3): measured absorption column density, in units of 10$^{22}$ cm$^{-2}$. 
Col. (4): power-law photon index. Col. (5): photon index of the soft power-law 
component.
% ($\Gamma_{SX}$),
%ionization parameter of the warm absorber ($\xi$).
%***Massimo: I don't understand the two lines above, which I have
% ``commented out,'' because I don't see in this column any values of
% \xi. Please explain.
Col. (6): temperature of the thermal component ($kT$) in units of keV.
Col. (7): equivalent width of the Fe~K line, in units of keV. 
Col. (8): reduced chi-squared and number of degrees of freedom.
Col. (9)--(10): observed fluxes in the soft (0.5--2 keV) and hard (2--10 keV) 
X-ray bands, in units of 10$^{-13}$ erg cm$^{-2}$ s$^{-1}$.
Col. (11)--(12): Log of the absorption-corrected luminosities in the soft 
(0.5--2 keV) and hard (2--10 keV) X-ray bands 
(computed using distances from Table 1).

\begin{table}
\centerline{\bf Table 4: Compton-thick/thin Seyfert 2 candidates}
\small{
\begin{center}
\begin{tabular}{lcccc}
\hline
\hline
\multicolumn{1}{c}{Name} &
\multicolumn{1}{c}{$\Gamma_{\rm 2-10~keV}$} &
\multicolumn{1}{c}{EW$_{\rm Fe~K}$} &
\multicolumn{1}{c}{Flux diag.} &
\multicolumn{1}{c}{Thick?} \\
\multicolumn{1}{c}{(1)} &
\multicolumn{1}{c}{(2)} &
\multicolumn{1}{c}{(3)} &
\multicolumn{1}{c}{(4)} &
\multicolumn{1}{c}{(5)} \\
\hline
\hline
NGC 676  &       &       &Thick   & $\surd$      \\   
NGC 1058 &       &       &Thick/SB& ?            \\
NGC 1068 & Thick &Thick  &Thick   & $\surd$      \\
NGC 2685 & Thick &       &Thin    & ?            \\
NGC 3079 &       &Thick  &Thick   & $\surd$      \\
NGC 3185 &       &       &Thick   & $\surd$      \\
NGC 3486 & Thick &       &Thick/SB& ?            \\
NGC 3941 &       &       &Thin    & $\times$     \\
NGC 4138 & Thick &Thin   &Thin    & $\times$     \\
NGC 4258 & Thick &Thin   &Thin    & $\times$     \\
NGC 4388 &       &Thin   &Thin    & $\times$     \\
NGC 4472 &       &       &Thin    & ?            \\
NGC 4477 &       &       &Thin    & $\times$     \\
NGC 4501 & Thin  &       &   Thin & $\times$     \\
NGC 4565 &       &Thin   &   Thin & $\times$     \\
NGC 4698 &       &       &Thin    & $\times$     \\
NGC 4725 &       &       &Thin    & $\times$     \\
NGC 5194 &       &Thick  & Thick  & $\surd$      \\
\hline
\end{tabular}
\end{center}
Notes: ``Thick'' = Compton-thick candidate, ``Thin'' = Compton-thin candidate,
``SB'' = starburst candidate on the basis of spectral diagnostics (cols. 2 and 3) 
and flux diagnostics (col. 4). Final classification (col. 5): 
``$\surd$'' = Compton-thick candidates, ``$\times$'' = Compton-thin candidate,
and ``?'' = classification uncertain (likely to contain a starburst, but CT nature 
cannot be excluded).} 
\label{ct}
\end{table}

%***Massimo: in Table 4, I don't see any ``SB'' listed, so why do we
% mention it in the notes?

\vfill\eject

\begin{figure}[!]
\parbox{7cm}{
\psfig{file=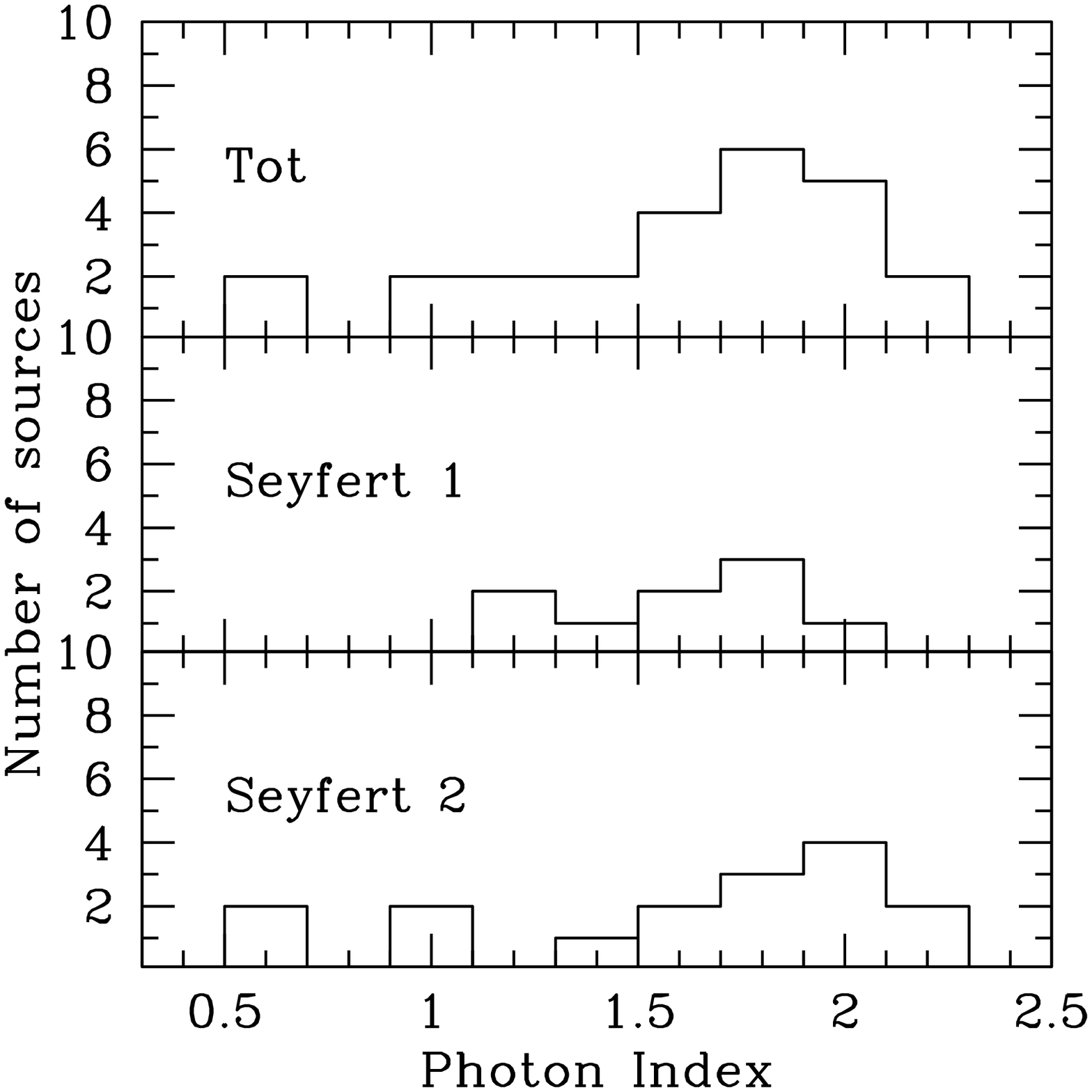,width=8cm,height=8cm,angle=0}}
\hspace{0.1cm} \
\parbox{7cm}{
\psfig{file=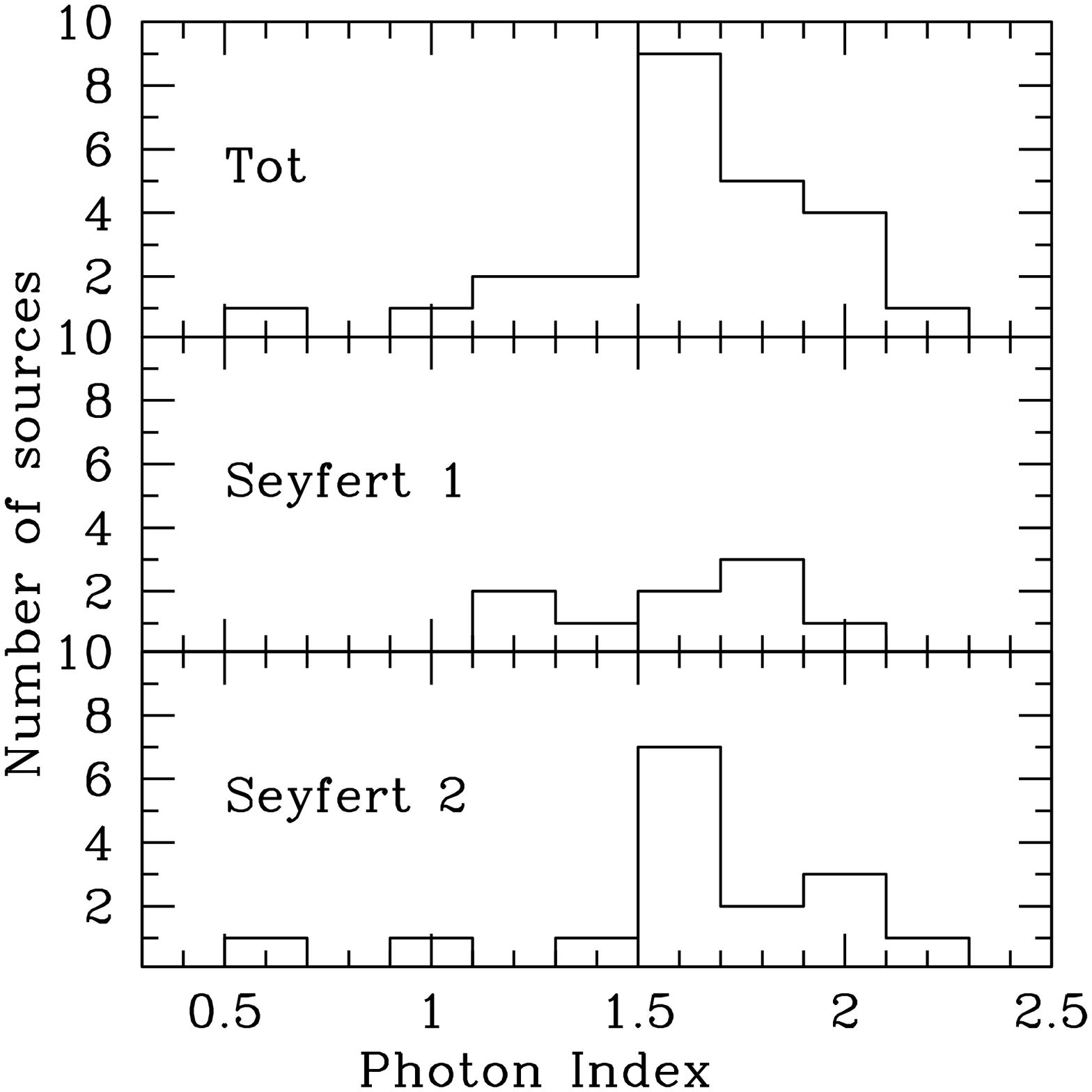,width=8cm,height=8cm,angle=0}}
\caption{Distribution of the photon index before (left) 
and after (right) correction for Compton-thick candidates (after 
Table 4) for which a value of $\Gamma$=1.62 is assumed (see text for details).}
\end{figure}

\begin{figure}[!]
\parbox{7cm}{
\psfig{file=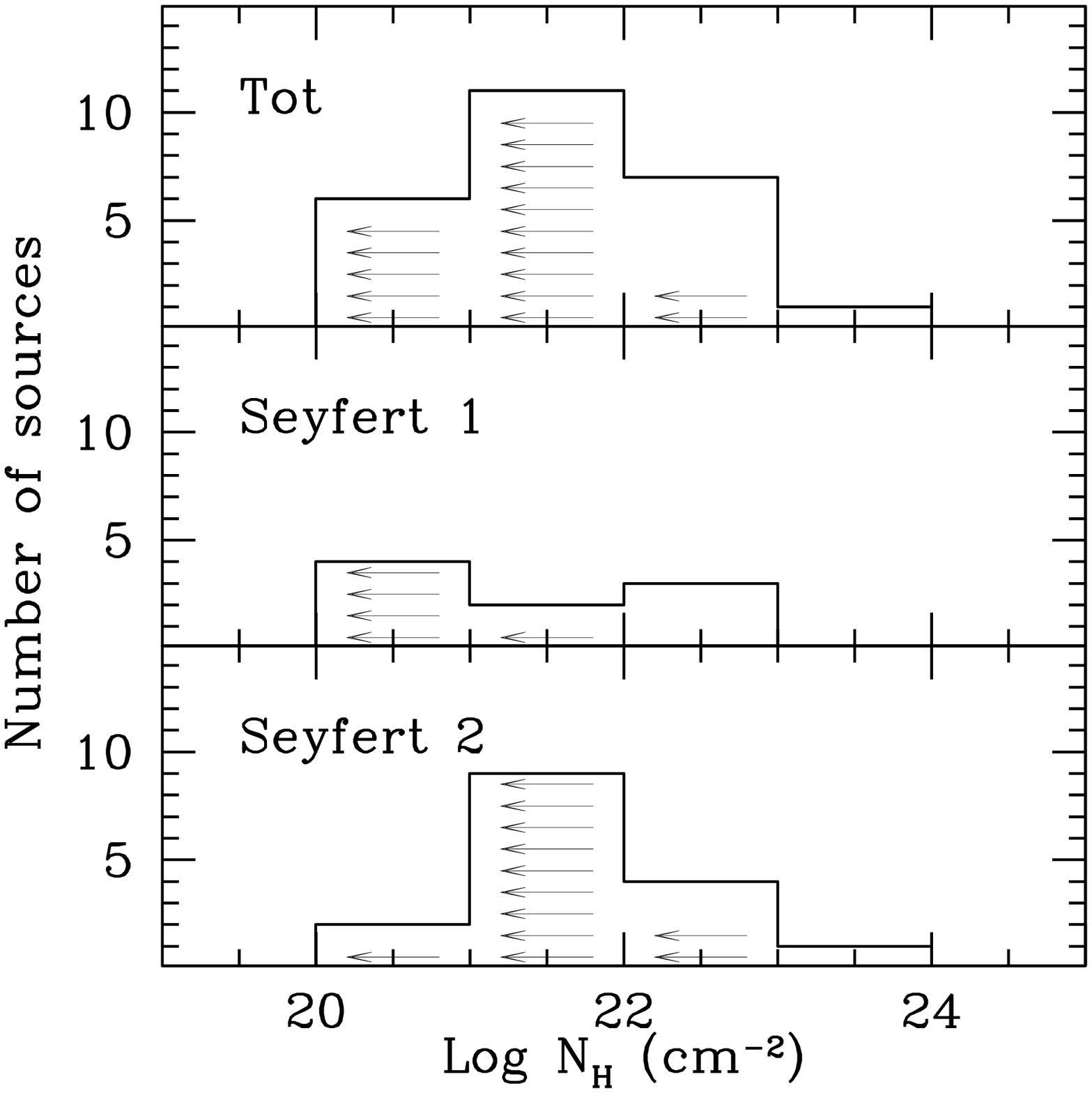,width=8cm,height=8cm,angle=0}}
\hspace{0.1cm} \
\parbox{7cm}{
\psfig{file=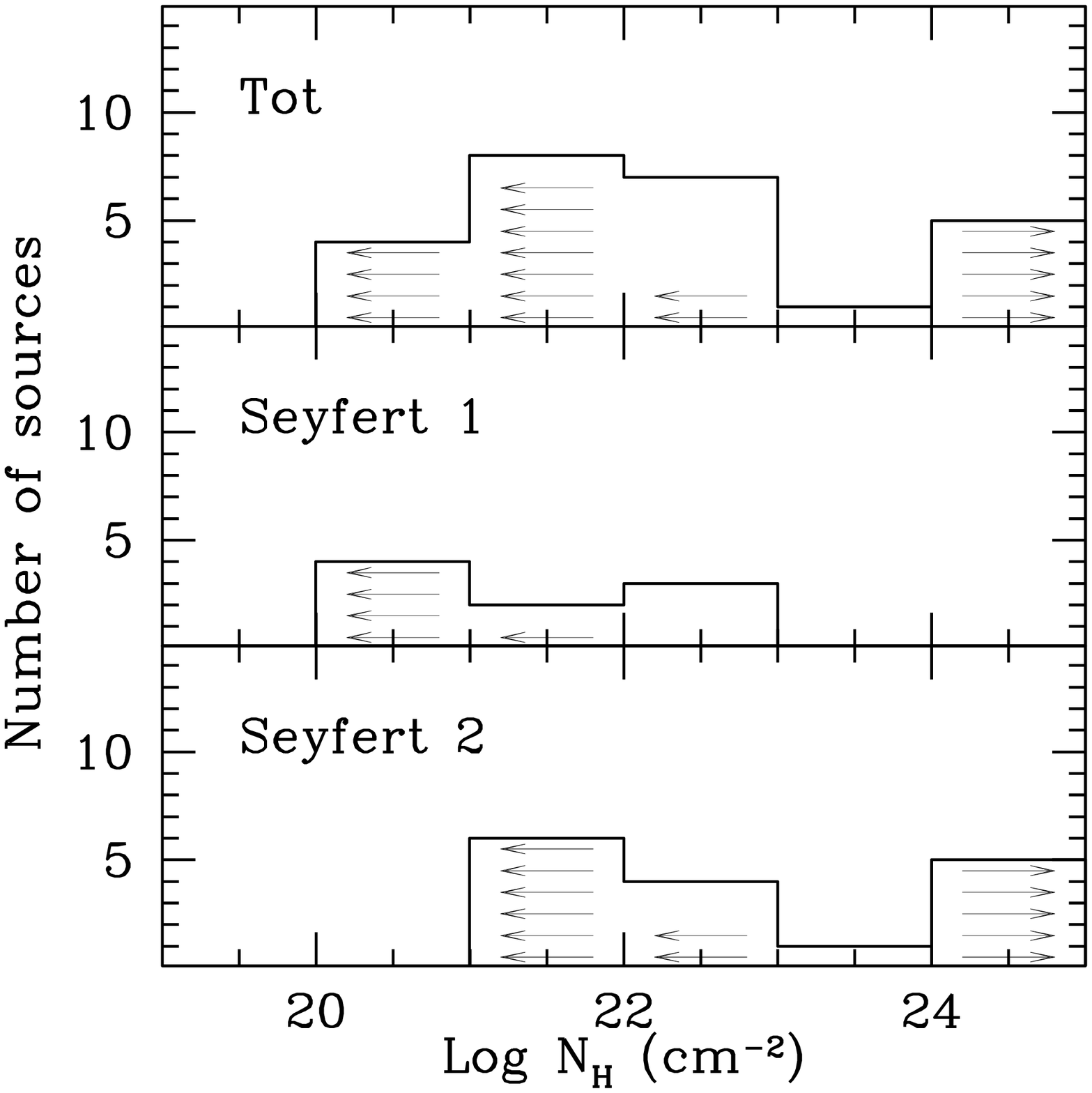,width=8cm,height=8cm,angle=0}}
\caption{Distribution of the absorption column densities before (left) 
and after (right) correction for Compton-thick candidates (after 
Table 4). Upper and lower limits are indicated with arrows.}
\end{figure}

\begin{figure}[!]
\parbox{7cm}{
\psfig{file=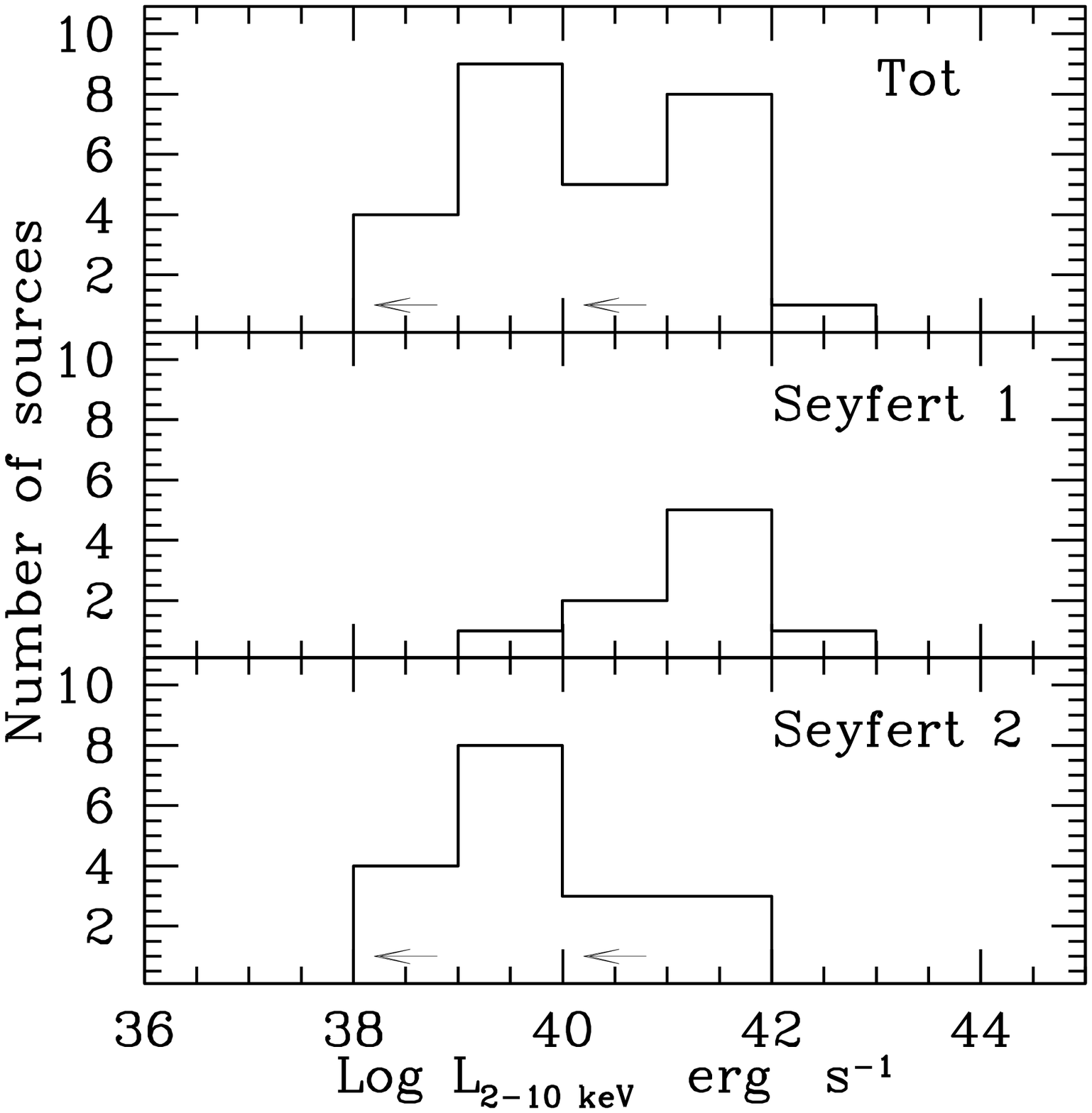,width=8cm,height=8cm,angle=0}}
\hspace{0.1cm} \
\parbox{7cm}{
\psfig{file=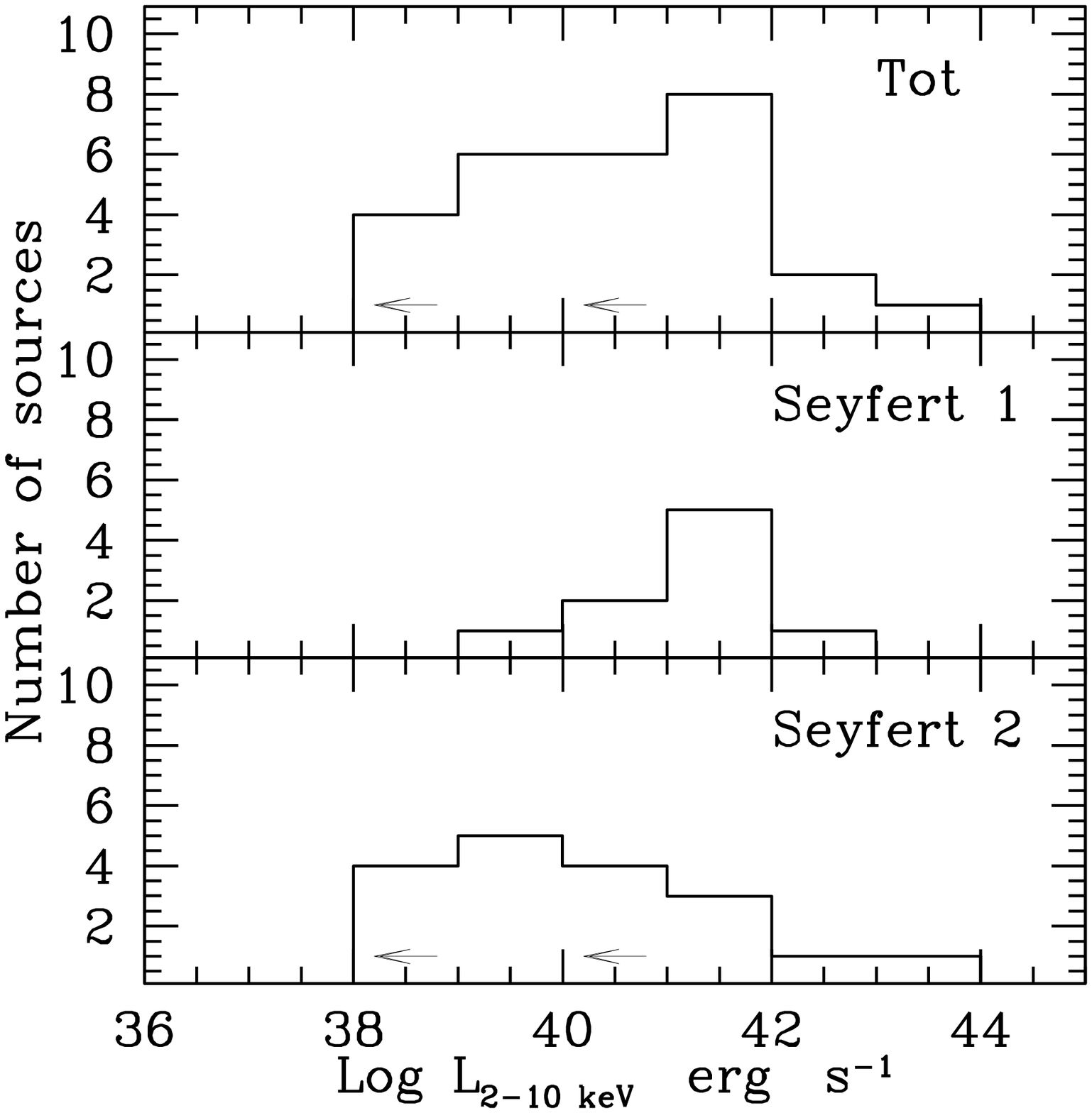,width=8cm,height=8cm,angle=0}}
\caption{Distribution of the 2--10~keV luminosities before (left) 
and after (right) correction for Compton-thick candidates (after 
Table 4). Upper limits have been indicated with arrows.}
\end{figure}

\begin{figure}[!]
\psfig{file=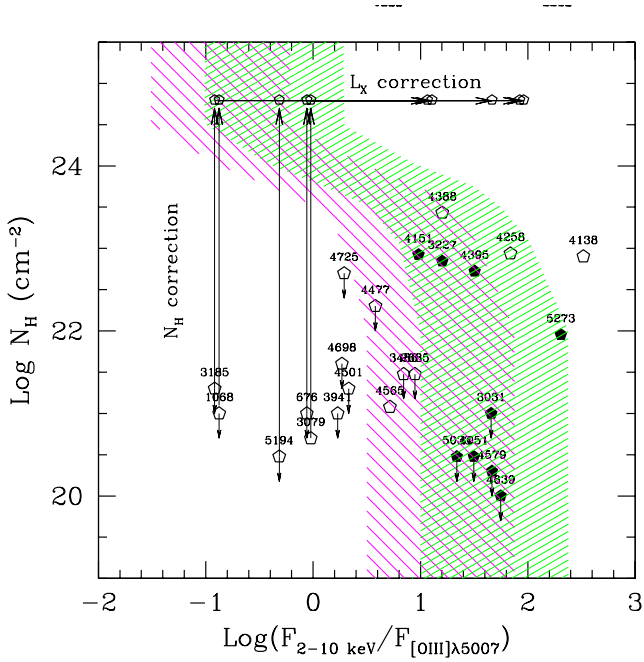,width=9cm,height=9cm,angle=0}
\caption{Diagram of the absorbing column density $N_{\rm H}$ versus
the ratio between the observed 2--10~keV flux and the reddening-corrected 
[O~III] flux taken from Panessa (2004). Filled polygons are type 1 
Seyferts and open polygons are type 2 Seyferts. The shaded region 
(lower-left to upper-right diagonals) indicates the expected correlation 
under the assumption that $L_{2-10~\rm keV}$ is absorbed by the $N_{\rm H}$ 
reported on the ordinate, starting from the average 
$F_{\rm HX}$/$F_{\rm [O~III]}$ ratio observed in type 1 Seyfert galaxies
and by assuming a 1\% reflected component. Also, the shaded region 
(upper-left to lower-right diagonals) obtained by Maiolino et al. 
(1998) is shown.}
\end{figure}

\
\vfill\eject
\vfill\eject
\
\vspace{10cm}

\appendix

\section{Atlas of X-ray spectra} 

Spectra in order of increasing NGC number are shown.
For each object, the upper panel shows the unfolded spectrum and the 
baseline parameterization, together with the contributions to the model of 
the various additive components. Residuals are shown in the lower panel in 
units of $\sigma$. Spectral analysis was performed following general 
``recipes'' described in \S~4, and as described in the notes on individual 
sources given in Appendix B.

\begin{figure}[!]
\vspace{0.5cm}{
\parbox{8cm}{
\hspace{5.5cm}{\Large NGC 676}\\
\vspace{-1.25cm}
\psfig{file=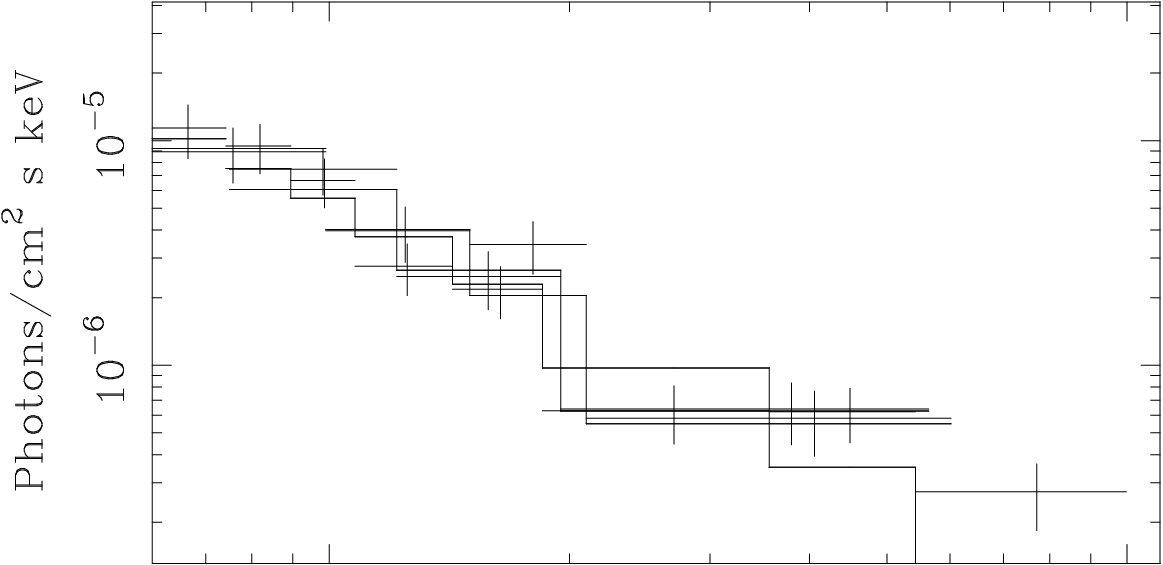,width=8cm,height=3.5cm,angle=-90}
\psfig{file=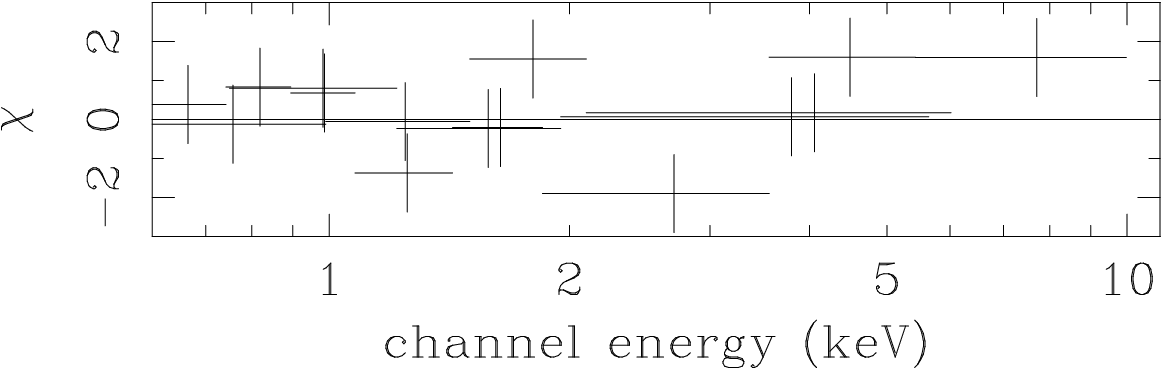,width=8cm,height=2cm,angle=-90}}
\hspace{0.1cm} \
\parbox{8cm}{
\hspace{5.5cm}{\Large NGC 1058}\\
\vspace{-1.25cm}
\psfig{file=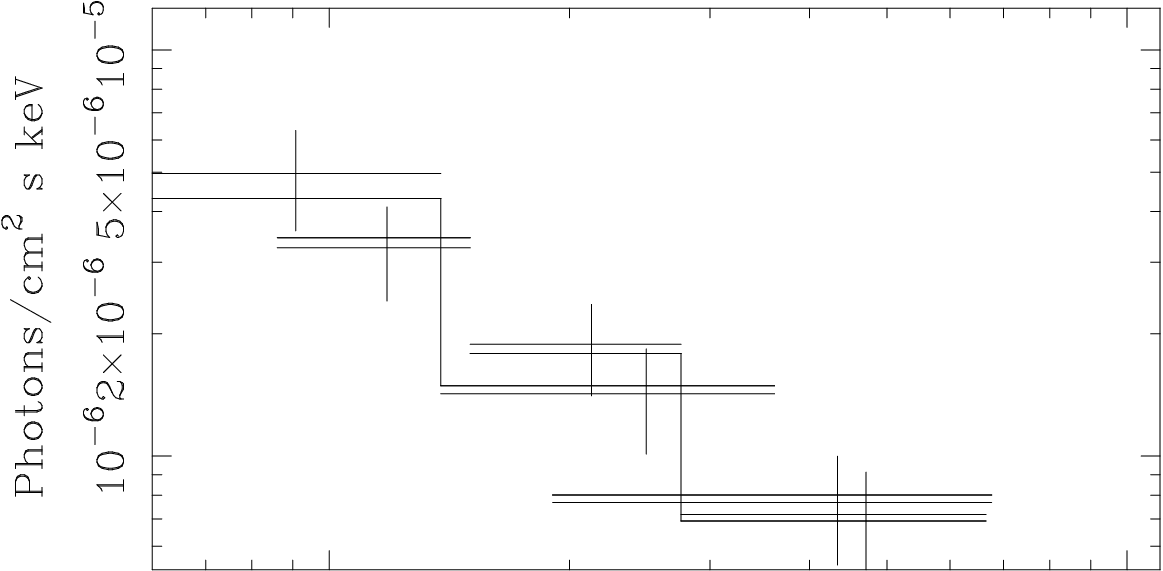,width=8cm,height=3.5cm,angle=-90}
\psfig{file=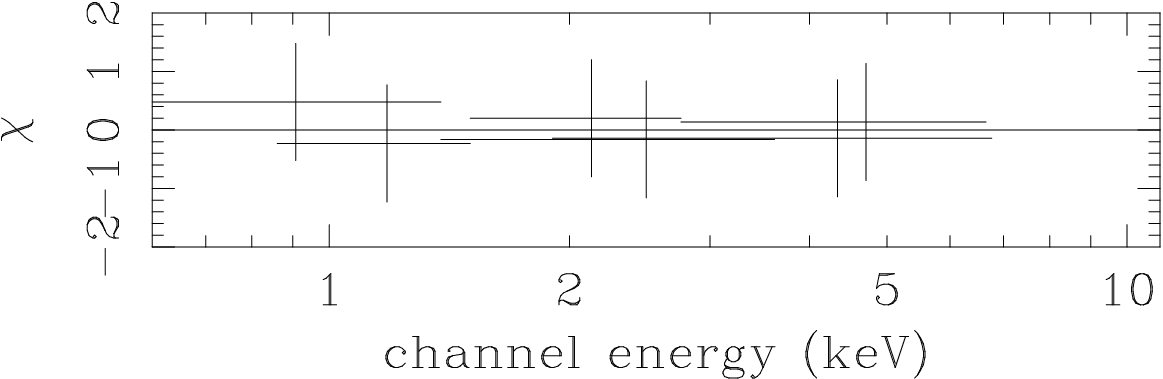,width=8cm,height=2cm,angle=-90}}
}
\\ \\ \\ \\
\parbox{8cm}{
\hspace{5.5cm}{\Large NGC 1068}\\
\vspace{-1.25cm}
\psfig{file=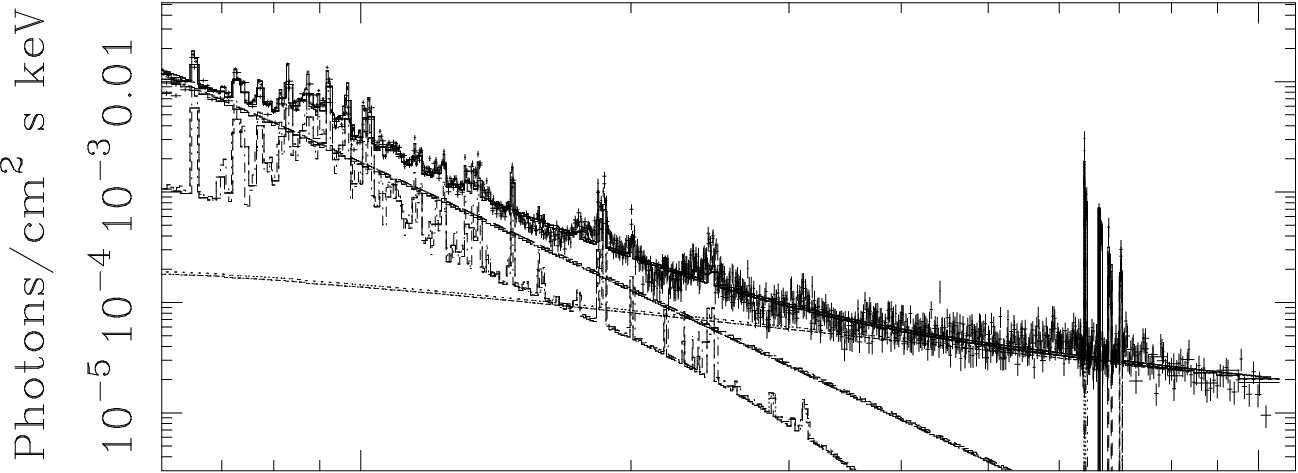,width=8cm,height=3.5cm,angle=-90}
\psfig{file=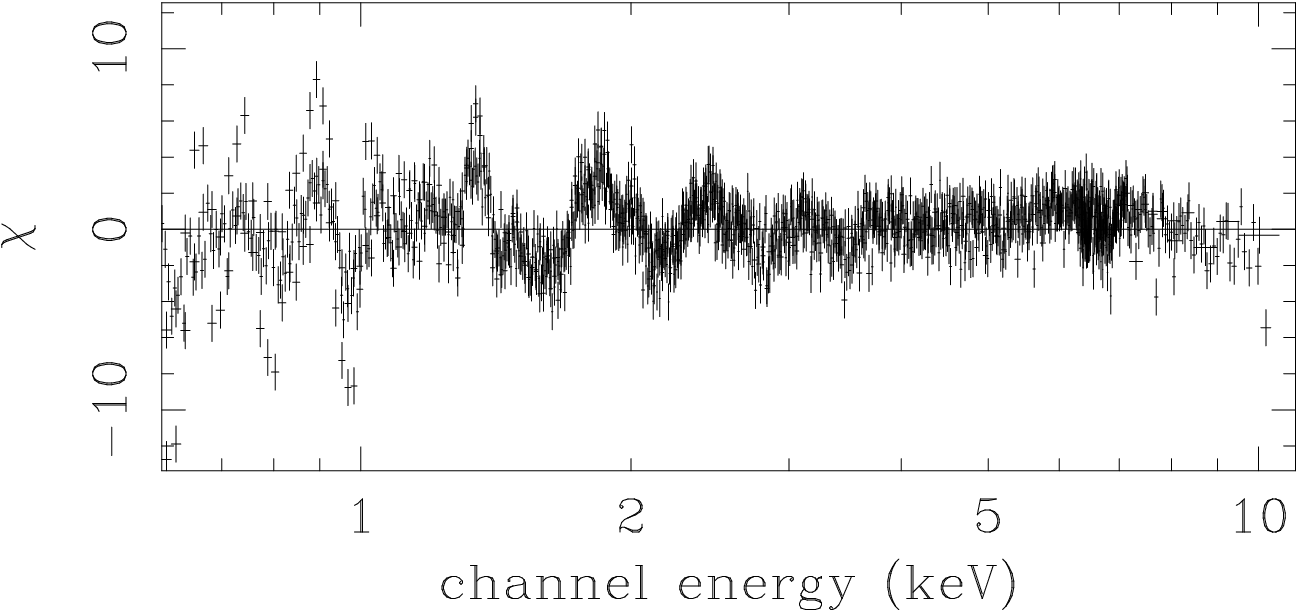,width=8cm,height=2cm,angle=-90}}
\hspace{0.1cm} \
\parbox{8cm}{
\hspace{5.5cm}{\Large NGC 2685}\\
\vspace{-1.25cm}
\psfig{file=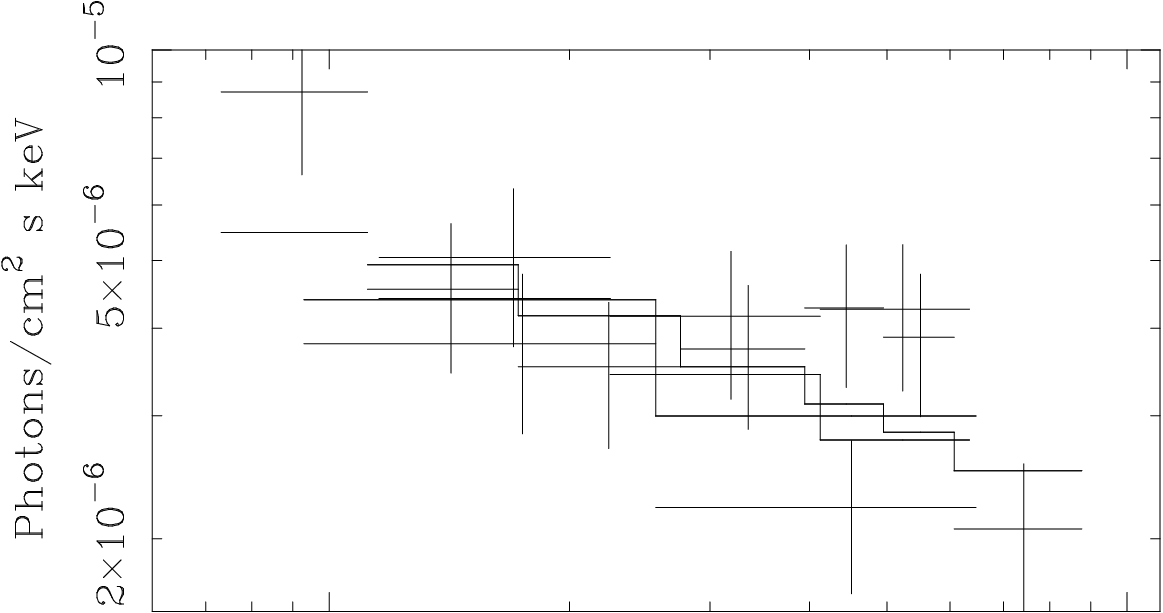,width=8cm,height=3.5cm,angle=-90}
\psfig{file=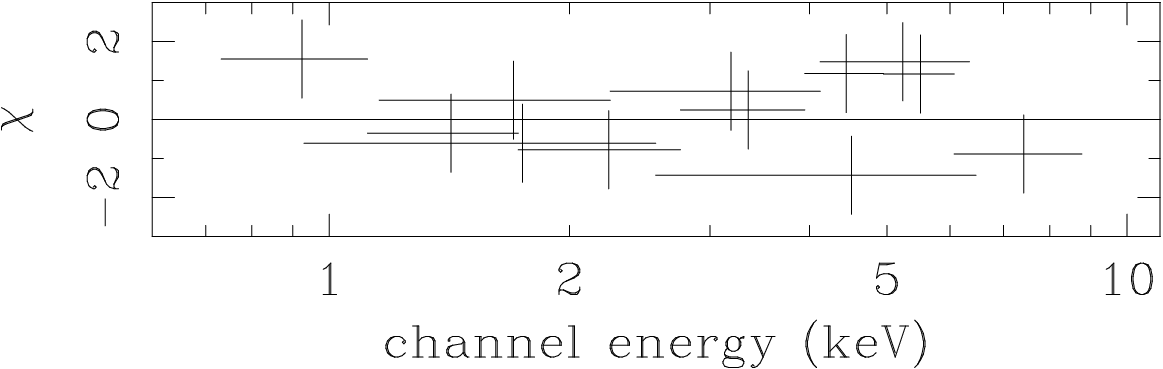,width=8cm,height=2cm,angle=-90}}
\\ 
\end{figure}

\begin{figure}[!]
\vspace{0.5cm}{
\parbox{8cm}{
\hspace{5.5cm}{\Large NGC 3031}\\
\vspace{-1.25cm}
\psfig{file=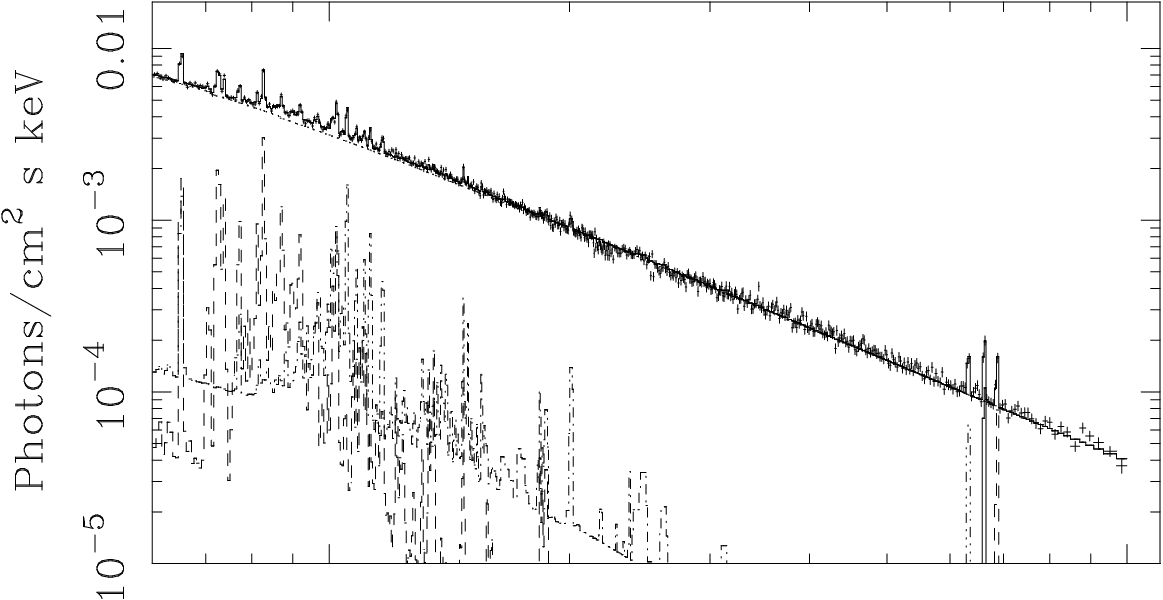,width=8cm,height=3.5cm,angle=-90}
\psfig{file=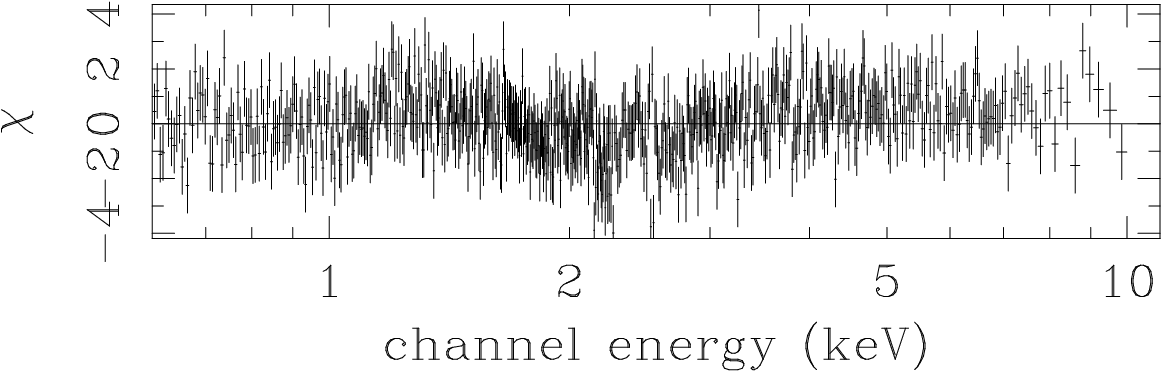,width=8cm,height=2cm,angle=-90}}
\hspace{0.1cm} \
\parbox{8cm}{
\hspace{5.5cm}{\Large NGC 3079}\\
\vspace{-1.25cm}
\psfig{file=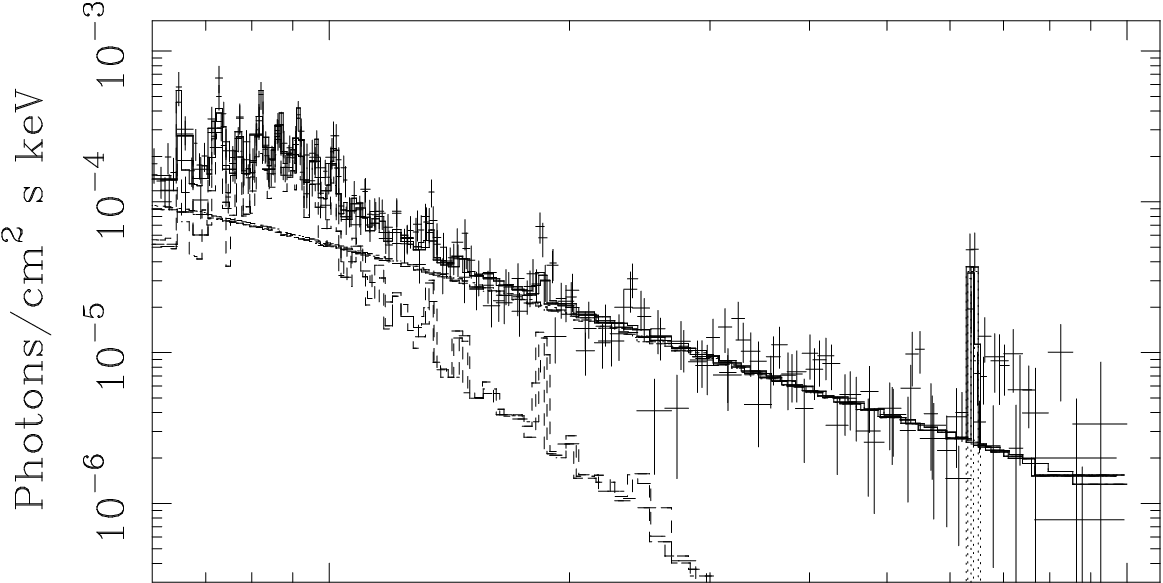,width=8cm,height=3.5cm,angle=-90}
\psfig{file=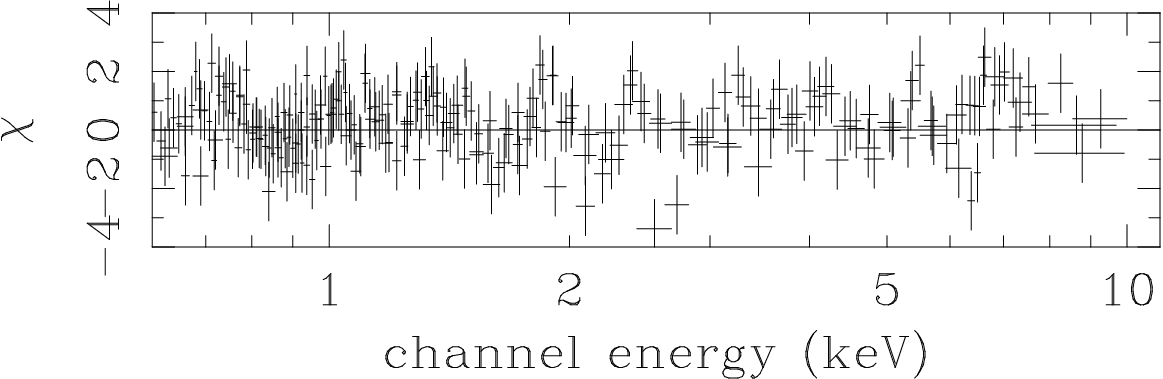,width=8cm,height=2cm,angle=-90}}
\\ \\ \\ \\
\parbox{8cm}{
\hspace{5.5cm}{\Large NGC 3185}\\
\vspace{-1.25cm}
\psfig{file=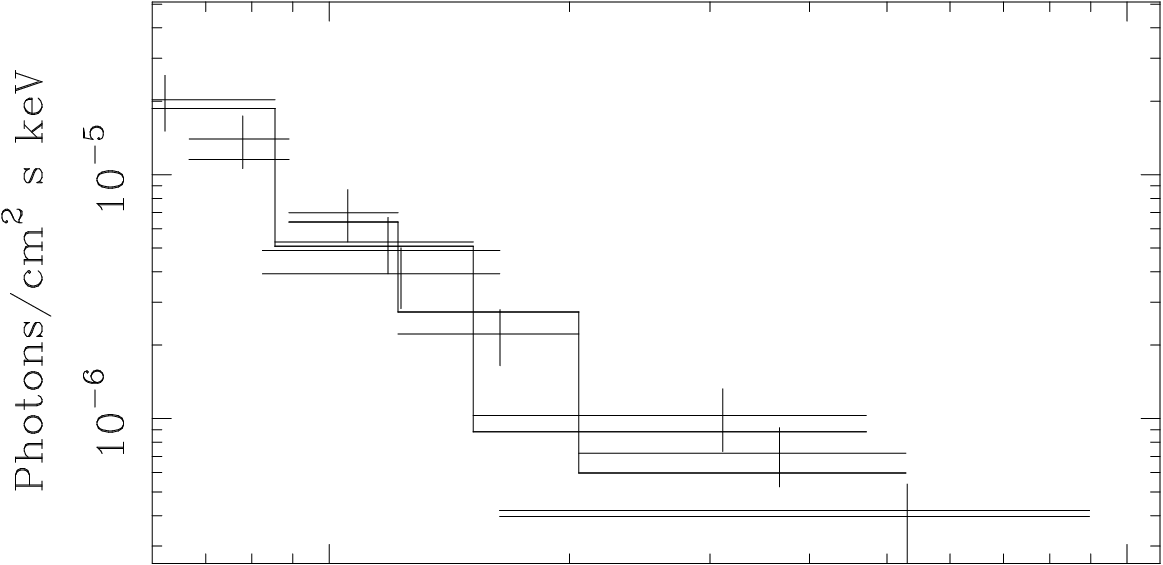,width=8cm,height=3.5cm,angle=-90}
\psfig{file=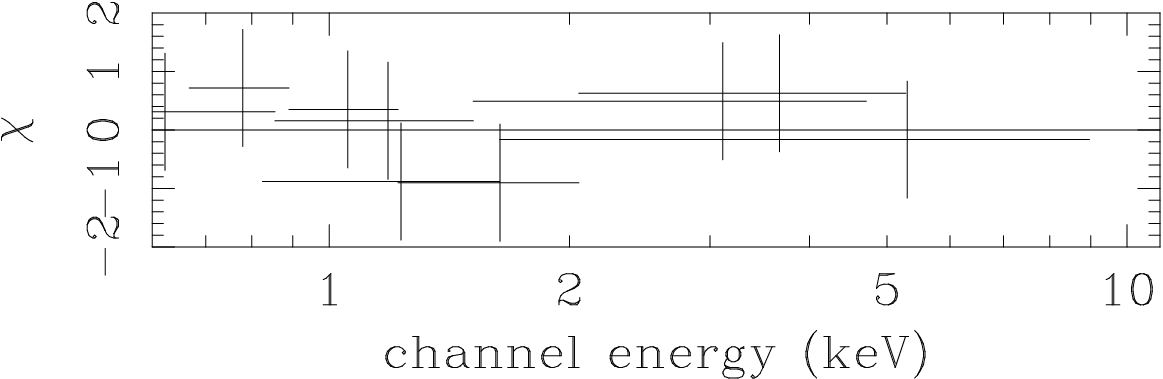,width=8cm,height=2cm,angle=-90}}
\hspace{0.1cm} \
\parbox{8cm}{
\hspace{5.5cm}{\Large NGC 3227}\\
\vspace{-1.25cm}
\psfig{file=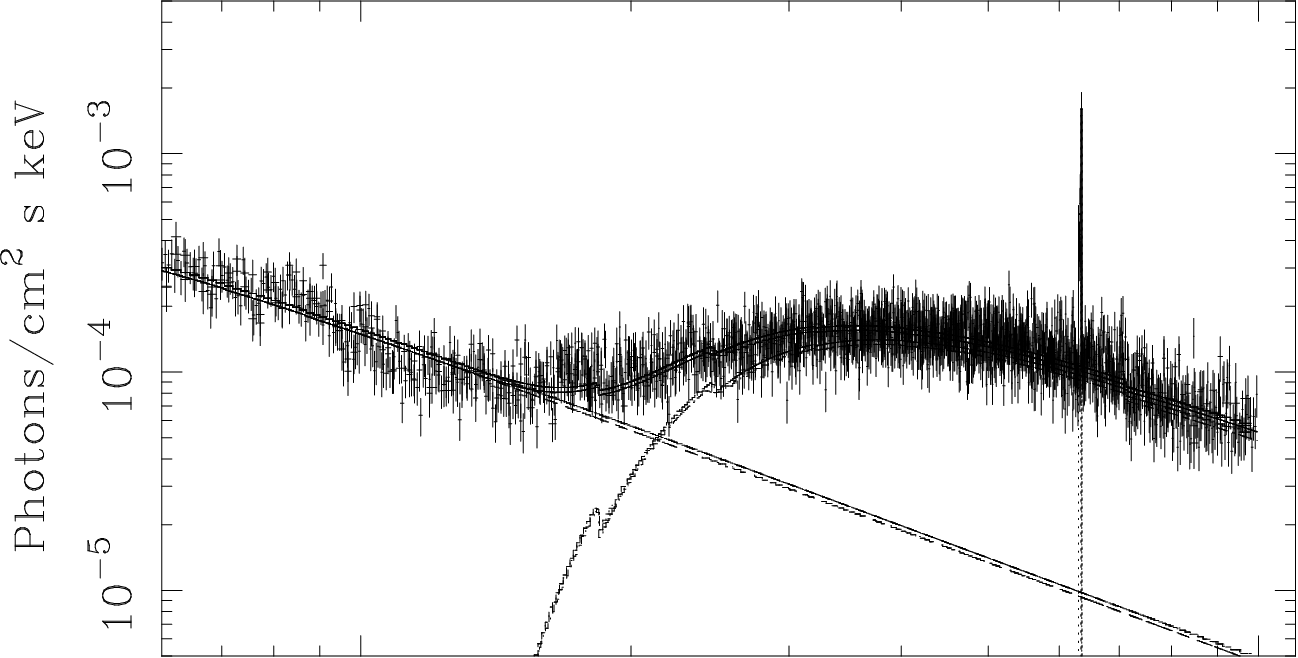,width=8cm,height=3.5cm,angle=-90}
\psfig{file=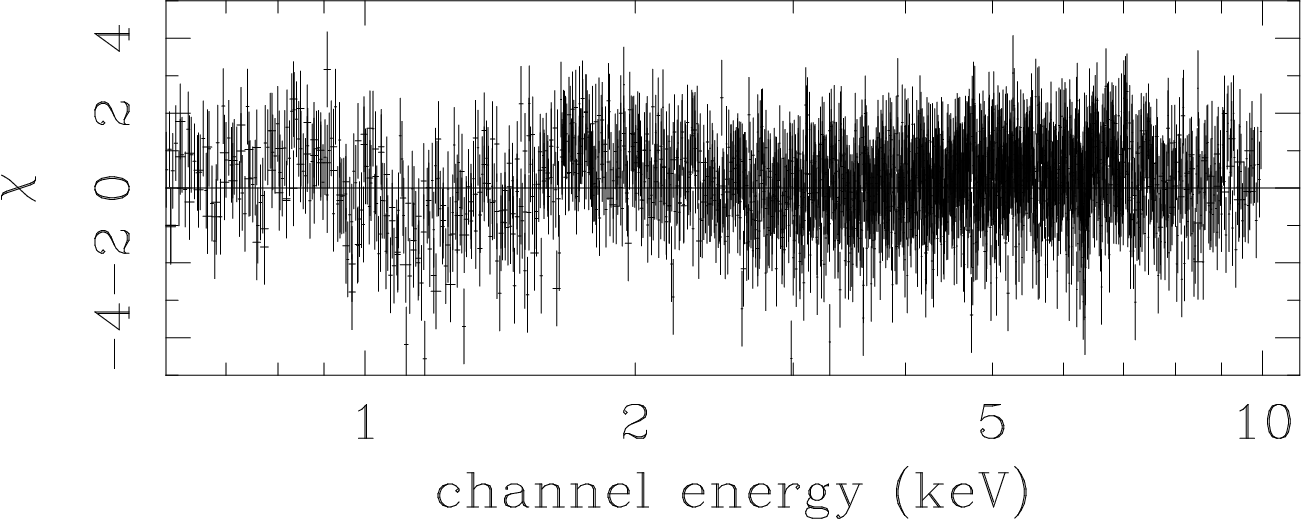,width=8cm,height=2cm,angle=-90}}
}\\ 
\end{figure}

\begin{figure}[!]
\vspace{0.5cm}{
\parbox{8cm}{
\hspace{5.5cm}{\Large NGC 3486}\\
\vspace{-1.25cm}
\psfig{file=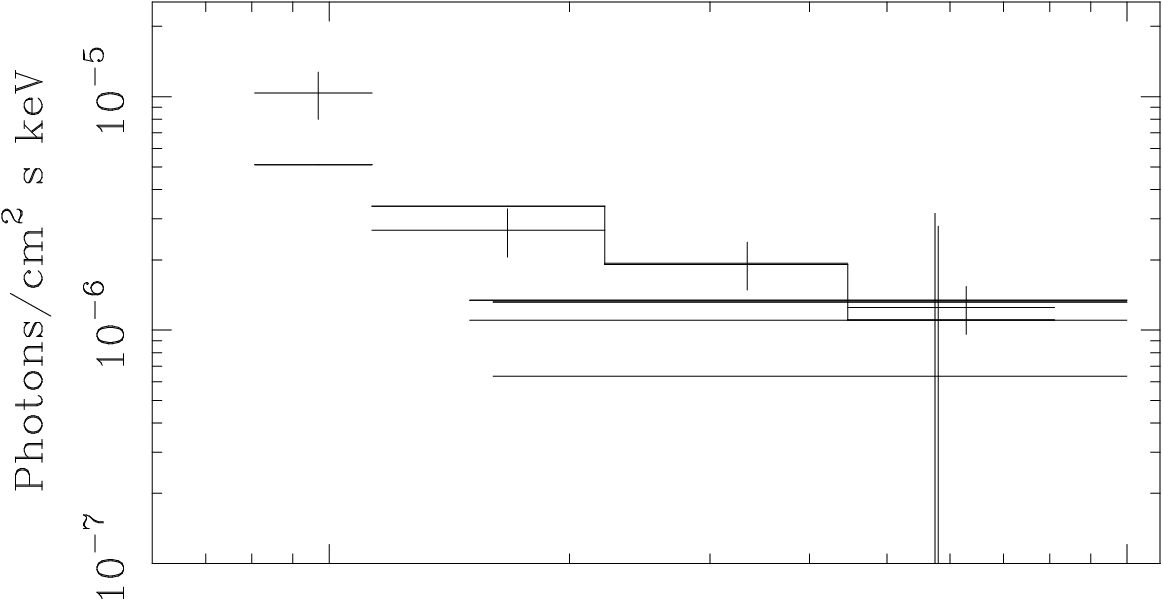,width=8cm,height=3.5cm,angle=-90}
\psfig{file=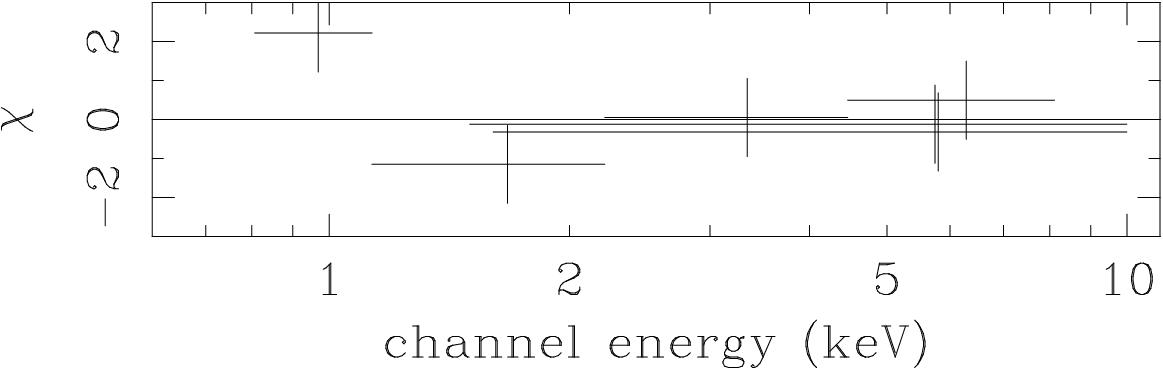,width=8cm,height=2cm,angle=-90}}
\hspace{0.1cm} \
\parbox{8cm}{
\hspace{5.5cm}{\Large NGC 3941}\\
\vspace{-1.25cm}
\psfig{file=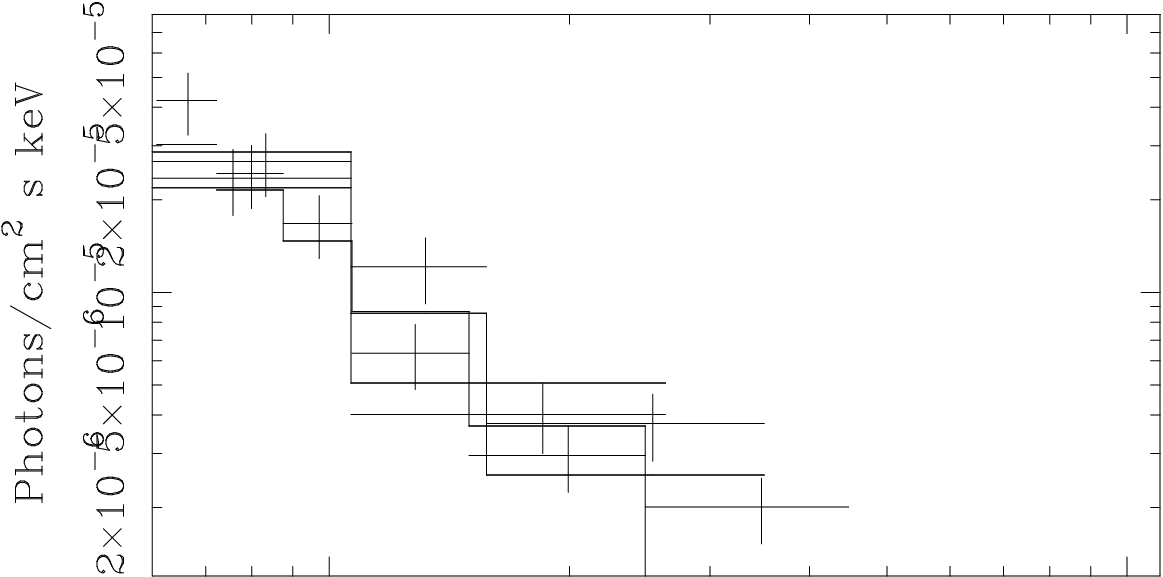,width=8cm,height=3.5cm,angle=-90}
\psfig{file=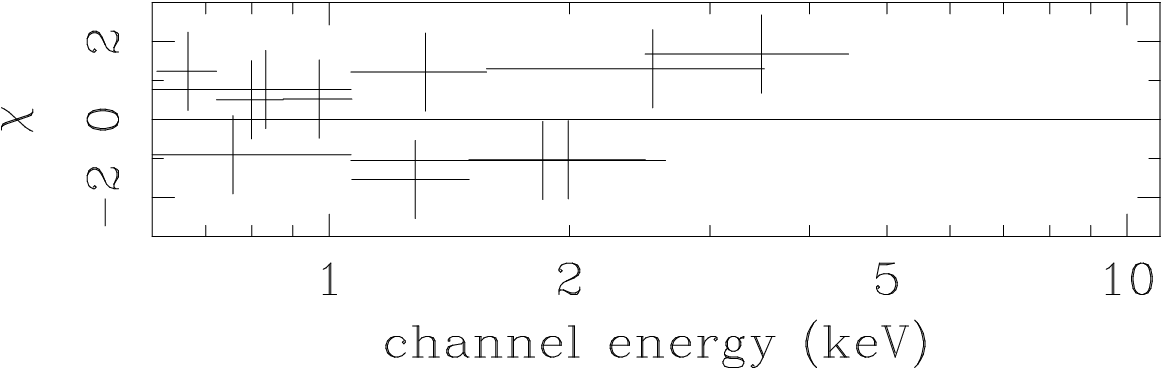,width=8cm,height=2cm,angle=-90}}
}
\\ \\ \\ \\
\parbox{8cm}{
\hspace{5.5cm}{\Large NGC 4051}\\
\vspace{-1.25cm}
\psfig{file=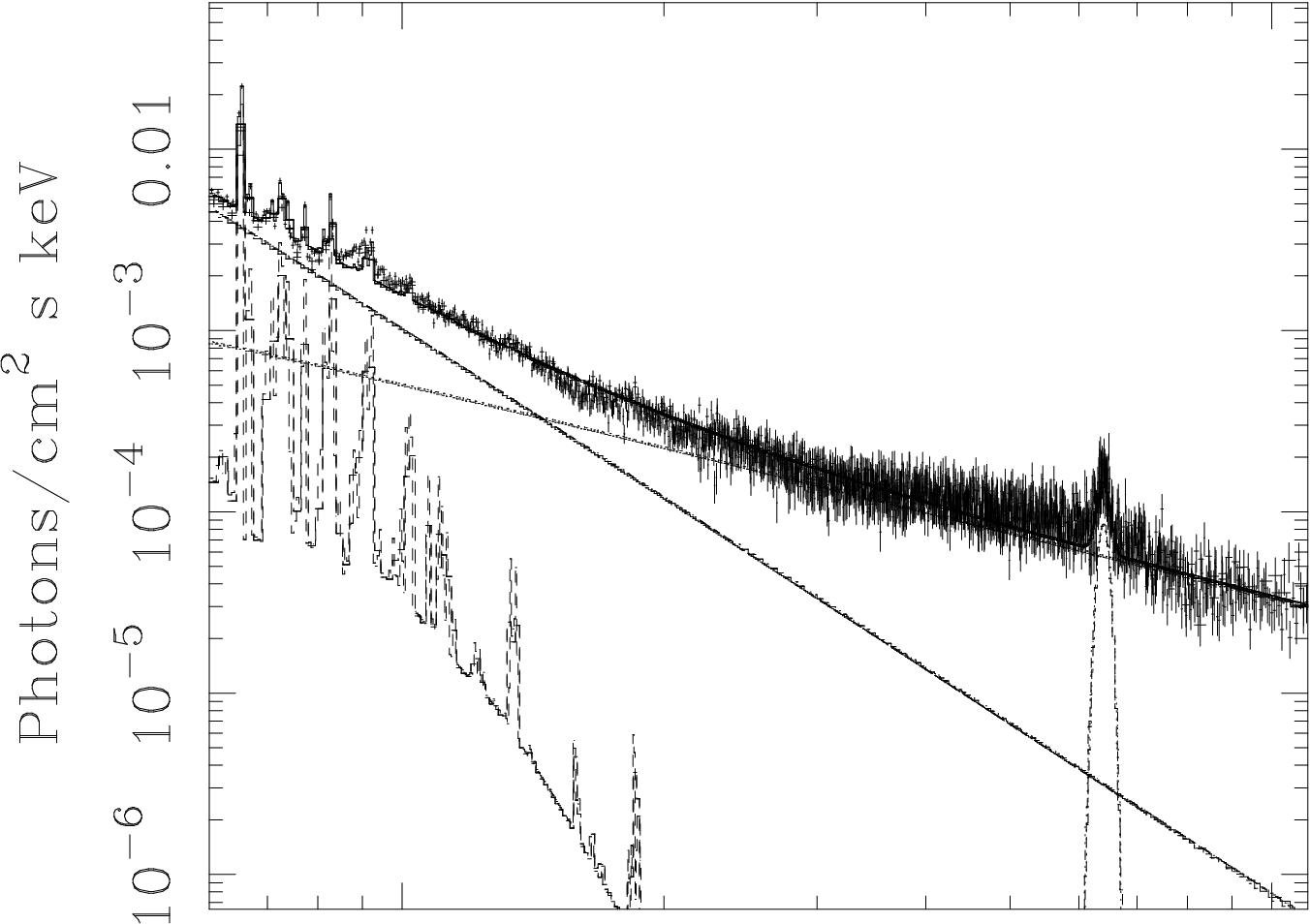,width=8cm,height=3.5cm,angle=-90}
\psfig{file=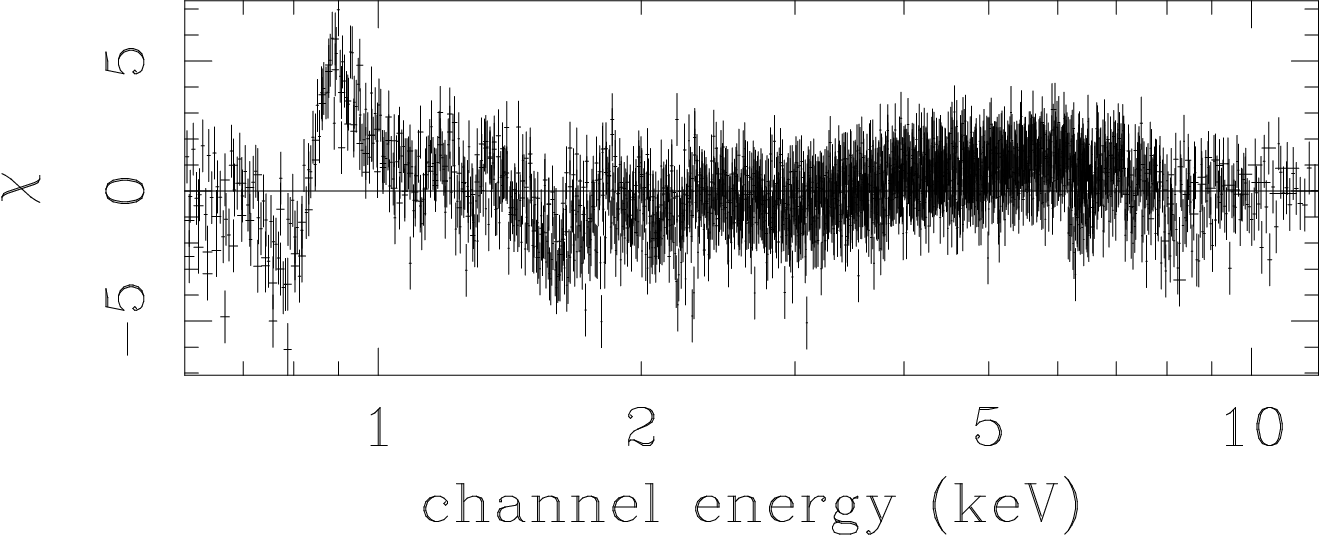,width=8cm,height=2cm,angle=-90}}
\hspace{0.1cm} \
\parbox{8cm}{
\hspace{5.5cm}{\Large NGC 4138}\\
\vspace{-1.25cm}
\psfig{file=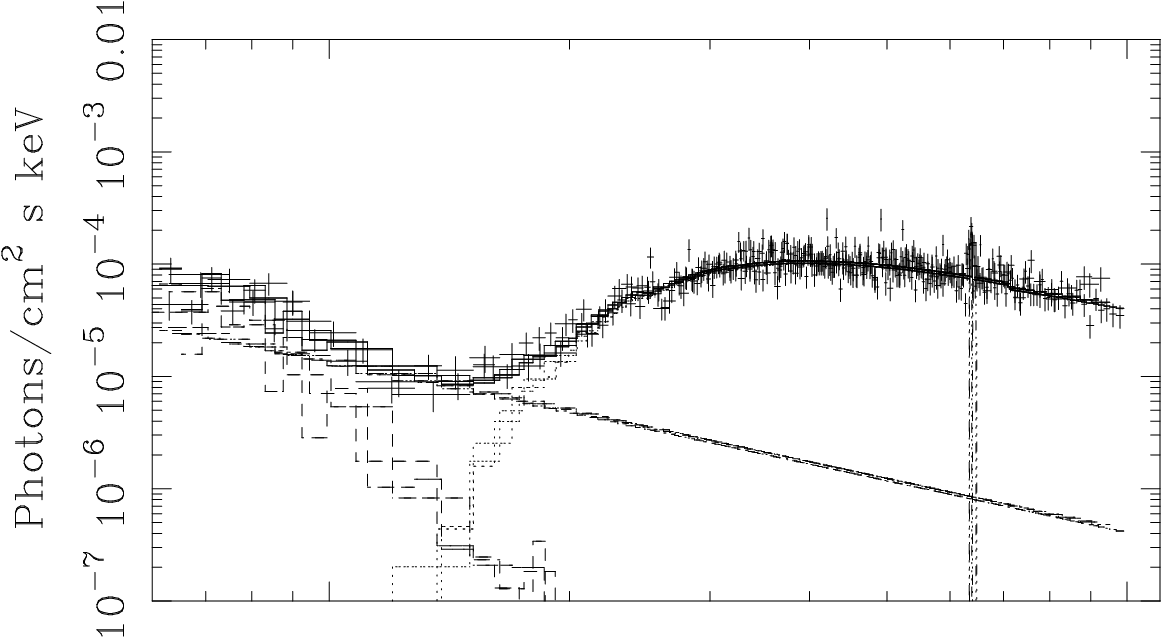,width=8cm,height=3.5cm,angle=-90}
\psfig{file=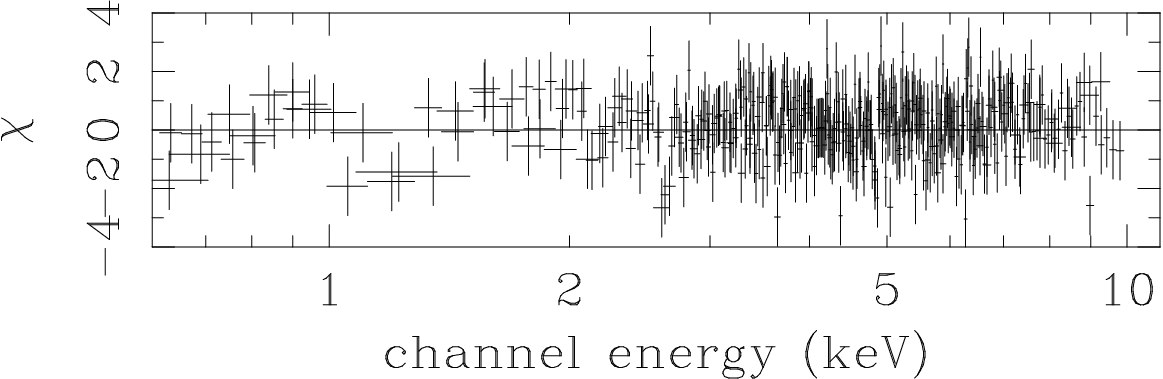,width=8cm,height=2cm,angle=-90}}
\\ 
\end{figure}

\begin{figure}[!]
\vspace{0.5cm}{
\parbox{8cm}{
\hspace{5.5cm}{\Large NGC 4151}\\
\vspace{-1.25cm}
\psfig{file=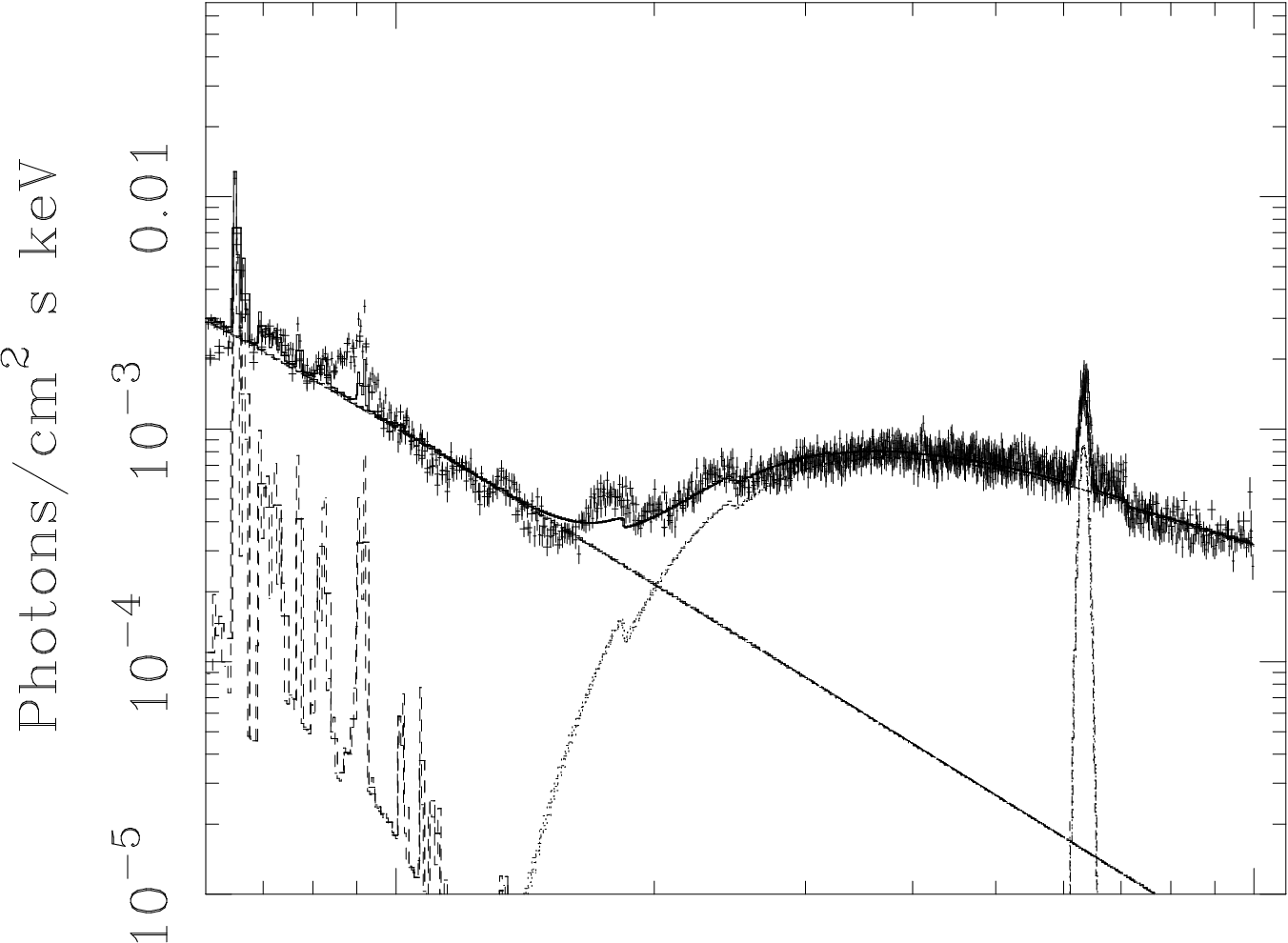,width=8cm,height=3.5cm,angle=-90}
\psfig{file=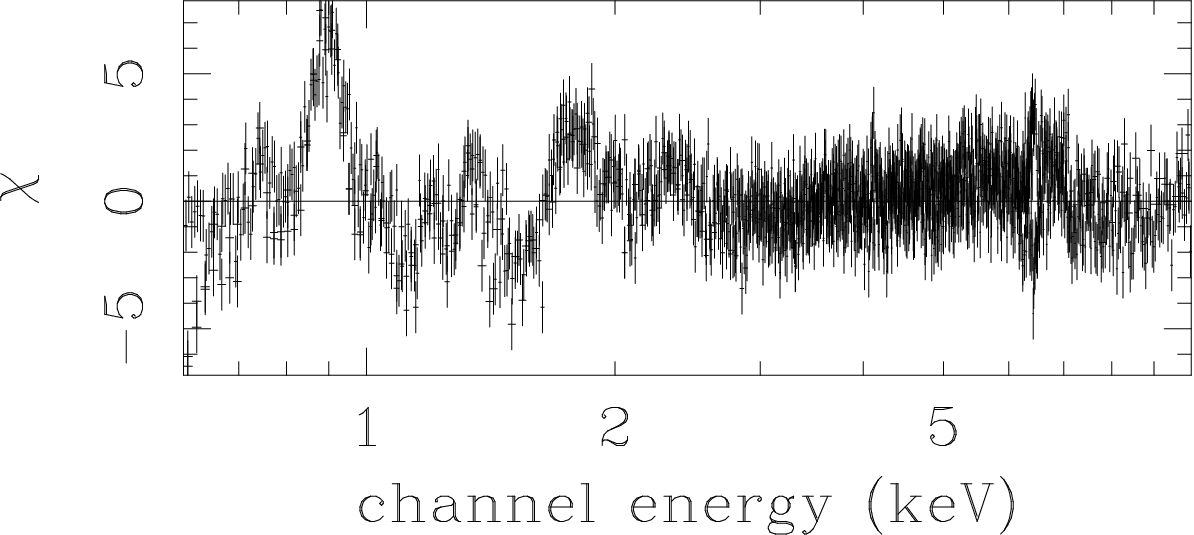,width=8cm,height=2cm,angle=-90}}
\hspace{0.1cm} \
\parbox{8cm}{
\hspace{5.5cm}{\Large NGC 4258}\\
\vspace{-1.25cm}
\psfig{file=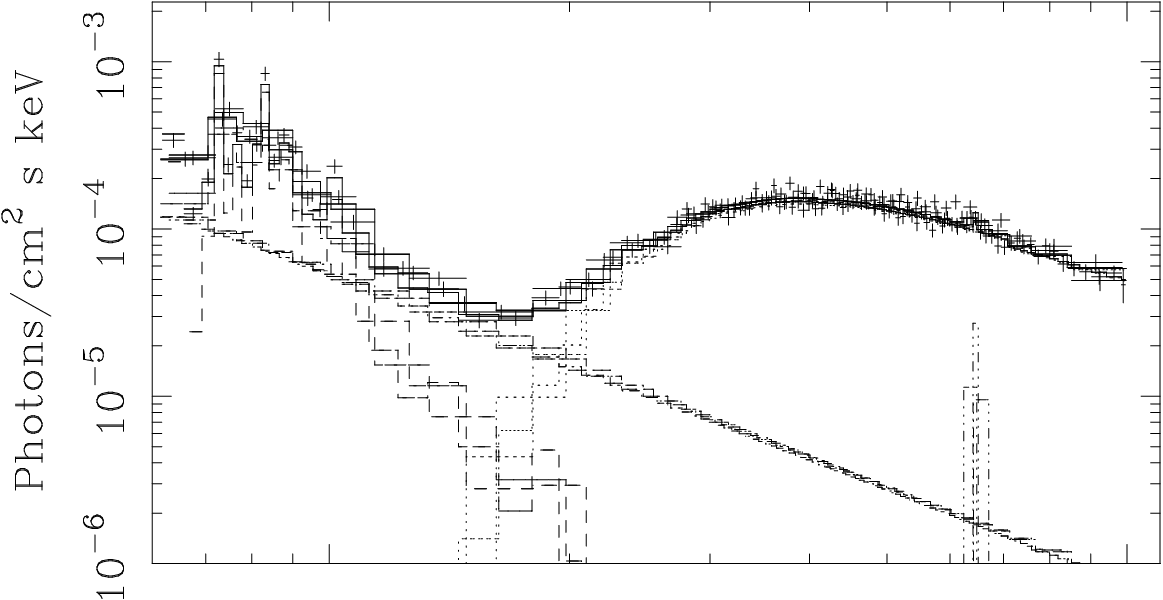,width=8cm,height=3.5cm,angle=-90}
\psfig{file=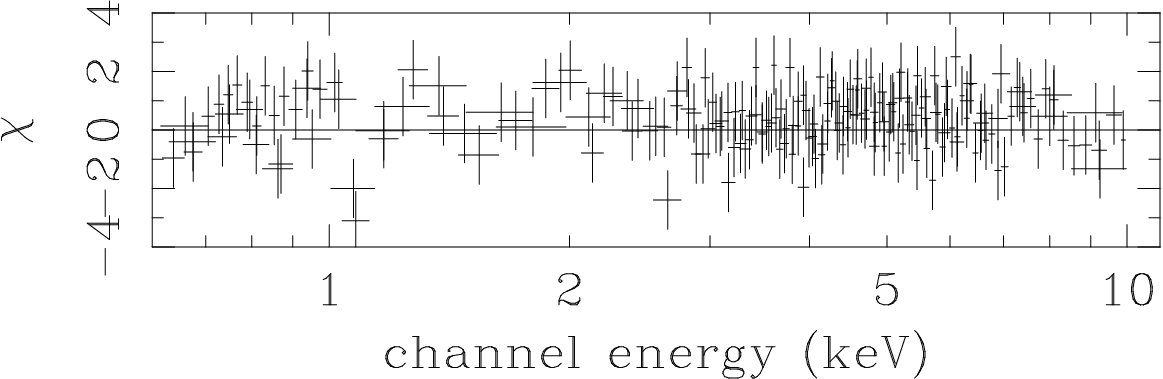,width=8cm,height=2cm,angle=-90}}
%\hspace{5.5cm}{\Large NGC 4168}\\
%\vspace{-1.25cm}
%\psfig{file=/galileo_data1/cappi/xmm/gt/ngc4168/analisi_m12_pn/all/ufspec.ps,width=8cm,height=3.5cm,angle=-90}
%\psfig{file=/galileo_data1/cappi/xmm/gt/ngc4168/analisi_m12_pn/all/del.ps,width=8cm,height=2cm,angle=-90}}
}
\\ \\ \\
\parbox{8cm}{
\hspace{5.5cm}{\Large NGC 4388}\\
\vspace{-1.25cm}
\psfig{file=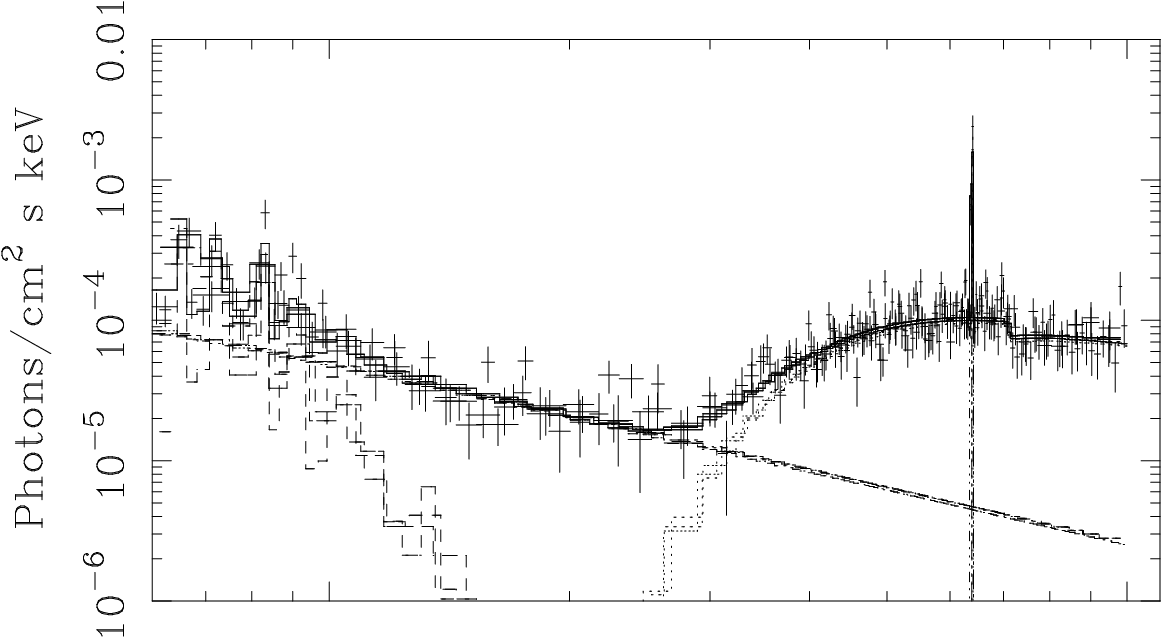,width=8cm,height=3.5cm,angle=-90}
\psfig{file=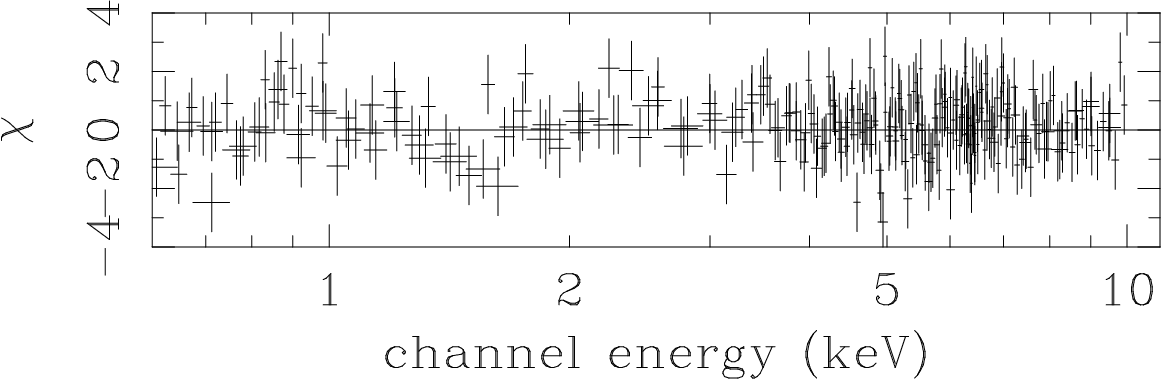,width=8cm,height=2cm,angle=-90}}
\hspace{0.1cm} \
\parbox{8cm}{
\hspace{5.5cm}{\Large NGC 4395}\\
\vspace{-1.25cm}
\psfig{file=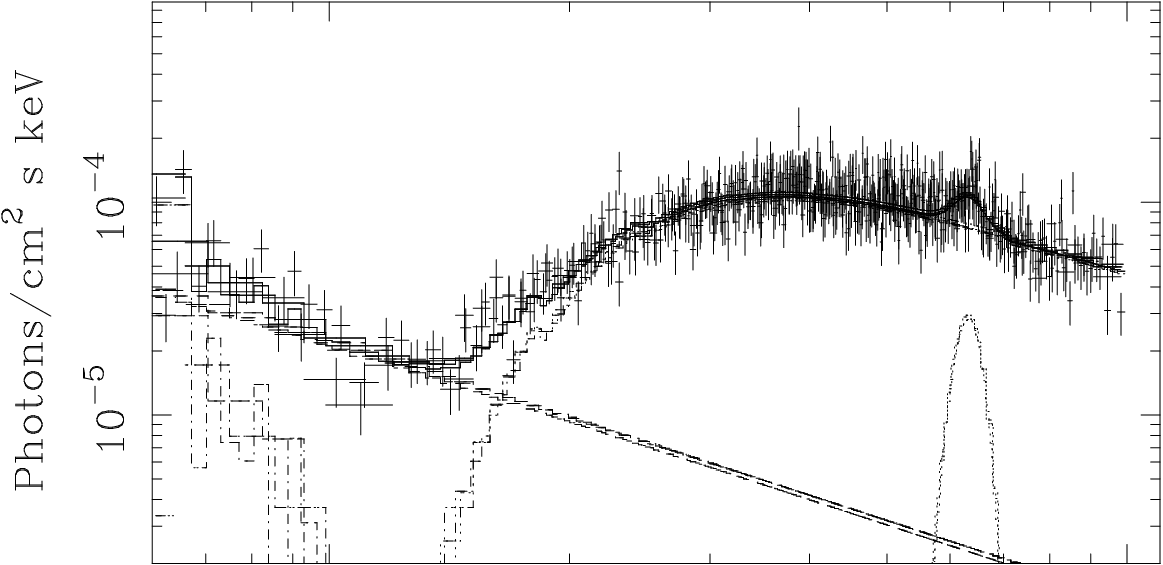,width=8cm,height=3.5cm,angle=-90}
\psfig{file=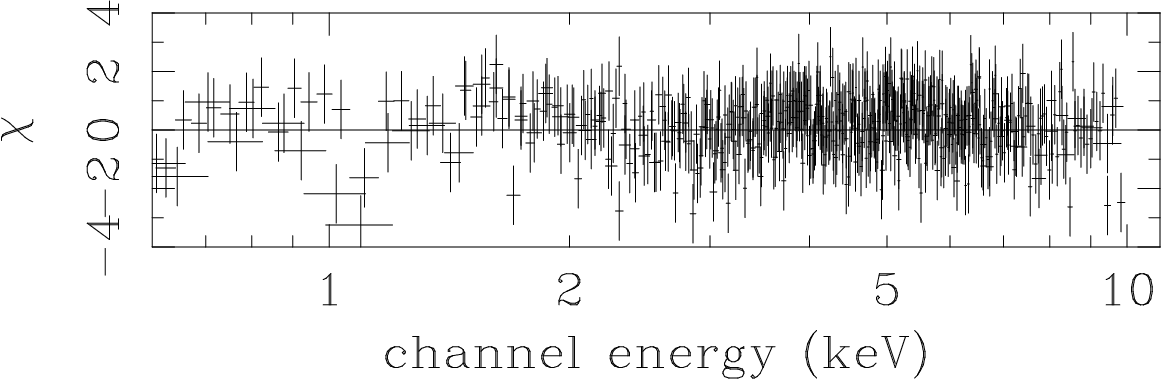,width=8cm,height=2cm,angle=-90}}
\\
\end{figure}

\begin{figure}[!]
\vspace{0.5cm}{
\parbox{8cm}{
\hspace{5.5cm}{\Large NGC 4472}\\
\vspace{-1.25cm}
\psfig{file=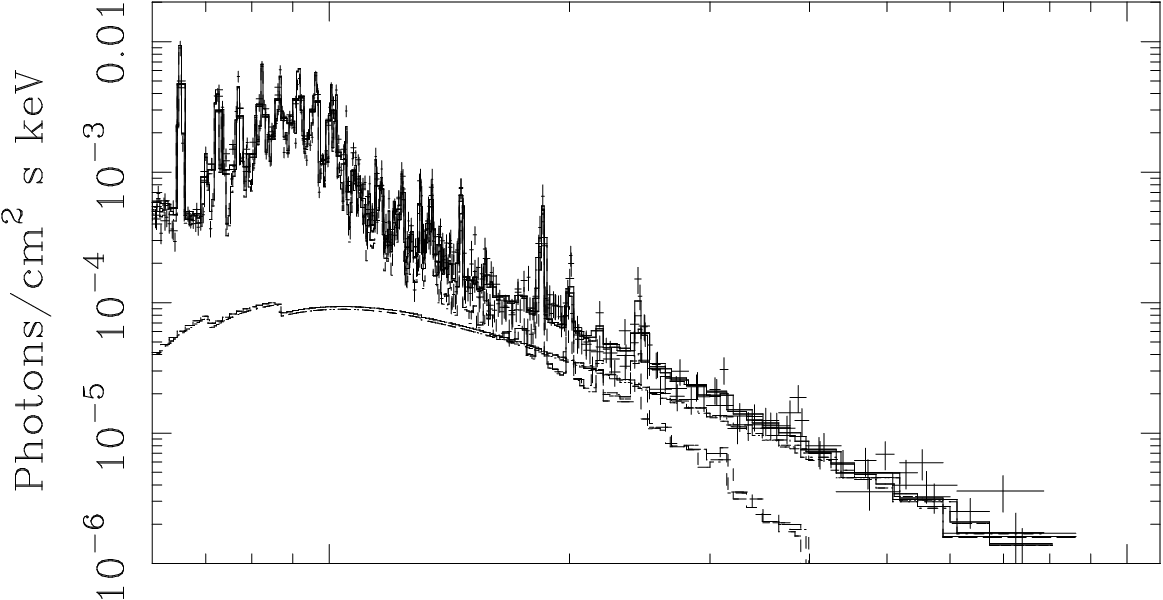,width=8cm,height=3.5cm,angle=-90}
\psfig{file=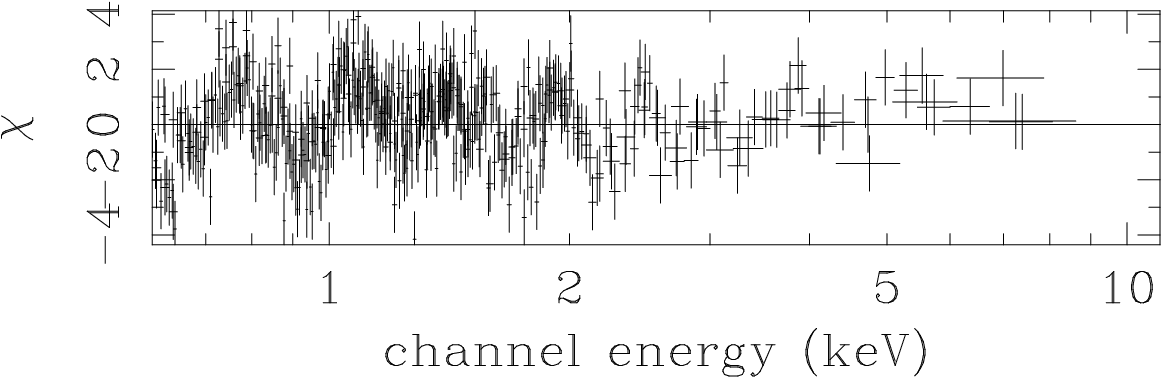,width=8cm,height=2cm,angle=-90}}
\hspace{0.1cm} \
\parbox{8cm}{
\hspace{5.5cm}{\Large NGC 4477}\\
\vspace{-1.25cm}
\psfig{file=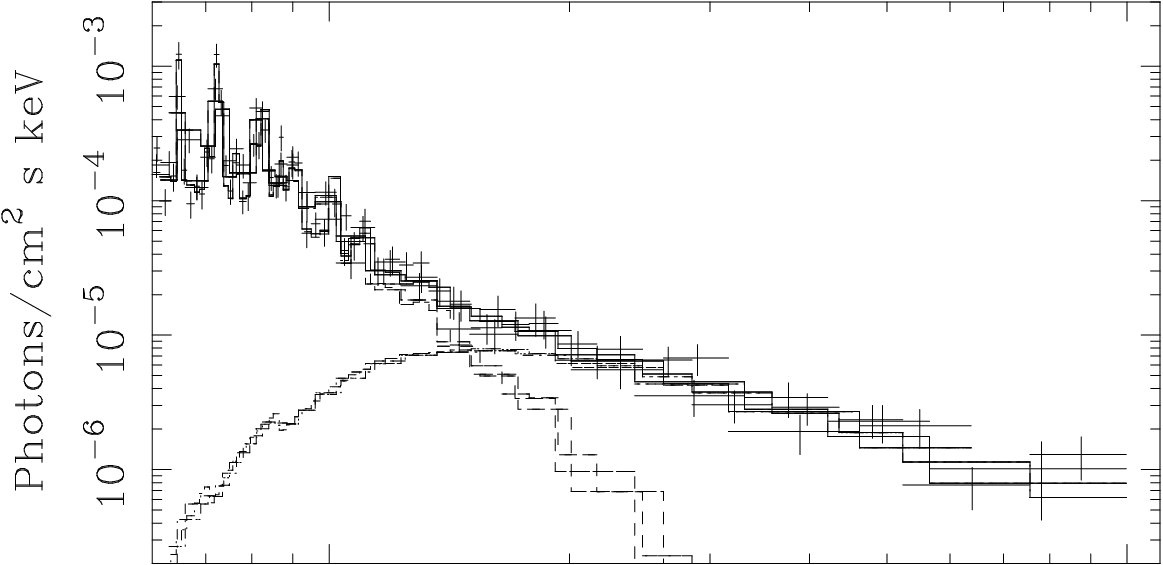,width=8cm,height=3.5cm,angle=-90}
\psfig{file=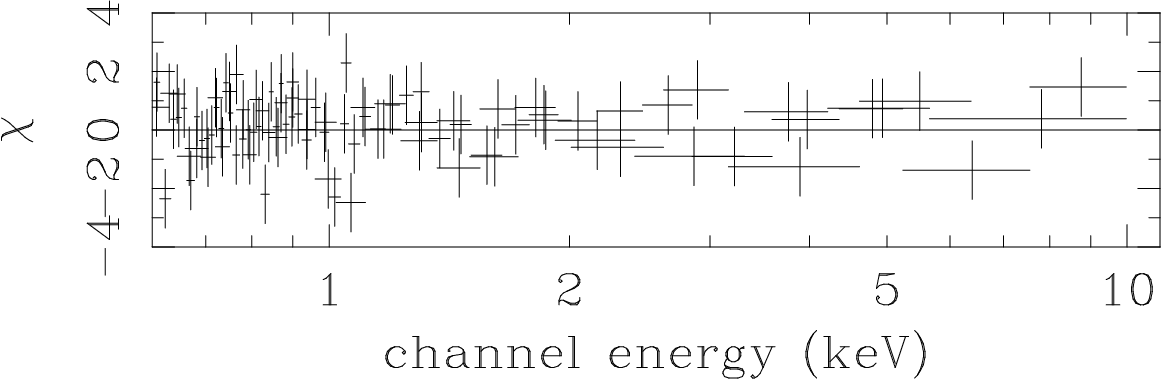,width=8cm,height=2cm,angle=-90}}
\vspace{0.5cm}{
\parbox{8cm}{
\hspace{5.5cm}{\Large NGC 4501}\\
\vspace{-1.25cm}
\psfig{file=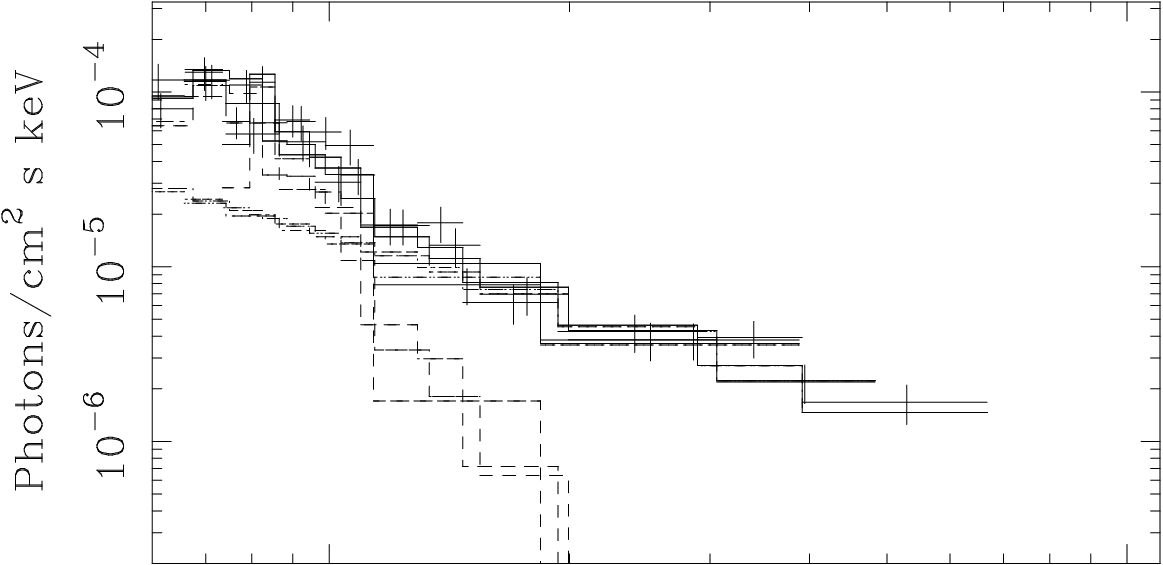,width=8cm,height=3.5cm,angle=-90}
\psfig{file=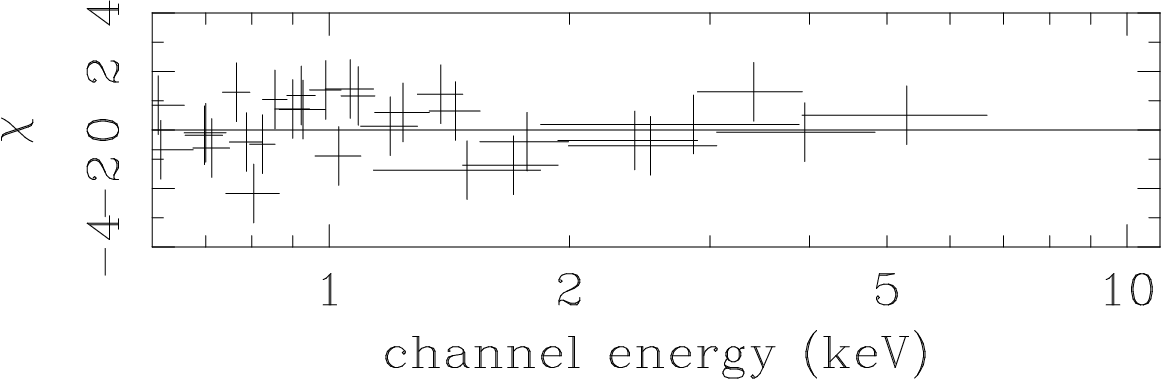,width=8cm,height=2cm,angle=-90}}
\hspace{0.1cm} \
\parbox{8cm}{
\hspace{5.5cm}{\Large NGC 4565}\\
\vspace{-1.25cm}
\psfig{file=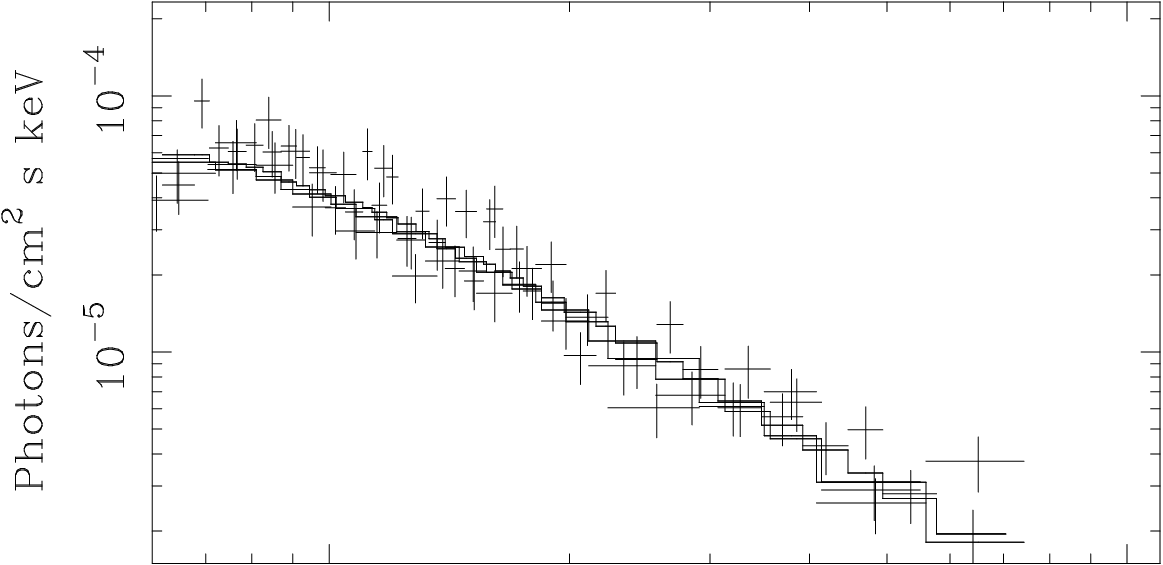,width=8cm,height=3.5cm,angle=-90}
\psfig{file=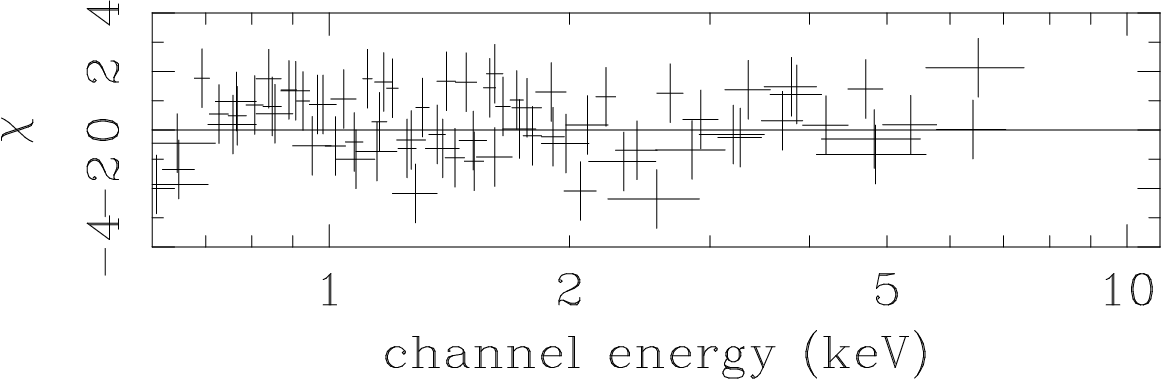,width=8cm,height=2cm,angle=-90}}
}}\\
\end{figure}

\begin{figure}[!]
\vspace{0.5cm}{
\parbox{8cm}{
\hspace{5.5cm}{\Large NGC 4579}\\
\vspace{-1.25cm}
\psfig{file=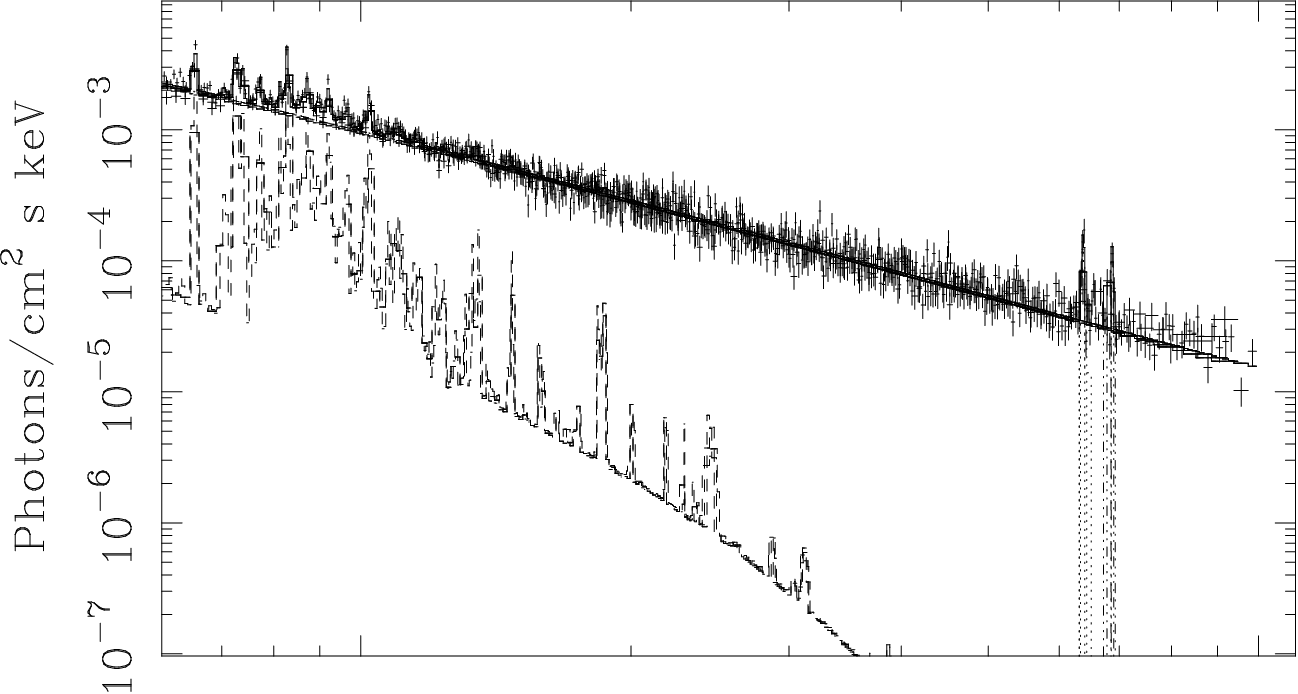,width=8cm,height=3.5cm,angle=-90}
\psfig{file=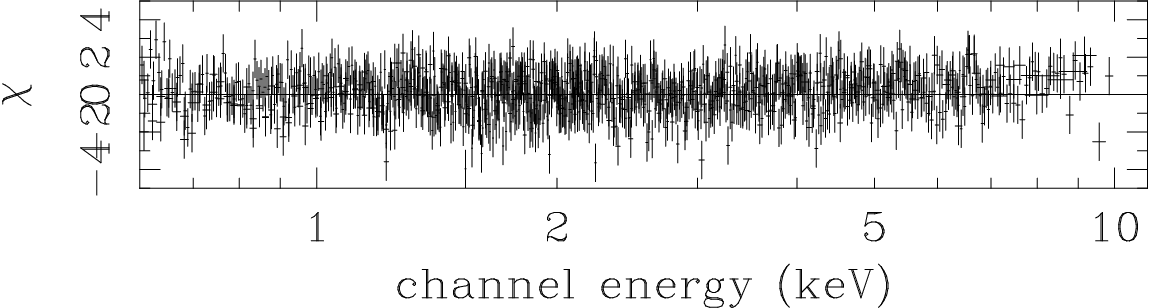,width=8cm,height=2cm,angle=-90}}
\hspace{0.1cm} \
\parbox{8cm}{
\hspace{5.5cm}{\Large NGC 4639}\\
\vspace{-1.25cm}
\psfig{file=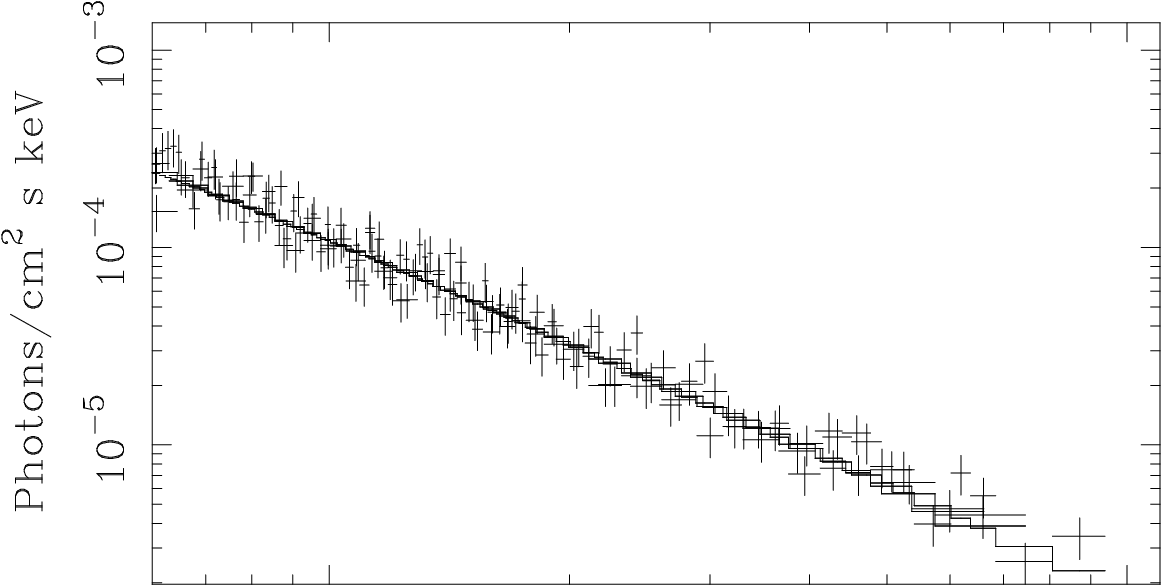,width=8cm,height=3.5cm,angle=-90}
\psfig{file=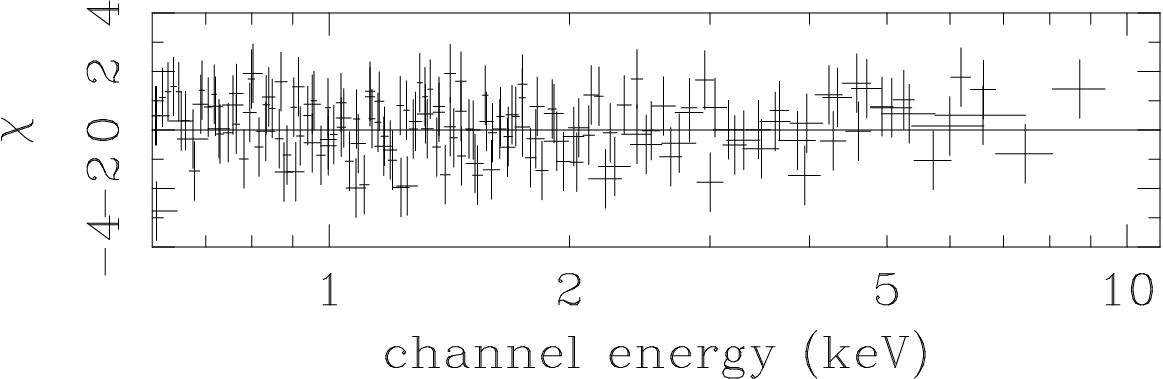,width=8cm,height=2cm,angle=-90}}
\\ \\ \\ \\
\parbox{8cm}{
\hspace{5.5cm}{\Large NGC 4698}\\
\vspace{-1.25cm}
\psfig{file=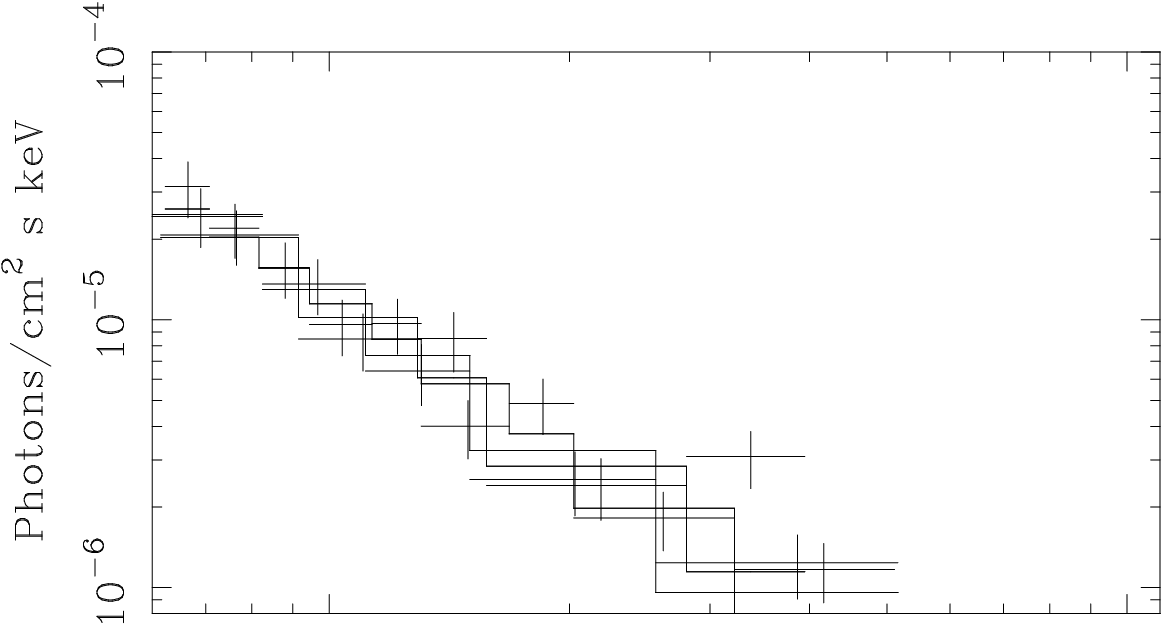,width=8cm,height=3.5cm,angle=-90}
\psfig{file=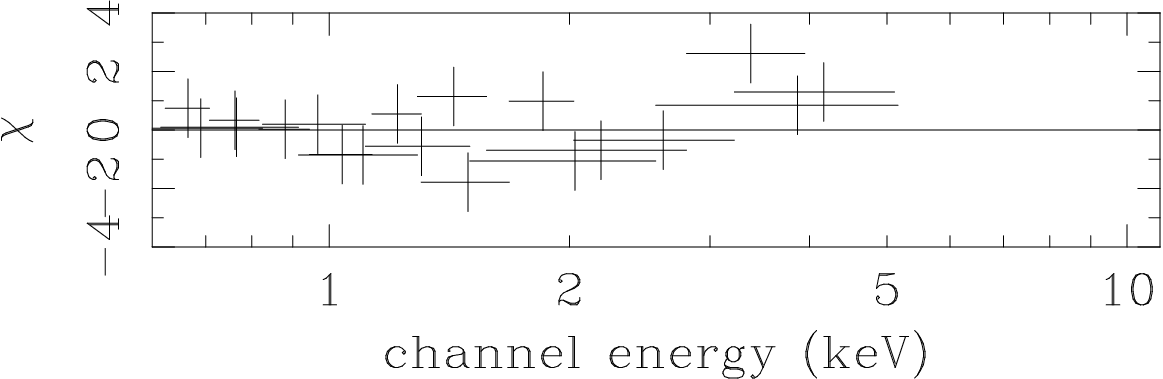,width=8cm,height=2cm,angle=-90}}
\hspace{0.1cm} \
\parbox{8cm}{
\hspace{5.5cm}{\Large NGC 4725}\\
\vspace{-1.25cm}
\psfig{file=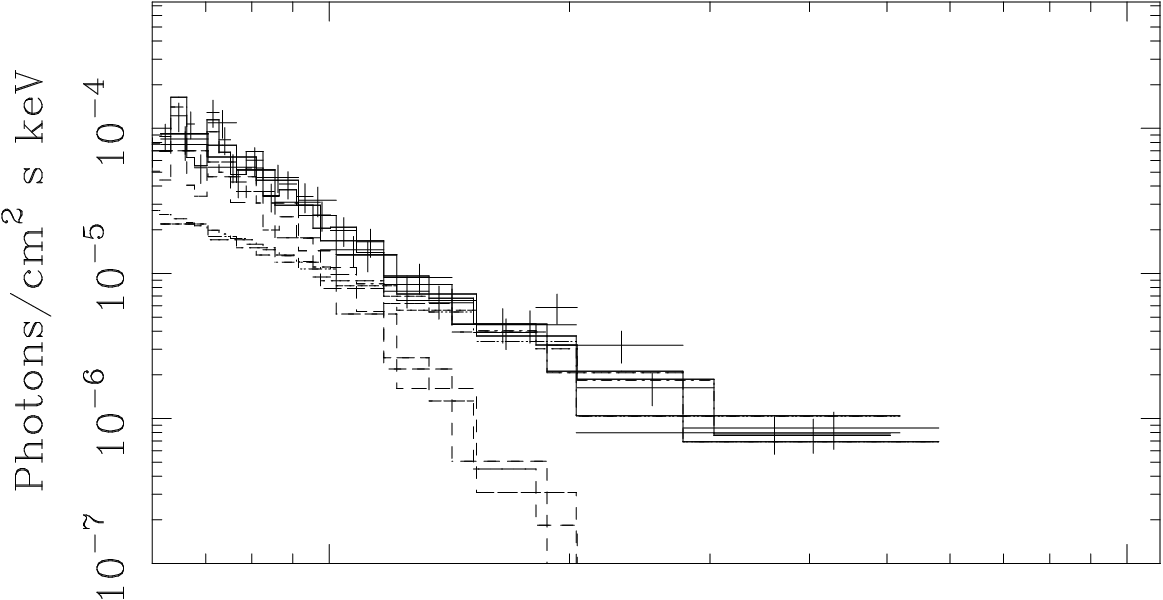,width=8cm,height=3.5cm,angle=-90}
\psfig{file=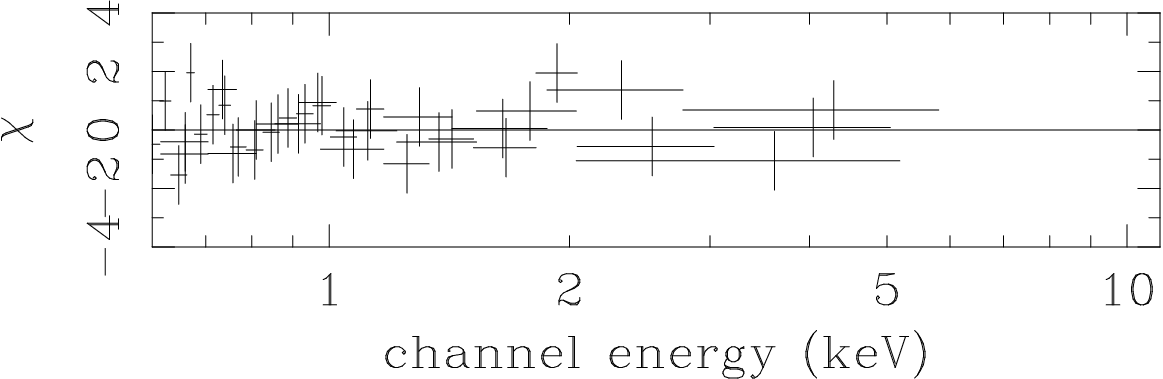,width=8cm,height=2cm,angle=-90}}
}
\\
\end{figure}

\begin{figure}[!]
\vspace{0.5cm}{
\parbox{8cm}{
\hspace{5.5cm}{\Large NGC 5033}\\
\vspace{-1.25cm}
\psfig{file=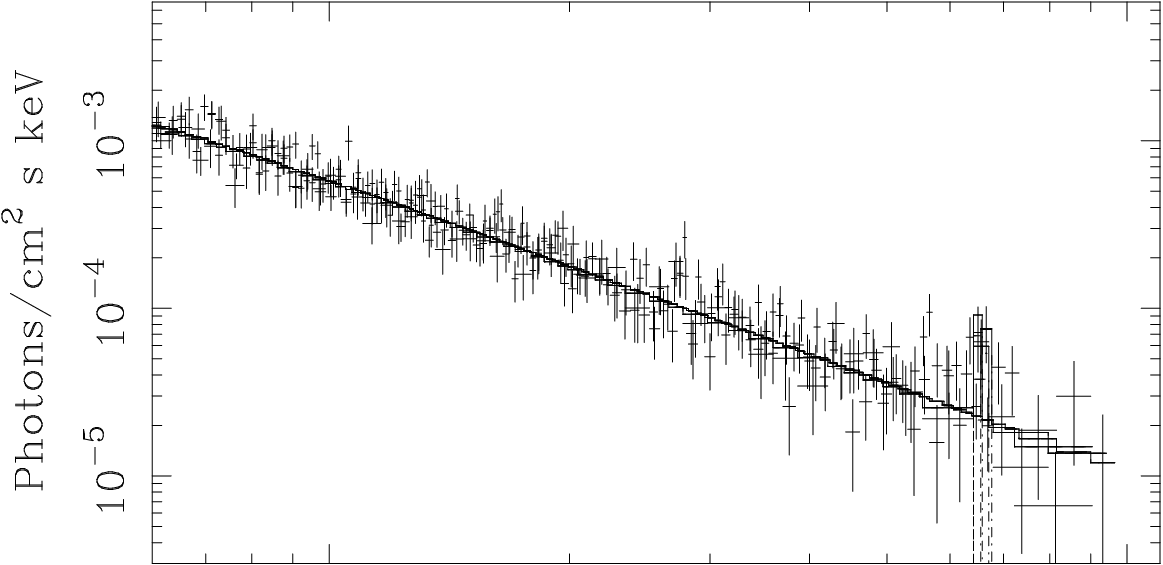,width=8cm,height=3.5cm,angle=-90}
\psfig{file=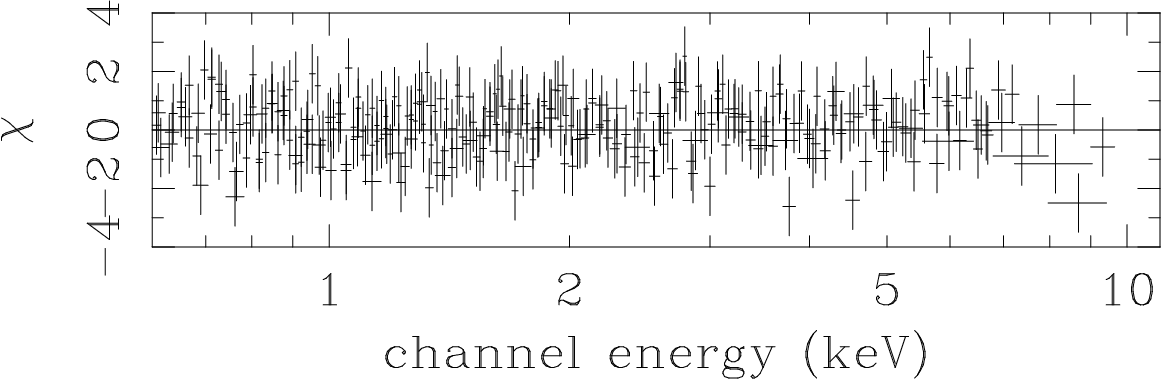,width=8cm,height=2cm,angle=-90}}
\hspace{0.1cm} \
\parbox{8cm}{
\hspace{5.5cm}{\Large NGC 5194}\\
\vspace{-1.25cm}
\psfig{file=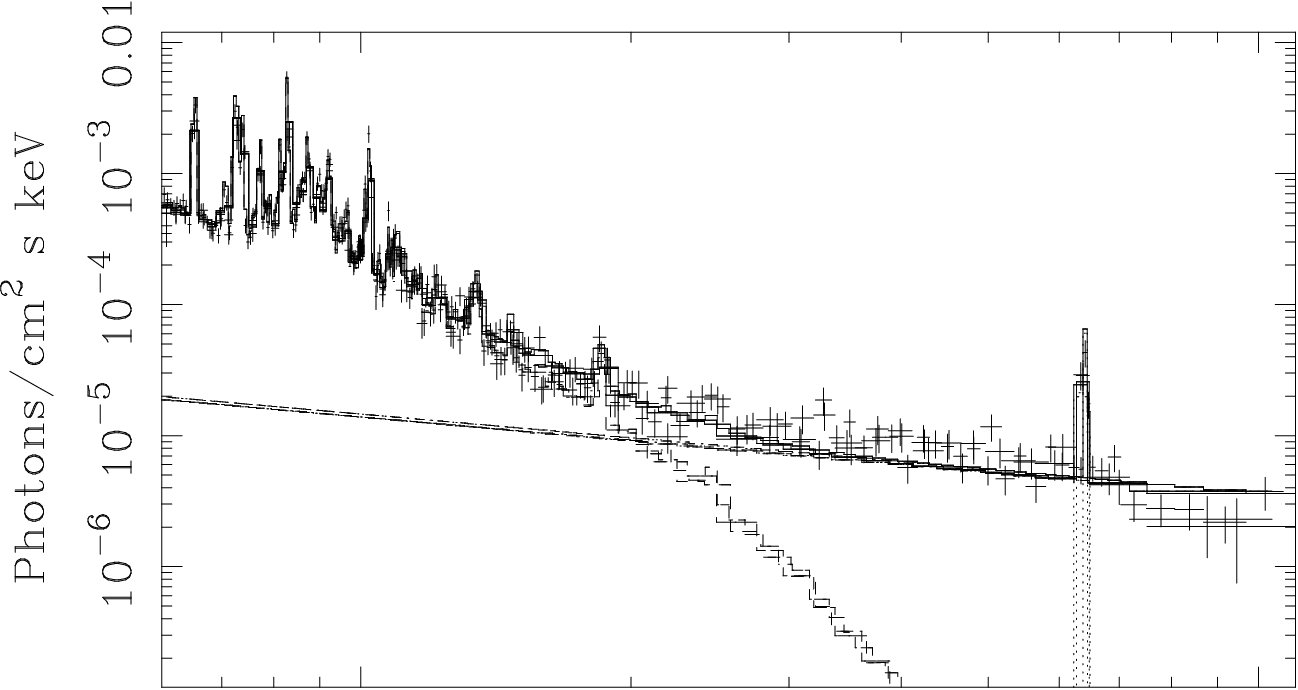,width=8cm,height=3.5cm,angle=-90}
\psfig{file=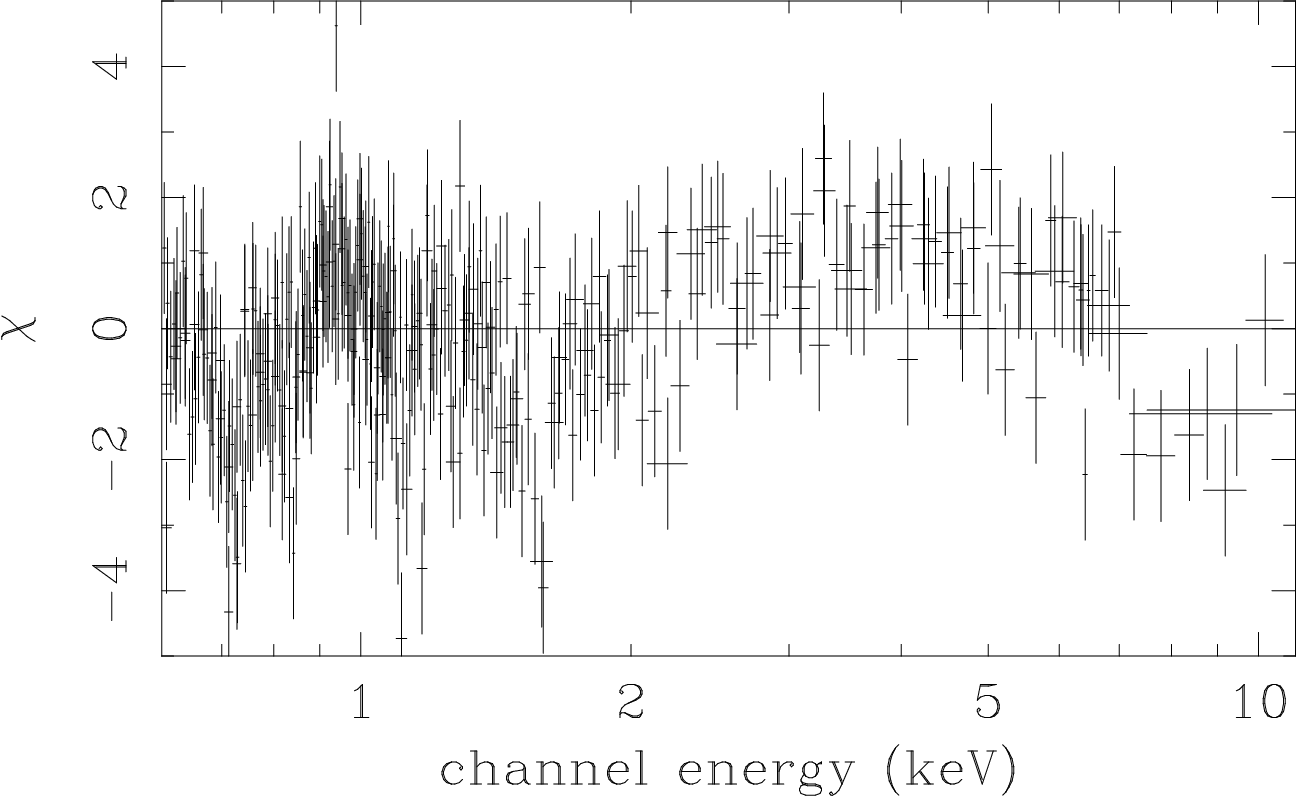,width=8cm,height=2cm,angle=-90}}
\\ \\ \\ \\
\parbox{8cm}{
\hspace{5.5cm}{\Large NGC 5273}\\
\vspace{-1.25cm}
\psfig{file=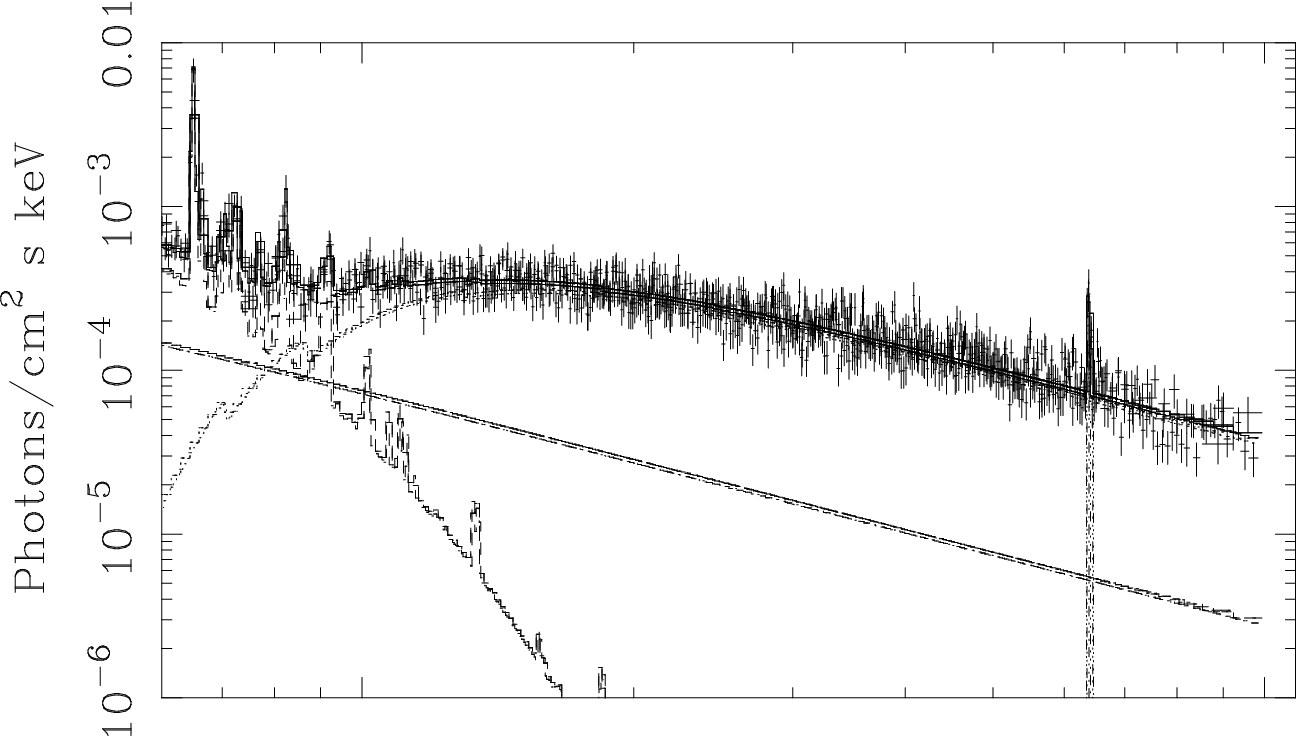,width=8cm,height=3.5cm,angle=-90}
\psfig{file=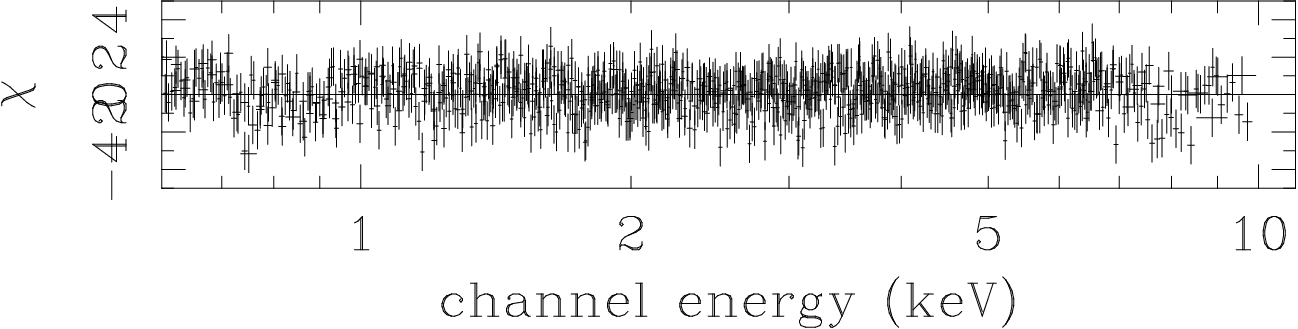,width=8cm,height=2cm,angle=-90}}
}
\end{figure}

\vfill\eject
\
\vfill\eject

\onecolumn

\section{Notes on Individual Objects} 

In this section we give notes on individual galaxies. In particular,
we include a description of (i) the nuclear X-ray morphologies,
(ii) the \xmm spectral results, and (iii) results from the literature. 
Spectral best-fit results are discussed only for spectra with more than 
100 counts. Bright objects have been observed several times with different 
X-ray telescopes, but here only the most pertinent literature references
are reported. In some cases, {\it Chandra} images are also used for 
comparison with the \xmm ones and for their superior spatial resolution. 
An atlas of both {\it Chandra} and \xmm images is given in Panessa (2004).

\
\medskip

%***Massimo: Below (NGC 676), the uncertainty in Gamma is given as 0.1, 
%whereas in Table 3 it is given as 0.3. Which is correct? Change the other.

{\it NGC 676 (S2:)}: This is the first X-ray detection of NGC 676. The 
\xmm image shows the presence of emission associated with the nuclear optical 
position and unresolved surrounding emission which might be associated with 
the nearby off-nuclear sources. The spectrum has poor statistics, but it is 
described by a power law ($\Gamma$ = 1.9 $\pm$ 0.1) with no absorption in 
excess of the Galactic value. This source has not been detected at radio
wavelengths (HU01). \\

{\it NGC 1058 (S2)}: There is no strong nuclear core in this object, 
in agreement with the absence of a radio core detection (HU01). 
Comparison of the \xmm image with the \cha one shows that the 0.5--10~keV 
flux obtained from the \xmm data likely suffers from contamination from 
an off-nuclear source and thus must be considered as an upper limit.\\

%The upper limit on the 2-10 keV luminosity reported 
%for this source has been derived from the \cha observation 
%assuming $\Gamma$ = 1.8, in agreement with the value 
%reported by Ho et al. (2001) on the same data set. 

{\it NGC 1068 (S1.9)}: \xmm images of this source reveal complex
nuclear structure: a compact hard nucleus, coincident with the radio 
core position, embedded in diffuse soft emission. 
The X-ray spectrum has been thoroughly studied; in particular, the use
of the {\it BeppoSAX}/PDS instrument has confirmed that this is
a Compton-thick source (Matt et al. 1997, and references therein).
The \xmm observation of this source was published by 
Kinkhabwala et al. (2002), who focused their analysis on the RGS study, and 
by Matt et al. (2004), who focused their analysis on the $E > 4$~keV data.
Here we give only a very rough, approximate description of 
the 0.5--10~keV spectrum in terms of a soft thermal component, plus a 
scattered power-law component and a flat power law for the hard X-ray 
continuum, plus a strong (EW $\approx$ 1--2 keV) Fe complex between 
6 and 7~keV [which includes the Fe~K$\alpha$ line at 6.4~keV, plus 
recombination/resonant emission lines from He-like (6.7~keV) and H-like 
(6.96~keV) iron, consistent with Matt et al. 2004]. Both the flat hard X-ray 
continuum and the strong Fe line are clear spectral signatures of the 
Compton-thick nature of this active nucleus.\\

{\it NGC 2685 (S2/T2:)}: Only weak nuclear emission is revealed in the 
\xmm image, with some faint extended emission associated with the 
galaxy. The spectrum has poor statistics, yielding a flat power law 
with low absorption and large uncertainties. The core has not been 
detected in the radio (HU01).\\

{\it NGC 3031 (S1.5/L1.5)}: This is an X-ray bright galactic nucleus 
extensively studied by most X-ray satellites (Pellegrini et al. 2000, 
Terashima et al. 2002). A single power law with Galactic absorption 
plus soft thermal emission gives a good parameterization of the \xmm 
data. Further absorption and/or emission structures are found in the 
soft energy band, indicating the presence of a complex photoionized 
and/or thermal plasma. Three Fe lines at different energies ($\sim$6.4, 
6.7, and 6.9~keV) are detected with equivalent widths on the order of 
40--50~eV each. Detailed analysis and interpretation of the complex 
lines in the \xmm dataset are given in two different studies by 
Dewangan et al. (2004) and Page et al. (2004). A bright and variable 
radio core is detected (HU01).\\

%***Massimo: Below (N3079), it is not clear whether the 10'' region is 
%diameter or radius; I have inserted ``radius'' but please change if
% this is wrong.

{\it NGC 3079 (S2)}: The \xmm images of this source show a complex
and unresolved structure which extends for $\sim 30''$ around the nuclear
position. The hard spectrum is described by a power law ($\Gamma$ $\approx$
1.7 $\pm$ 0.1) modified by little absorption and a strong Fe~K line with 
an EW of almost 2~keV. The $10''$-radius region around the nucleus is resolved 
in the \cha image: the strong nuclear source is embedded in a bubble of 
diffuse emission. However, the diffuse emission contributes less than 
10\% of the nuclear emission at $E \gsimeq 2$~keV. A \cha and {\it HST} 
study of the superbubble by Cecil, Bland-Hawthorn, \& Veilleux (2002) 
shows that the optical and X-ray emissions match. The radio core position 
is coincident with the 2--10~keV peak. We also extracted the \cha spectrum 
from a circular region of $2''$ radius. The spectral parameters are not 
well constrained due to the poor photon statistics, but the results are 
in good agreement with those from {\it XMM--Newton}. The strong Fe~K line 
at 6.4~keV (detected at much greater than 99\% significance with 
{\it XMM--Newton}) suggests that this source is heavily absorbed and 
confirms the {\it BeppoSAX} results which indicate that it is Compton 
thick (Iyomoto et al. 2001). \\

{\it NGC 3185 (S2:)}: The \xmm observation of this object shows
weak nuclear emission. An extraction radius of $20''$ was chosen in
order to separate the nuclear emission from a nearby ULX (Foschini et
al. 2002). The spectrum is described by a power law ($\Gamma$ = 2.1 $\pm$ 0.1) 
with absorption lower than 2 $\times$ 10$^{21}$ cm$^{-2}$.
Radio emission is detected only marginally (HU01). \\

%***Massimo: Below (N3227), the uncertainty in Gamma is given as 0.02, whereas
%in Table 3 it is given as 0.1. Which is correct? Change the other.

{\it NGC 3227 (S1.5)}: This type 1.5 Seyfert is known to show significant 
spectral variability in the X-ray band (George et al. 1998) and a warm absorber 
(Gondoin et al. 2003). The \xmm 0.5--10~keV spectrum is parameterized here with a 
single power law plus a scattered component, with the soft and hard X-ray power laws 
having identical slopes ($\Gamma$ = 1.5 $\pm$ 0.02, $N_{\rm H}$ = 6.8 $\pm$ 0.3 $\times$
10$^{22}$ cm$^{-2}$). Addition of a warm absorber would modify only slightly 
these continuum parameters. The Fe~K line is detected at 6.4~keV with EW = 190 
$\pm$ 40~eV. Spectral parameters are in agreement with the best-fit results from 
Gondoin et al. (2003) and Lamer, Uttley, \& McHardy (2003).\\

{\it NGC 3486 (S2)}: Emission from the nuclear
region has been detected in the \xmm observation,
surrounded by extended emission. An extraction radius of $20''$ was
used in order to separate the nucleus from a nearby ULX (Foschini et al. 
2002). The very poor statistics of the spectrum yield a best fit with
a power law ($\Gamma$ = 0.9 $\pm$ 0.2) and Galactic absorption.
The \xmm 2--10~keV flux was measured assuming the above best-fit
model, and it turns out to be consistent with the \cha limit 
(Ho et al. 2001).

{\it NGC 3941 (S2:)}: The \xmm observation of this object shows a bright
off-nuclear source $40''$ from the nuclear position. The spectrum 
has very poor statistics and is best fitted with an unabsorbed 
power law with $\Gamma$ = 2.1 $\pm$ 0.3. The source is marginally 
detected in the radio band (HU01).\\

%{\it NGC 3982} - This source has been observed by ASCA.
%The 2-10 keV flux has been derived from Moran et al. (2001).
%An unresolved radio core has been detected (HU01).\\

{\it NGC 4051 (S1.5)}: Detailed analyses of this observation have been 
presented by Uttley et al. (2003) and Pounds et al. (2004), and have been 
compared to the earlier observation performed in 2001 (Lamer et al. 2003).
Major spectral variations between the high-flux and low-flux states have been 
interpreted either in terms of variations in the ionization state of the 
absorption column density (Pounds et al. 2004) or in terms of large 
changes in the power-law slope (Uttley et al. 2003).
Here, we fitted the 0.5--10~keV integrated spectrum of the low-flux state 
with a soft thermal component plus a steep scattered power law, and a very 
flat hard power law ($\Gamma$ = 1.2$ \pm 0.1$). The flat power-law shape, 
together with the strong Fe~K emission line at 6.4~keV with EW $\approx$ 
250~eV and the two absorption edges marginally detected at 7.2 and 7.9~keV, 
suggest the presence of both a strong reflection component and a heavy warm 
absorber, in agreement with Pounds et al. (2004). Further detailed temporal 
analysis is beyond the scope of the present work. NGC 4051 shows complex 
radio structure (HU01).\\

%Two further emission features are found significant at E$\sim$ 5.5 keV 
%(EW$\sim$30 eV) and 6.1 keV (EW$sim$40 eV).

{\it NGC 4138 (S1.9)}: The \xmm image of this object shows a bright
nuclear source with marginal evidence of a surrounding diffuse component.
Detailed imaging and spectral fitting of the same dataset is 
reported by Foschini et al. (2002). The best-fit model is 
obtained with an intrinsic absorbed power-law component 
[$\Gamma$ = 1.5 $\pm$ 0.1, $N_{\rm H}$ = (8 $\pm$ 1) $\times$ 10$^{22}$ 
cm$^{-2}$] plus an Fe line at 6.4~keV (EW = 83 $\pm$ 30~eV). We add a 
scattered component plus a thermal component to fit the data below 1.5~keV 
($kT = 0.3 \pm 0.05$~keV, $\Gamma_{\rm SX} \equiv \Gamma_{\rm HX}$).
A weak, compact radio core is seen at the position of the optical nucleus 
(HU01).\\

{\it NGC 4151 (S1.5)}: 
EPIC images show a bright, point-like nuclear source. 
The spectrum below 1.5~keV is complex and can be described in terms of 
a warm absorber, scattered radiation, and additional spectral components 
(see, e.g., Yang, Wilson, \& Ferruit 2001, Ogle et al. 2000). 
We concentrate here on the 2--10~keV spectrum and obtain a good 
description with an absorbed power law ($\Gamma$ = 1.6 $\pm$ 0.1, $N_{\rm H}$ =
8.4 $\pm$ 4 $\times$ 10$^{22}$ cm$^{-2}$) plus a soft scattered component with 
$\Gamma$ = 1.8 $\pm$ 0.5. A strong Fe~K line is detected with EW $\approx$ 300 
$\pm$ 30~eV and 41 $\pm$ 15~eV for the K$\alpha$ and K$\beta$ components. 
More detailed analysis of the same dataset is given by Schurch et al. (2004). 
A complex radio structure is present in this source (HU01).\\

%Recently, a detailed study on \cha data has been published by Yang,
%Wilson and Ferruit (2001) (see also Ogle et al. 2000). Only the 0.1 s 
%frame time observation is not affected by pile-up. Unfortunately,
%alternating mode observations give sometimes problems in
%the determination of the exposure times as in this case,
%so we consider the \xmm observation for the spectral analysis.

%{\it NGC 4168} - This source was observed for the first time in X-rays
%by \xmm. The EPIC images show emission from the nuclear region and weak 
%off-nuclear emission which extends
%for 20" from the optical position. The poor statistics spectrum 
%is fitted with a single power-law model ($\Gamma$ = 2.0$\pm$0.2, $N_{\rm H}$ 
%$\leq$ 3$\times$10$^{20}$cm$^{-2}$). 
%The radio core has been detected (HU01).\\

%{\it NGC 4235} - The ASCA hard X-ray luminosity has been obtained 
%by modeling the spectrum with an absorbed power-law 
%($\Gamma$ = 1.57$\pm$0.06, N$_{H}$ = 1.5$\times$10$^{21}$).
%The ASCA data are taken from the HEASARC archive. 
%An unresolved radio core is present in the source (HU01).\\

{\it NGC 4258 (S1.9)}:  
The \xmm image of this source shows a bright, hard, point-like nucleus and 
unresolved diffuse emission. The \xmm spectral results are in good 
agreement with those shown by Pietsch \& Read (2002). 
The \xmm hard spectrum is modeled by a
power law ($\Gamma$ = 1.7 $\pm$ 0.1) with high absorption ($N_{\rm H}$ =
8.7 $\pm$ 0.3 $\times$ 10$^{22}$ cm$^{-2}$). A narrow Fe~K$\alpha$
emission line is marginally detected (EW = 27 $\pm$ 20~eV).
The \xmm hard luminosity is a factor of 2 lower than the \cha
luminosity (Young \& Wilson 2004).  
This difference is probably due to intrinsic variability,
already found between previous {\it ASCA} and {\it BeppoSAX} observations
(Terashima et al. 2002, Risaliti 2002).
The same is true for the Fe~K$\alpha$ line which had been
detected in previous {\it ASCA} and {\it BeppoSAX} observations 
but which is not significantly detected here.\\

%***Massimo: Below (N4388), Gamma is given as 1.2, whereas
%in Table 3 it is given as 1.3. Which is correct? Change the other.

{\it NGC 4388 (S1.9)}:
The 2--10~keV \xmm data show a bright nucleus embedded in 
diffuse emission extending to a radius of $\sim 20''$ in the full-band 
image. An unusually high background level precludes detailed 
analysis of the extended component. The spectrum from the nuclear 
region, with the background determined locally, is well fitted 
by a complex (thermal plus scattered) soft component plus 
a hard ($\Gamma$ = 1.2 $\pm$ 0.2), heavily absorbed ($N_{\rm H}$ =
2.7 $\times$ 10$^{23}$ cm$^{-2}$), power-law component. 
A strong Fe~K$\alpha$ line at 6.4~keV is detected (EW $\approx$ 450 
$\pm$ 70~eV). Our spectral results are in good agreement with those 
reported by Iwasawa et al. (2003).\\

{\it NGC 4395 (S1.5)}: This source is well known for showing 
large amplitude, complex flux and spectral variations (see, e.g., 
Iwasawa et al. 2000, Shih, Iwasawa \& Fabian 2003, Moran et al. 2005). Detailed 
analysis of the \xmm observations is given by Vaughan et al. (2005) 
and Iwasawa et al. (2005). For a simpler comparison with other sources, 
we consider here only its time-averaged properties and parameterize its 
spectrum with a hard power-law model ($\Gamma$ = 1.2) absorbed by a 
column density of $N_{\rm H}$ $\approx$ (5.3 $\pm$ 0.3) $\times$ 
10$^{22}$ cm$^{-2}$, plus a soft complex component which includes 
thermal emission and a scattered power law. An iron line is detected 
at 6.4~keV (EW $\approx$ 100 $\pm$ 25~eV). A second line is found at 
$\sim$6.2~keV, with EW = 45 $\pm$ 30~eV.\\

%***Massimo: Below, the Soldatenkov et al. reference needs to be put in
% the bibliography.

{\it NGC 4472 (S2::)}: This is a giant elliptical galaxy. The \xmm
image reveals strong, soft, diffuse emission. A \cha 2--10~keV image 
(Loewenstein et al. 2001, Soldatenkov, Vikhlinin, \& Pavlinsky 2003) 
shows complex structure (including diffuse and off-nuclear point
sources) without any evidence for a dominant core emission.
For this reason, the hard X-ray flux and luminosity are treated here 
as upper limits. The \xmm spectrum is mostly thermal and poorly 
parameterized here by a single thermal component plus a hard power-law tail. 
The spectrum is shown in Appendix A, but it is not used in our analysis.
This source is marginally detected in the radio band (HU01). \\

%since probably contaminated.A
%thermal component is required to fit the data below 2 keV (kT =
%0.8$\pm$0.1 keV). The weak hard component is fitted with a power-law
%($\Gamma$ = 2.7$\pm$0.3) with low absorption (N$_{H}$ $\leq$
%3$\times$10$^{21}$ cm$^{-2}$). The soft \cha spectrum is in good
%agreement with the XMM results (kT = 0.62$\pm$0.05 keV), however the
%spectral shape of the power-law is poorly constrained, so we fixed it at the
%XMM value ($\Gamma$ = 3). Our \cha 2-10 keV upper limit on the flux is in agreement
%with Loewenstein et al.(2001), while \xmm soft and hard fluxes
%probably suffer from contamination (F$_{0.5-2 keV}$ =
%1.1$\times$10$^{-12}$ erg cm$^{-2}$ sec$^{-1}$ and F$_{2-10 keV}$ =
%1.7$\times$10$^{-13}$ erg cm$^{-2}$ sec$^{-1}$).

{\it NGC 4477 (S2)}: The X-ray spectrum obtained from \xmm is the first 
for this object. The EPIC images show that the nucleus seems to be
dominated by diffuse soft emission. The 2--10~keV image for this
source is not unambiguously compact and is very weak. The spectrum was 
extracted from a region of radius $25''$ to separate the nuclear emission 
from a possible off-nuclear source at a distance of $\sim 40''$ from the 
nucleus. The 0.5--10~keV spectrum appears to be dominated by the soft 
component ($kT = 0.4 \pm 0.1$~keV). Above 2~keV the statistics are very 
poor and the data can be fitted with an absorbed power law ($\Gamma$ = 
1.9 $\pm$ 0.3, $N_{\rm H}$ $\leq$ 2 $\times$ 10$^{22}$ cm$^{-2}$).
This source is marginally detected in the radio band (HU01).  \\

{\it NGC 4501 (S2)}: The 2--10~keV MOS1 and MOS2 images reveal the 
presence of a weak nuclear core. The spectrum was extracted from a region 
of $20''$ radius in order to exclude the emission of an off-nuclear 
source (Foschini et al. 2002). The 0.5--10~keV spectrum is described by
a soft thermal component ($kT = 0.4 \pm 0.1$~keV) plus a power law with
$\Gamma$ = 1.5 $\pm$ 0.3. Absorption in excess of the Galactic value is
not required by the fit, yielding an upper limit for the column density
($N_{\rm H}$ $\leq$ 2 $\times$ 10$^{21}$ cm$^{-2}$). This is consistent 
with previous $ASCA$ results reported by Terashima et al. (2002).
An unresolved radio core is also detected (HU01).\\

{\it NGC 4565 (1.9)}: Mizuno et al. (1999) revealed the presence of two
bright point-like sources in the $ASCA$ observation of this galaxy. A
\cha study of NGC 4565 by Wu et al. (2002) leads to a clear 
identification of the nuclear source which is separated from the 
off-nuclear source by $\sim 50''$. The parameters describing the 
\cha 0.5--10~keV spectrum, with poor statistics, are in good agreement 
with those found by Terashima \& Wilson (2003) using the same data set. 
Also, in the \xmm images the two sources are clearly separated (Foschini
et al. 2002). We extracted the spectrum from a region of radius $25''$. 
We find that a power law with a low amount of absorption gives a good fit for the
spectrum ($\Gamma$ = 1.8 $\pm$ 0.2, $N_{\rm H}$ $\leq$ 1.2 $\times$ 10$^{21}$
cm$^{-2}$). The \xmm and \cha fluxes are in good agreement. The radio 
nucleus has been detected and it is possibly variable (HU01).  \\

%***Massimo: below, I inserted ``in radius'' after 40'' -- please check.

{\it NGC 4579 (S1.5/L1.5)}: 
This object has been observed by \cha for $\sim$35~ks
(Eracleous et al. 2002) and $\sim$3~ks (Ho et al. 2001, Terashima \&
Wilson 2003). \cha images show a hard compact nucleus
surrounded by soft diffuse emission which extends for $\sim 40''$
in radius. The \xmm spectrum is best fitted with a soft thermal component 
($kT \approx 0.6$~keV), a power law with $\Gamma$ $\approx$ 1.8, plus 
two iron Gaussian lines, one at $\sim$6.4~keV with EW $\approx$ 170 $\pm$ 
50~eV and one at $\sim$6.85~keV with EW $\approx$ 140 $\pm$ 60~eV.
We find no absorption in excess of the Galactic value.
The \xmm fluxes and luminosities are in good agreement
with the Eracleous et al. (2002) and Ho et al. (2001) results. 
A radio core is detected (HU01). \\

{\it NGC 4639 (S1.5)}: 
The \xmm image shows a bright nucleus surrounded by weak diffuse emission 
which is less prominent in the 2--10~keV image. This is similar to what 
was obtained with the \cha images (Ho et al. 2001), thus excluding 
contamination from unresolved sources. The \xmm spectrum is well fitted 
by a simple power law with $\Gamma$ $\approx$ 1.8 and no absorption in 
excess of the Galactic value.  The \xmm observation yields a hard X-ray 
flux ($F_{2-10~\rm keV}$ $\approx$ 2.3 $\times$ 10$^{-12}$ erg s$^{-1}$ 
cm$^{-2}$) about two times higher than the $ASCA$ flux (Ho et al. 1999, 
Terashima et al. 2002).
The source is marginally detected in the radio band (HU01).  \\

%The 0.5-10 keV
%EPIC spectrum has been extracted from a region of 25". It is described
%by an unabsorbed power-law ($\Gamma$ = 1.8$\pm$0.05) .  
%The spectral shape is in good agreement with what found with ASCA, 
%but the 2-10 keV flux is $\sim$ 3 times lower (Ho et al. 1999). 
%The 4' radius of ASCA observation are contaminated by the presence 
%of two ULXs (Foschini et al. 2002). 

{\it NGC 4698 (S2)}: The X-ray nuclear region of this object has been
studied in detail using \cha data which reveal a few 
possible ULXs within NGC 4698. Two of these are as close as $\sim 30''$
from the optical-radio nuclear position (Georgantopoulos \& Zesas 2003). 
The \xmm images show a weak nucleus. Its spectrum was extracted from a 
region of radius $25''$ to exclude any possible contamination from the 
nearest two ULXs, which are marginally detected. The \xmm flux is in 
good agreement with the \cha one. A power-law fit with $\Gamma$ = 2.0 
$\pm$ 0.2 gives a good fit to the \xmm data. We measured very low intrinsic 
absorption ($N_{\rm H}$ $\leq$ 4 $\times$ 10$^{21}$ cm$^{-2}$).
NGC 4698 is marginally detected in the radio band (HU01).  \\

%***Massimo: Below (N4725), the uncertainty in Gamma is given as 0.3,
%whereas in Table 3 it is given as 0.5. Also, N_H is given here as <4,
%whereas in Table 3 it is <5. Which are correct? Fix.

{\it NGC 4725 (S2:)}: This source has not been detected in the radio band
(HU01), so we use the optical position to determine
the nuclear position. The \xmm image reveals the presence of a nuclear core
and of several nearby off-nuclear sources positioned at $\gsimeq 30''$ from 
the nucleus. The spectrum was extracted with a radius of $20''$ to avoid 
contamination from these off-nuclear sources. 
Despite the poor statistics, the spectrum is best fitted with two distinct
components: a soft thermal component with $kT = 0.3 \pm 0.1$~keV, and a
hard power-law component with $\Gamma$ = 1.9 $\pm$ 0.3 with absorption 
$\leq$ 4 $\times$ 10$^{22}$ cm$^{-2}$. The \xmm hard flux is a factor 
of $\sim$2 times lower than the \cha flux (Ho et al. 2001), indicating 
possible intrinsic variability of this object.\\

%The core
%emission shows weak structures and the nuclear brightness is comparable to
%that of an off-nuclear source positioned at 20" from the nucleus (class II). 
%Structures close to the
%nucleus are also visible from the EPIC images; we use an extraction
%radius of 20" to avoid contamination from off-nuclear sources.

{\it NGC 5033 (S1.5)}: 
Despite the presence of very high background flares during the 
\xmm observation, the source is clearly detected. We use an extraction 
radius of $20''$ and a local background subtraction for the analysis. 
The spectrum is best modeled with a power law ($\Gamma$ = 1.7 $\pm$ 0.1),
with no absorption in excess of the Galactic value. We detect an iron line 
at 6.4~keV (with EW = 470 $\pm$ 215~eV). These results agree well with 
those from {\it ASCA} (Terashima et al. 1999, 2002).
A radio core is detected (HU01).\\

{\it NGC 5194 (S2)}: 
In the \xmm images, this galaxy (also called M51) shows a complex 
nuclear region characterized by extended features and off-nuclear 
sources in the soft band, but the nucleus is compact in the hard 
band. This is similar to, and consistent with, what is found in 
the \cha images by Terashima \& Wilson (2001).
The bright nucleus is seen in the optical position
coincident with the radio core position (HU01).
The soft emission has been modeled with a thermal plasma with 
$kT = 0.5 \pm 0.1$~keV, and the hard component with a very flat 
power law with photon index $\Gamma$ $\approx$ 0.6. The iron line 
detected at $\sim$6.4~keV is very strong, with EW $\approx$ 1~keV, 
which is a clear indication of the Compton-thick nature 
of this source. This has also been confirmed by a {\it BeppoSAX} 
observation of M51 which has shown that the nucleus is absorbed by 
a column density of 5.6 $\times$ 10$^{24}$ cm$^{-2}$ (Fukazawa et al. 
2001). \\

%***Massimo: Below (N5273), you give Gamma of 1.4, whereas in
% Table 3 it is listed as 1.5. Which one is correct? Fix.

{\it NGC 5273 (S1.5)}: 
The \xmm observation shows a bright compact nucleus in both 
the soft and hard energy bands. Higher angular resolution
\cha images reveal the presence of a compact core coincident 
with the radio position (HU01) and the presence of 
a comparatively weak off-nuclear source $20''$ from the nucleus.
The \xmm spectrum is properly fit by a power-law model with 
$\Gamma$ $\approx$ 1.5 and an absorption column of 0.9 $\times$ 
10$^{22}$ cm$^{-2}$, plus a soft thermal component with $kT \approx 
0.2$~keV, and an Fe~K line at 6.4~keV with EW = 226 $\pm $75~eV. 
There is also a marginal detection of an absorption edge at 
$\sim$7.8~keV, indicating the possible presence of an ionized 
absorber.\\

\end{document}